\documentclass[american,nofootinbib]{revtex4-2}
\usepackage{lmodern}

\usepackage[T1]{fontenc}
\usepackage{geometry}
\usepackage{babel}
\usepackage{enumitem}
\usepackage{amsmath}
\usepackage{amsthm}
\usepackage{amssymb}
\usepackage{dsfont}
\usepackage[normalem]{ulem}

\usepackage[unicode=true,pdfusetitle,
 bookmarks=true,bookmarksnumbered=true,bookmarksopen=true,bookmarksopenlevel=1,
 breaklinks=false,pdfborder={0 0 0},pdfborderstyle={},backref=false,colorlinks=false]
 {hyperref}
\usepackage{color}
\usepackage{xcolor}

\makeatletter
\theoremstyle{plain}
\newtheorem{prop}{\protect\propositionname}
\theoremstyle{plain}
\newtheorem{lem}{\protect\lemmaname}

\date{\today}

\makeatother

\providecommand{\lemmaname}{Lemma}
\providecommand{\propositionname}{Proposition}

\begin{document}
\title{Gauge invariant perturbations of static spatially compact LRS II spacetimes}
\author{Paulo Luz}
\email{paulo.luz@tecnico.ulisboa.pt}

\affiliation{Centro de Astrof\'{\i}sica e Gravita\c{c}\~{a}o - CENTRA, Departamento
de F\'{\i}sica, Instituto Superior T\'{e}cnico - IST, Universidade
de Lisboa - UL, Av. Rovisco Pais 1, 1049-001 Lisboa, Portugal,}
\affiliation{Departamento de Matem\'{a}tica, ISCTE - Instituto Universit\'{a}rio
de Lisboa, Portugal}
\author{Sante Carloni}
\email{sante.carloni@unige.it}

\affiliation{Institute of Theoretical Physics, Faculty of Mathematics and Physics,
Charles University, Prague, V Hole\v{s}ovi\v{c}k\'ach 2, 180 00
Prague 8, Czech Republic,}
\affiliation{DIME, Universit\`a di Genova, Via all'Opera Pia 15, 16145 Genova,
Italy,}
\affiliation{INFN Sezione di Genova, Via Dodecaneso 33, 16146 Genova, Italy}
\begin{abstract}
We present a framework to describe completely general first-order perturbations
of static, spatially compact, and locally rotationally symmetric class II spacetimes within
the theory of general relativity. The perturbation variables are by construction
covariant and identification gauge invariant and encompass the geometry and the thermodynamics of the fluid sources. The new equations are then applied to the study of isotropic, adiabatic perturbations.
We discuss how the choice of frame in which perturbations are described can significantly simplify the
mathematical analysis of the problem and show that it is possible to change frames directly from the linear
level equations. We find explicitly that the case of isotropic, adiabatic
perturbations can be reduced to a singular Sturm-Liouville eigenvalue problem, 
and lower bounds for the values of the eigenfrequencies can be derived. These
results lay the theoretical groundwork to analytically describe linear,
isotropic, and adiabatic perturbations of static, spherically symmetric
spacetimes.
\end{abstract}
\maketitle

\section{Introduction}
In the last few years, we have witnessed a true renaissance of relativistic astrophysics. The detection of gravitational waves \cite{LIGOScientific:2016aoc} and the first images of the shadows of supermassive black holes \cite{EventHorizonTelescope:2019dse} have brought a new wealth of data to the research community. It has, therefore, become paramount to devise new tools that allow for a clear interpretation of the latest data and deepen our understanding of relativistic astrophysical systems. 

Due to the fundamental nonlinear nature of the field equations of the theory of general relativity (GR), finding exact solutions for these systems, even assuming highly symmetric setups with simple source fields, is a formidable task. Indeed, although several exact solutions have been found in vacuum or in the presence of matter, many of the latter type of solutions are, in general, not suitable to accurately model physically meaningful setups, especially in the so-called strong field regime, where relativistic gravitational effects play a pivotal role in the dynamics of the matter fields.

To circumvent the limitations in the applicability of idealized solutions, various perturbative schemes were developed to linearize the field equations and study perturbations of exact solutions. Indeed, the study of perturbations of black holes has been an area of intense development in the past few decades. Combining the advent of numerical relativity techniques with the analytical results from linear perturbation theory for vacuum solutions ultimately allowed for identifying patterns in the data and detecting gravitational waves. On the other hand, in the case of compact stars, several open problems remain in developing analytic tools to study perturbations of this type of objects. Especially in the strong field regime, we have so far relied almost exclusively on numerical methods to evolve the full nonlinear equations of GR. These methods allow us to understand oscillations of compact stellar objects in great generality, but carry the inevitable limitation of purely numerical approaches. Hence, there is a need for analytic methods that could complement the numerics in a synergic way.

The problem of linear perturbations of massive astrophysical objects is markedly different from black hole perturbations. Black holes are described by vacuum solutions of the Einstein field equations. In contrast, compact stellar objects are described by solutions of GR with matter fluid sources characterized by perfect fluids with complex equations of state, fluids with non-trivial anisotropic pressure terms, or non-perfect multifluid models. Understandably, the evolution of the perturbations of these solutions is strongly tied to the properties of the matter fluid where, in general, anisotropies and momentum flows may be generated, even at a linear level, which then act as sources of shear and vorticity.

In 1964, by encoding the spacetime perturbations directly in the choice of gauge, Chandrasekhar derived for the first time in Refs.~\cite{Chandrasekhar_1964_PRL,Chandrasekhar_1964_ApJ} an equation to describe isotropic, adiabatic perturbations of static self-gravitating perfect fluids with a barotropic equation of state. In the 1990s, in Refs.~\cite{Chandrasekhar_Ferrari_1991a, Ipser_Price_1991}, structure equations were derived for nonradial perturbations of non-rotating perfect fluids. In Refs.~\cite{Carter_1973, Chandrasekhar_Ferrari_1991b, Kojima_1992, Ruoff_et_al_2001}, various extensions were proposed to study perturbations of slowly rotating stars.
However, understanding the perturbative properties of relativistic stars using those frameworks has remained challenging because of fundamental mathematical limitations and the underlying methods used by those approaches. For instance, the original radial pulsation equation by Chandrasekhar for isotropic, adiabatic perturbations relies on the introduction of auxiliary trial functions, making it impossible to assert the stability of the background solution unequivocally. Moreover, it was later noticed that the choice of gauge in Ref.~\cite{Chandrasekhar_Ferrari_1991a} leads to a higher-order system of equations when compared to the choice of another gauge because of an extraneous degree of freedom, adding unnecessary complication to the equations~\cite{Price_Ipser_1991}.
Several works have tried to improve these results in the following years, but all versions of those equations remain gauge-dependent. 

In geometric gravity theories, the identification gauge problem in perturbation theory arises because, even if the choice of perturbation variables is physically reasonable, those might depend on the mapping between the equilibrium and the perturbed manifold. Hence, their values and rates of change are ambiguous. Therefore, to develop a rigorous perturbative framework in geometric theories of gravity, the perturbation variables have to be methodically chosen \cite{Stewart_Walker_1974, Sonego_Bruni_1998}.

In the case of vanishing or slow rotation, a Locally Rotationally Symmetric (LRS) metric can be used to successfully describe the geometry of the spacetime for the interior of compact stellar objects. This class of spacetimes was first identified and classified according to their extra symmetries by Ellis \cite{Ellis:1966ta} and Ellis and Stewart \cite{Stewart_Ellis_1968} at the end of the 1960s and is characterized by a local rotational symmetry at every event. Many spacetimes of interest in astrophysics and cosmology belong to the LRS class. Examples are the Bianchi cosmologies~\cite{Ellis:1968vb}, the Lema\^{\i}tre-Tolman-Bondi spacetime~\cite{Lemaitre:1933gd,Tolman:1934za,Bondi:1947fta}, the Oppenheimer-Snyder spacetime~\cite{Oppenheimer:1939ue}, and the vacuum Schwarzschild spacetime.

Recently, a new formalism has been proposed that is especially suitable to deal with LRS spacetimes: the 1+1+2 covariant approach~\cite{Clarkson_Barrett_2003,Betschart_Clarkson_2004,Clarkson_2007}. This formalism, analogous to the better-known covariant 1+3 formalism extensively used in cosmology~\cite{Ellis:1998ct,Ellis:book, Luz_Lemos_2023}, is based on the procedure of covariant spacetime threading. It can also be considered a semi-tetradic approach to the description of spacetimes, which makes full use of the symmetries of the spacetime. The covariant approaches have two important properties: (i) they allow the maintenance of covariance at all stages of the calculations, and (ii) they enable the description of spacetimes in terms of well-defined physical quantities. 

Analyzing LRS spacetimes with the 1+1+2 formalism can reveal many important aspects of these spacetimes and their physical processes. For instance, in Refs.~\cite{Carloni_Vernieri_2018a, Carloni_Vernieri_2018b} it was found a covariant formulation of the Tolman-Oppenheimer-Volkoff equation for perfect fluids and fluids with non-trivial anisotropic pressure terms, allowing for the derivation of new solutions. In Ref.~\cite{Carloni_Rosa_2019}, using this formalism, it was proved a generalization of Derrick's theorem, showing that the conclusion of the original theorem holds independently of the geometric properties of the spacetime. In Ref.~\cite{Luz_Carloni_2019}, the 1+1+2 formalism was applied to the Einstein-Cartan theory to derive the general structure equations for static, isotropic spacetimes, which were then used to find the first known regular solutions suitable to model the interior of massive astrophysical objects. However, the power of covariant formalisms is truly revealed in the context of perturbation theory. Indeed, these approaches are the cornerstone for constructing a covariant, gauge-invariant theory of perturbations, which can be employed in many contexts. 
In particular, the 1+1+2 formalism has been used to describe tensor perturbations of Schwarzschild black holes~\cite{Clarkson_Barrett_2003}, to describe complex interaction between gravitational and electromagnetic degrees of freedom~\cite{Marklund_2003}, or to study cosmological perturbations~\cite{Bradley_et_el_2012, Tornkvist_Bradley_2019}.

The scope of this work is to construct a completely general covariant and gauge-invariant perturbation theory of non-vacuum, static, spatially compact LRS II spacetimes. Moreover, as a first application, we aim to describe the dynamics of adiabatic isotropic perturbations. Using the covariant nature of the equations, we will be able to describe the evolution of the perturbation equations from the point of view of an observer locally comoving with the fluid and one which is static with respect to an observer at spatial infinity. For both cases, we propose a method to find a family of exact solutions in the form of a power series for a wide variety of background solutions. We will also prove that the perturbation equations with appropriate boundary conditions constitute a singular Sturm-Liouville eigenvalue problem with a limit-point-non-oscillating endpoint, which, to our knowledge, has not been rigorously proven before. Using this property, we will be able to establish lower bounds for the absolute value of the fundamental eigenfrequency in terms of quantities of the background equilibrium solution. 

The paper is organized as follows: Section~\ref{sec:1p1p2_decomposition} summarizes the 1+1+2  formalism for a generic spacetime. In Section~\ref{sec:Linearized_equations}, we derive the linearized covariant gauge-invariant perturbation equations for a static, spatially compact LRSII background, and we perform a harmonic decomposition, in particular,
isolating even and odd perturbations. In Section~\ref{sec:Adiabatic-isotropic-perturbations}, we focus on adiabatic and isotropic
perturbations, writing the equations in a comoving frame and a static frame, and for each 
frame, we properly define the boundary value problem. Here, we show that the perturbation 
equations in the static frame form a Sturm-Liouville eigenvalue problem and derive lower 
bounds for the values of the fundamental eigenvalue. Last, we summarize and draw some 
conclusions in Section~\ref{Conclusions}. The paper also contains four appendices. In 
Appendix~\ref{Appendix:General_1p1p2_eqs}, we present the Einstein field equations in the 
language of the 1+1+2 formalism, and in 
Appendices~\ref{Appendix:Linear_eqs_angular_variables} and 
\ref{Appendix:Linear_eqs_dot-derivatives_variables} we display the linearized field 
equations for the two sets of gauge invariant quantities adopted in the body of the text. 
In Appendix~\ref{Appendix:Harmonics_properties}, we present the definitions and properties 
of the eigenfunctions of the covariant Laplace-Beltrami operator on 2-hypersurfaces. In 
Appendix~\ref{Appendix:isotropic frame transformations}, we discuss the effects on the 
equation of state of a generalized Lorentz boost between two frames.
 
Throughout the article, we will work in the geometrized unit system
where $8\pi G=c=1$, and consider the metric signature $(-+++)$.

\section{\label{sec:1p1p2_decomposition}The 1+1+2 decomposition}

To construct a general set of gauge-invariant, covariant equations
for the perturbations of locally rotationally symmetric class II
spacetimes, from hereon LRS II~\citep{Stewart_Ellis_1968}, we will
adopt the language of the 1+1+2 covariant formalism~\citep{Clarkson_Barrett_2003,Betschart_Clarkson_2004,Clarkson_2007}.
In this section, we will then introduce the basic quantities and conventions used throughout the article.

\subsection{Projectors and the Levi-Civita volume form}

Consider a Lorentzian manifold of dimension 4, $\left(\mathcal{M},g\right)$,
where $g$ represents the metric tensor, admitting in some open neighborhood
the existence of a congruence of timelike curves with tangent vector
field $u$. We will assume that the congruence to be affinely parameterized
and $u_{\alpha}u^{\alpha}=-1$. Without loss of generality, we can locally
foliate the manifold in 3-surfaces, $V$, orthogonal at each point
to the curves of the congruence, such that all tensor quantities are
defined by their behavior along the direction of $u$ and in $V$.
This procedure is usually called 1+3 spacetime decomposition. Such
decomposition of the spacetime manifold relies on the existence of
a pointwise projector to the cotangent space of $V$, which can be
naturally defined as 
\begin{equation}
h_{\alpha\beta}=g_{\alpha\beta}+u_{\alpha}u_{\beta}\,,\label{Def_eq:projector_h_definition}
\end{equation}
where $g_{\alpha\beta}$ represents the components of the metric tensor
in some local coordinate system, with the following properties
\begin{equation}
\begin{aligned}h_{\alpha\beta} & =h_{\beta\alpha}\,, &  &  & h_{\alpha\beta}h^{\beta\gamma} & =h_{\alpha}{}^{\gamma}\,,\\
h_{\alpha\beta}u^{\alpha} & =0\,, &  &  & h_{\alpha}{}^{\alpha} & =3\,.
\end{aligned}
\end{equation}

The 1+1+2 decomposition builds from the 1+3 decomposition by defining
a congruence of spacelike curves with tangent vector field $e$ such
that any tensor quantity defined in the submanifold $V$ is defined
by its behavior along $e$ and the 2-surfaces $W$, orthogonal to
both $u$ and $e$ at each point. We shall refer to each surface $W$
as ``sheet''. We will consider that the spacelike congruence is
affinely parameterized and $e_{\alpha}e^{\alpha}=1$. We can then
define a projector onto $W$ by
\begin{equation}
N_{\alpha\beta}=h_{\alpha\beta}-e_{\alpha}e_{\beta}\,,\label{Def_eq:projector_N_definition}
\end{equation}
verifying

\begin{equation}
\begin{aligned}N_{\alpha\beta} & =N_{\beta\alpha}\,, &  &  & N_{\alpha\beta}N^{\beta\gamma} & =N_{\alpha}{}^{\gamma}\,,\\
N_{\alpha\beta}u^{\alpha} & =N_{\alpha\beta}e^{\alpha}=0\,, &  &  & N_{\alpha}{}^{\alpha} & =2\,.
\end{aligned}
\label{Def_eq:projector_N_properties}
\end{equation}

It is useful to introduce the following tensors derived from the covariant
Levi-Civita tensor $\varepsilon_{\alpha\beta\gamma\sigma}$,
\begin{equation}
\begin{aligned}\varepsilon_{\alpha\beta\gamma} & =\varepsilon_{\alpha\beta\gamma\sigma}u^{\sigma}\,,\\
\varepsilon_{\alpha\beta} & =\varepsilon_{\alpha\beta\gamma}e^{\gamma}\,,
\end{aligned}
\label{Def_eq:volume_forms}
\end{equation}
 with the following properties
\begin{equation}
\begin{aligned}\varepsilon_{\alpha\beta\gamma} & =\varepsilon_{\left[\alpha\beta\gamma\right]}\,, &  &  & \varepsilon_{\alpha\beta} & =\varepsilon_{\left[\alpha\beta\right]}\,,\\
\varepsilon_{\alpha\beta\gamma}u^{\gamma} & =0\,, &  &  & \varepsilon_{\alpha\beta}u^{\alpha} & =\varepsilon_{\alpha\beta}e^{\alpha}=0\,,\\
\varepsilon_{\alpha\beta\gamma}\varepsilon^{\mu\nu\sigma} & =6h^{\mu}{}_{\left[\alpha\right.}h^{\nu}{}_{\beta}h_{\left.\gamma\right]}{}^{\sigma} &  &  & \varepsilon_{\alpha\beta}\varepsilon^{\mu\nu} & =N^{\mu}{}_{\alpha}N^{\nu}{}_{\beta}-N^{\mu}{}_{\beta}N^{\nu}{}_{\alpha}\\
\varepsilon_{\alpha\beta\gamma}\varepsilon^{\mu\nu\gamma} & =h_{\alpha}{}^{\mu}h_{\beta}{}^{\nu}-h_{\beta}{}^{\mu}h_{\alpha}{}^{\nu}\,, &  &  & \varepsilon_{\alpha}{}^{\gamma}\varepsilon_{\beta\gamma} & =N_{\alpha\beta}\,,\\
\varepsilon_{\alpha\mu\nu}\varepsilon^{\beta\mu\nu} & =2h_{\alpha}{}^{\beta}\,, &  &  & \varepsilon_{\alpha\beta\gamma} & =e_{\alpha}\varepsilon_{\beta\gamma}-e_{\beta}\varepsilon_{\alpha\gamma}+e_{\gamma}\varepsilon_{\alpha\beta}\,,
\end{aligned}
\label{Def_eq:volume_forms_properties}
\end{equation}
where in the right-hand side of the relation for $\varepsilon_{\alpha\beta\gamma}\varepsilon^{\mu\nu\sigma}$
the anti-symmetrization is to be considered on all, and only, the lower
indices. We will adopt the convention to indicate the symmetric and
anti-symmetric part of a tensor using parentheses and brackets, such
that for a 2-tensor $\chi$
\begin{equation}
\begin{aligned}\chi_{\left(\alpha\beta\right)} & =\frac{1}{2}\left(\chi_{\alpha\beta}+\chi_{\beta\alpha}\right)\,,\\
\chi_{\left[\alpha\beta\right]} & =\frac{1}{2}\left(\chi_{\alpha\beta}-\chi_{\beta\alpha}\right)\,.
\end{aligned}
\end{equation}

\subsection{\label{subsec:Covariant_derivatives_decomposition}Covariant derivatives
of $u$ and $e$}

\subsubsection{Decomposition of the covariant derivatives on the sheet $W$}

Using the definitions of the projector operators onto the surfaces
$V$ and $W$, the covariant derivatives of the tangent vector fields
$u$ and $e$ can be, at each point, uniquely decomposed in their
components along $u$, $e$ and in $W$.\footnote{Although it is an abuse of language, we will refer to the tensor fields $h$
and $N$ as projectors onto the space $V$ or $W$, being implicit that we mean that
the projection is onto the respective tangent and cotangent spaces.}

For notational convenience, given a tensor quantity $\chi$, throughout the article, we will use the compact notation
	\begin{equation}
		\begin{aligned}
			\dot{\chi} & :=u^{\mu} \nabla_\mu \chi\,, &  &  & \widehat{\chi} & :=e^{\mu} \nabla_\mu \chi\,,
		\end{aligned}
	\end{equation}
	to represent the derivative along the integral curves of the vector field $u$ and the derivative along the integral curves of the vector field $e$, respectively.

The covariant derivatives of the tensor fields $u$ and $e$ on the
sheet can be uniquely decomposed as

\begin{equation}
\delta_{\alpha}u_{\beta}\equiv N_{\alpha}{}^{\mu}N_{\beta}{}^{\nu}\nabla_{\mu}u_{\nu}=\frac{1}{2}N_{\alpha\beta}\tilde{\theta}+\Sigma_{\alpha\beta}+\varepsilon_{\alpha\beta}\Omega\,,\label{Def_eq:delta_operator_definition}
\end{equation}
where
\begin{equation}
\begin{aligned}\tilde{\theta} & =\delta_{\alpha}u^{\alpha}\,,\\
\Sigma_{\alpha\beta} & =\delta_{\{\alpha}u_{\beta\}}\,,\\
\Omega & =\frac{1}{2}\varepsilon^{\mu\nu}\delta_{\mu}u_{\nu}\,,
\end{aligned}
\label{Def_eq:W_projected_cov_u_quantities}
\end{equation}
and
\begin{equation}
\delta_{\alpha}e_{\beta}=\frac{1}{2}N_{\alpha\beta}\phi+\zeta_{\alpha\beta}+\varepsilon_{\alpha\beta}\xi\,,\label{Def_eq:W_projected_cov_space-like_e}
\end{equation}
with
\begin{equation}
\begin{aligned}\phi & =\delta_{\mu}e^{\mu}\,,\\
\zeta_{\alpha\beta} & =\delta_{\{\alpha}e_{\beta\}}\,,\\
\xi & =\frac{1}{2}\varepsilon^{\mu\nu}\delta_{\mu}e_{\nu}\,,
\end{aligned}
\label{Def_eq:W_projected_cov_e_quantities}
\end{equation}
where the curly brackets represent the projected symmetric part without
trace of a tensor in $W$, that is, for a tensor $\chi_{\alpha\beta}$,
\begin{equation}
\chi_{\left\{ \alpha\beta\right\} }=\left[N^{\mu}{}_{(\alpha}N_{\beta)}{}^{\nu}-\frac{N_{\alpha\beta}}{2}N^{\mu\nu}\right]\chi_{\mu\nu}\,.\label{Def_eq:curly_notation_definition}
\end{equation}
Using the 2-form $\varepsilon_{\alpha\beta}$, a completely anti-symmetric
tensor defined on the sheet, $\chi_{\alpha\beta}=\chi_{\left[\alpha\beta\right]}=N^{\mu}{}_{[\alpha}N_{\beta]}{}^{\nu}\chi_{\mu\nu}$,
can be written as 
\begin{equation}
\chi_{\left[\alpha\beta\right]}=\varepsilon_{\alpha\beta}\left(\frac{1}{2}\varepsilon^{\mu\nu}\chi_{\mu\nu}\right)\,.
\end{equation}

\subsubsection{Decomposition of the covariant derivatives on $V$}

The decomposition of the projected covariant derivatives of $u$ onto
$V$ is given by
\begin{equation}
D_{\alpha}u_{\beta}=h_{\alpha}{}^{\mu}h_{\beta}{}^{\nu}\nabla_{\mu}u_{\nu}=\frac{1}{3}h_{\alpha\beta}\theta+\sigma_{\alpha\beta}+\omega_{\alpha\beta}\,,\label{Def_eq:H_decomposition_cov_u}
\end{equation}
with 
\begin{align}
\theta & =h^{\mu\nu}D_{\mu}u_{\nu}\,,\label{Def_eq:H_decomposition_cov_expansion}\\
\sigma_{\alpha\beta} & =D_{\left\langle \alpha\right.}u_{\left.\beta\right\rangle }\,,\\
\omega_{\alpha\beta} & =h^{\mu}{}_{[\alpha}h_{\beta]}{}^{\nu}D_{\mu}u_{\nu}\,,\label{Def_eq:H_decomposition_cov_vorticity}
\end{align}
where we used the angular brackets to represent the projected symmetric
part without trace of a tensor on $V$, that is, for a tensor, $\chi_{\alpha\beta}$,
\begin{equation}
\chi_{\left\langle \alpha\beta\right\rangle }=\left[h^{\mu}{}_{(\alpha}h_{\beta)}{}^{\nu}-\frac{h_{\alpha\beta}}{3}h^{\mu\nu}\right]\chi_{\mu\nu}\,.\label{Def_eq:angular_brackets_definition}
\end{equation}
The scalar and tensor quantities in Eqs.~(\ref{Def_eq:H_decomposition_cov_expansion})--(\ref{Def_eq:H_decomposition_cov_vorticity})
can themselves be further decomposed in their contributions exclusively
on $W$ and along $e$, such that
\begin{equation}
\theta=\tilde{\theta}+\vartheta\,,
\end{equation}
where $\tilde{\theta}$ is defined in Eq.~(\ref{Def_eq:W_projected_cov_u_quantities})
and
\begin{equation}
\vartheta=-u^{\mu}\left(e^{\nu}D_{\nu}e_{\mu}\right)=-u^{\mu}\,\widehat{e}_{\mu}\,;
\end{equation}
\begin{equation}
\sigma_{\alpha\beta}=\Sigma_{\alpha\beta}+2\Sigma_{(\alpha}e_{\beta)}+\Sigma\left(e_{\alpha}e_{\beta}-\frac{1}{2}N_{\alpha\beta}\right)\,,
\end{equation}
with
\begin{equation}
\begin{aligned}\Sigma_{\alpha\beta} & =\sigma_{\left\{ \alpha\beta\right\} }\,,\\
\Sigma_{\alpha} & =N_{\alpha}{}^{\mu}e^{\nu}\sigma_{\mu\nu}\,,\\
\Sigma & =e^{\mu}e^{\nu}\sigma_{\mu\nu}=-N^{\mu\nu}\sigma_{\mu\nu}\,,
\end{aligned}
\end{equation}
and
\begin{equation}
	\omega_{\alpha\beta}=\varepsilon_{\alpha\beta\mu}\left(\Omega e^{\mu}+\Omega^{\mu}\right)\,,
\end{equation}
where $\Omega$ is given in Eq.~(\ref{Def_eq:W_projected_cov_u_quantities})
and
\begin{equation}
\Omega^{\alpha}=\frac{1}{2}N_{\gamma}{}^{\alpha}\varepsilon^{\mu\nu\gamma}D_{\mu}u_{\nu}\,.
\end{equation}

The quantities $\theta$, $\Sigma$, $\tilde{\theta}$ and $\vartheta$
are not independent, in fact: 
\begin{align}
\tilde{\theta} & =\frac{2}{3}\theta-\Sigma\,,\\
\vartheta & =\frac{1}{3}\theta+\Sigma\,;
\end{align}
as such, when setting up the 1+1+2 formalism, only two of those quantities are chosen.
The convention followed here uses the variables $\theta$ and $\Sigma$.

For the projected covariant derivative of the vector field $e$ on
$V$ we have 
\begin{equation}
D_{\alpha}e_{\beta}=h_{\alpha}{}^{\mu}h_{\beta}{}^{\nu}\nabla_{\mu}e_{\nu}=\delta_{\alpha}e_{\beta}+e_{\alpha}a_{\beta}\,,
\end{equation}
where $\delta_{\alpha}e_{\beta}$ is given by Eq.~(\ref{Def_eq:W_projected_cov_space-like_e})
and 
\begin{equation}
a_{\alpha}=e^{\mu}D_{\mu}e_{\alpha}=\widehat{e}_{\alpha}\,.
\end{equation}

\subsubsection{Decomposition of the covariant derivatives on the full manifold}

Finally, we can decompose the total covariant derivatives of $u^{\alpha}$
and $e^{\alpha}$, such that
\begin{equation}
\nabla_{\alpha}u_{\beta}=-u_{\alpha}\left(\mathcal{A}e_{\beta}+\mathcal{A}_{\beta}\right)+D_{\alpha}u_{\beta}\,,
\end{equation}
with
\begin{equation}
\begin{aligned}\mathcal{A} & =-u_{\mu}u^{\nu}\nabla_{\nu}e^{\mu}=-u_{\mu}\dot{e}^{\mu}\,,\\
\mathcal{A}_{\alpha} & =N_{\alpha\mu}\dot{u}^{\mu}\,,
\end{aligned}
\label{Def_eq:u_acceleration_cov_quantities}
\end{equation}
and
\begin{align}
\nabla_{\alpha}e_{\beta} & =D_{\alpha}e_{\beta}-u_{\alpha}\alpha_{\beta}-\mathcal{A}u_{\alpha}u_{\beta}+\left(\frac{1}{3}\theta+\Sigma\right)e_{\alpha}u_{\beta}+\left(\Sigma_{\alpha}-\varepsilon_{\alpha\mu}\Omega^{\mu}\right)u_{\beta}\,,
\end{align}
where
\begin{equation}
\alpha_{\alpha}=h_{\alpha}{}^{\mu}\dot{e}_{\mu}\,.
\end{equation}

\subsection{Covariant derivatives of the projectors and the Levi-Civita volume
form}

Gathering the results in subsection~\ref{subsec:Covariant_derivatives_decomposition},
the covariant derivatives of the tangent vector fields $u$ and $e$
can be written as

\begin{equation}
\begin{aligned}\delta_{\alpha}u_{\beta}=N_{\alpha}{}^{\mu}N_{\beta}{}^{\nu}\nabla_{\mu}u_{\nu} & =N_{\alpha\beta}\left(\frac{1}{3}\theta-\frac{1}{2}\Sigma\right)+\Sigma_{\alpha\beta}+\varepsilon_{\alpha\beta}\Omega\,,\\
D_{\alpha}u_{\beta}=h_{\alpha}{}^{\mu}h_{\beta}{}^{\nu}\nabla_{\mu}u_{\nu} & =\delta_{\alpha}u_{\beta}+\left(\frac{1}{3}\theta+\Sigma\right)e_{\alpha}e_{\beta}+2\Sigma_{(\alpha}e_{\beta)}-\varepsilon_{\alpha\mu}\Omega^{\mu}e_{\beta}+e_{\alpha}\varepsilon_{\beta\mu}\Omega^{\mu}\,,\\
\nabla_{\alpha}u_{\beta} & =D_{\alpha}u_{\beta}-u_{\alpha}\left(\mathcal{A}e_{\beta}+\mathcal{A}_{\beta}\right)\,,
\end{aligned}
\label{Def_eq:Cov_dev_vector_u}
\end{equation}
and 
\begin{equation}
\begin{aligned}\delta_{\alpha}e_{\beta}=N_{\alpha}{}^{\mu}N_{\beta}{}^{\nu}\nabla_{\mu}e_{\nu}= & \frac{1}{2}N_{\alpha\beta}\phi+\zeta_{\alpha\beta}+\varepsilon_{\alpha\beta}\xi\,,\\
D_{\alpha}e_{\beta}=h_{\alpha}{}^{\mu}h_{\beta}{}^{\nu}\nabla_{\mu}e_{\nu}= & \delta_{\alpha}e_{\beta}+e_{\alpha}a_{\beta}\,,\\
\nabla_{\alpha}e_{\beta}= & D_{\alpha}e_{\beta}-u_{\alpha}\alpha_{\beta}-\mathcal{A}u_{\alpha}u_{\beta}+\left(\frac{1}{3}\theta+\Sigma\right)e_{\alpha}u_{\beta}+\left(\Sigma_{\alpha}-\varepsilon_{\alpha\mu}\Omega^{\mu}\right)u_{\beta}\,.
\end{aligned}
\label{Def_eq:Cov_dev_vector_e}
\end{equation}

From these relations, we find the following expressions for the covariant
derivatives of the projector tensors:
\begin{equation}
\begin{aligned}\delta_{\alpha}N_{\beta\gamma} & =0\,, &  &  & \nabla_{\rho}\varepsilon_{\alpha\beta\gamma\delta} & =0\,,\\
e^{\mu}D_{\mu}N_{\alpha\beta} & =-2e_{(\alpha}a_{\beta)}\,, &  &  & u^{\mu}\nabla_{\mu}\varepsilon_{\alpha\beta\gamma} & =\varepsilon_{\alpha\beta\gamma\mu}\left(\mathcal{A}e^{\mu}+\mathcal{A}^{\mu}\right)\,,\\
u^{\mu}\nabla_{\mu}N_{\alpha\beta} & =2u_{(\alpha}\mathcal{A}_{\beta)}-2e_{(\alpha}\alpha_{\beta)}\,, &  &  & D_{\alpha}\varepsilon_{\beta\gamma\delta} & =0\,,\\
D_{\alpha}h_{\beta\gamma} & =0\,, &  &  & u^{\mu}\nabla_{\mu}h_{\alpha\beta} & =2u_{(\alpha|}\left(\mathcal{A}e_{|\beta)}+\mathcal{A}_{|\beta)}\right)\,.
\end{aligned}
\end{equation}

\subsection{Weyl and stress-energy tensors}

For the Levi-Civita connection, the Riemann tensor can be defined
by the Ricci identity valid for an arbitrary 1-form $\chi$:
\begin{equation}
R_{\alpha\beta\delta}{}^{\rho}\chi{}_{\rho}=\left(\nabla_{\alpha}\nabla_{\beta}-\nabla_{\beta}\nabla_{\alpha}\right)\chi_{\delta}\,.\label{Def_eq:Ricci_identity}
\end{equation}

In the case of a manifold of dimension 4, the components of the Riemann
curvature tensor, $R_{\alpha\beta\gamma\delta}$, can be written as
the following sum
\begin{equation}
R_{\alpha\beta\gamma\delta}=C_{\alpha\beta\gamma\delta}+R_{\alpha\left[\gamma\right.}g_{\left.\delta\right]\beta}-R_{\beta\left[\gamma\right.}g_{\left.\delta\right]\alpha}-\frac{1}{3}R\,g_{\alpha\left[\gamma\right.}g_{\left.\delta\right]\beta}\,,
\label{Def_eq:Weyl_tensor_definition}
\end{equation}
where $C_{\alpha\beta\gamma\delta}$ represent the components of the Weyl tensor, $R_{\alpha\beta}:=R_{\alpha\mu\beta}{}^{\mu}$
the components of the Ricci tensor and $R$ the Ricci scalar. The Weyl tensor plays
a pivotal role in relativistic gravity, describing the tidal forces
and the properties of gravitational waves. Remarkably, the Weyl 4-tensor
can be completely characterized by two 2-tensors, defined as
\begin{align}
E_{\alpha\beta} & =C_{\alpha\mu\beta\nu}u^{\mu}u^{\nu}\,,\label{Def_eq:Weyl_tensor_electric}\\
H_{\alpha\beta} & =\frac{1}{2}\varepsilon_{\alpha}{}^{\mu\nu}C_{\mu\nu\beta\delta}u^{\delta}\,,
\end{align}
respectively referred as the ``electric'' and ``magnetic'' part
of the Weyl tensor, both symmetric and traceless tensors, such that
\begin{equation}
C_{\alpha\beta\gamma\delta}=-\varepsilon_{\alpha\beta\mu}\varepsilon_{\gamma\delta\nu}E^{\nu\mu}-2u_{\alpha}E_{\beta\left[\gamma\right.}u_{\left.\delta\right]}+2u_{\beta}E_{\alpha\left[\gamma\right.}u_{\left.\delta\right]}-2\varepsilon_{\alpha\beta\mu}H^{\mu}{}_{\left[\gamma\right.}u_{\left.\delta\right]}-2\varepsilon_{\mu\gamma\delta}H^{\mu}{}_{\left[\alpha\right.}u_{\left.\beta\right]}\,.\label{Def_eq:Weyl_tensor_1+3_decomposition}
\end{equation}
In the 1+1+2 spacetime decomposition formalism, the components of
the Weyl tensor are given by
\begin{equation}
\begin{aligned}E_{\alpha\beta} & =\mathcal{E}\left(e_{\alpha}e_{\beta}-\frac{1}{2}N_{\alpha\beta}\right)+\mathcal{E}_{\alpha}e_{\beta}+e_{\alpha}\mathcal{E}_{\beta}+\mathcal{E}_{\alpha\beta}\,,\\
H_{\alpha\beta} & =\mathcal{H}\left(e_{\alpha}e_{\beta}-\frac{1}{2}N_{\alpha\beta}\right)+\mathcal{H}_{\alpha}e_{\beta}+e_{\alpha}\mathcal{H}_{\beta}+\mathcal{H}_{\alpha\beta}\,,
\end{aligned}
\end{equation}
where
\begin{equation}
\begin{aligned}\mathcal{E} & =E_{\mu\nu}e^{\mu}e^{\nu}=-N^{\mu\nu}E_{\mu\nu}\,, &  &  & \mathcal{H} & =e^{\mu}e^{\nu}H_{\mu\nu}=-N^{\mu\nu}H_{\mu\nu}\,,\\
\mathcal{E}_{\alpha} & =N_{\alpha}{}^{\mu}e^{\nu}E_{\mu\nu}=e^{\mu}N_{\alpha}{}^{\nu}E_{\mu\nu}\,, &  &  & \mathcal{H}_{\alpha} & =N_{\alpha}{}^{\mu}e^{\nu}H_{\mu\nu}=e^{\mu}N_{\alpha}{}^{\nu}H_{\mu\nu}\,,\\
\mathcal{E}_{\alpha\beta} & =E_{\left\{ \alpha\beta\right\} }\,, &  &  & \mathcal{H}_{\alpha\beta} & =H_{\left\{ \alpha\beta\right\} }\,.
\end{aligned}
\label{Def_eq:Weyl_tensor_components_definition}
\end{equation}

We shall also decompose  the metric stress-energy tensor, with components
$T_{\alpha\beta}$ in a local coordinate system, in terms of its pointwise projections onto $u$, $e$, and $W$:
\begin{equation}
T_{\alpha\beta}=\mu\,u_{\alpha}u_{\beta}+\left(p+\Pi\right)e_{\alpha}e_{\beta}+\left(p-\frac{1}{2}\Pi\right)N_{\alpha\beta}+2Qe_{\left(\alpha\right.}u_{\left.\beta\right)}+2Q_{\left(\alpha\right.}u_{\left.\beta\right)}+2\Pi_{\left(\alpha\right.}e_{\left.\beta\right)}+\Pi_{\alpha\beta}\,,\label{Def_eq:Stress-energy_tensor_decomposition}
\end{equation}
with
\begin{equation}
\begin{aligned}\mu & =u^{\mu}u^{\nu}T_{\mu\nu}\,, & Q_{\alpha} & =-N_{\alpha}{}^{\mu}u^{\nu}T_{\mu\nu}=-u^{\mu}N_{\alpha}{}^{\nu}T_{\mu\nu}\,, \\
p & =\frac{1}{3}\left(e^{\mu}e^{\nu}+N^{\mu\nu}\right)T_{\mu\nu}\,, & \Pi_{\alpha} & =N_{\alpha}{}^{\mu}e^{\nu}T_{\mu\nu}=e^{\mu}N_{\alpha}{}^{\nu}T_{\mu\nu}\,,\\
\Pi & =\frac{1}{3}T_{\alpha\beta}\left(2e^{\alpha}e^{\beta}-N^{\alpha\beta}\right)\,, & \Pi_{\alpha\beta} & =T_{\left\{ \alpha\beta\right\} }\,.\\
Q & =-e^{\mu}u^{\nu}T_{\mu\nu}=-u^{\mu}e^{\nu}T_{\mu\nu}\,,\\
\end{aligned}
\label{Def_eq:Energy_momentum_tensor_decomposition_quantities}
\end{equation}
Moreover, we have the following equations
\begin{equation}
\begin{aligned}p_{r} & =e^{\mu}e^{\mu}T_{\mu\nu}=p+\Pi\,,\\
p_{\perp} & =\frac{1}{2}N^{\mu\nu}T_{\mu\nu}=p-\frac{1}{2}\Pi\,,
\end{aligned}
\end{equation}
relating, respectively, the 1+1+2 variables $p$ and $\Pi$ with the
``radial'' and ``tangential'' components of the pressure.

\subsection{Commutation relations\label{subsec:Commutation-relations}}

Given a scalar field $\chi$ in $\left(\mathcal{M},g\right)$, using
the 1+1+2 spacetime decomposition, we have the following useful commutation
relations for its covariant derivatives:
\begin{equation}
\begin{aligned}\widehat{\dot{\chi}}-\dot{\hat{\chi}} & =\left(\frac{1}{3}\theta+\Sigma\right)\widehat{\chi}-\mathcal{A}\dot{\chi}+\left(\Sigma_{\mu}+\varepsilon_{\mu\nu}\Omega^{\nu}-\alpha_{\mu}\right)\delta^{\mu}\chi\,,\\
\delta_{\alpha}\dot{\chi}-N_{\alpha}{}^{\mu}\left(\delta_{\mu}\chi\right)^{\cdot} & =\left(\frac{1}{3}\theta-\frac{1}{2}\Sigma\right)\delta_{\alpha}\chi+\left(\Sigma_{\alpha\mu}+\varepsilon_{\alpha\mu}\Omega\right)\delta^{\mu}\chi+\left(\Sigma_{\alpha}-\varepsilon_{\alpha\mu}\Omega^{\mu}+\alpha_{\alpha}\right)\widehat{\chi}-\mathcal{A}_{\alpha}\dot{\chi}\,,\\
\delta_{\alpha}\widehat{\chi}-N_{\alpha}{}^{\mu}\left(\widehat{\delta_{\mu}\chi}\right) & =\frac{1}{2}\phi\delta_{\alpha}\chi+\left(\zeta_{\alpha}{}^{\mu}+\varepsilon_{\alpha}{}^{\mu}\xi\right)\delta_{\mu}\chi-2\varepsilon_{\alpha\mu}\Omega^{\mu}\dot{\chi}+\widehat{\chi}a_{\alpha}\,,\\
\delta_{\alpha}\delta_{\beta}\chi-\delta_{\beta}\delta_{\alpha}\chi & =2\varepsilon_{\alpha\beta}\Omega\dot{\chi}-2\varepsilon_{\alpha\beta}\xi\widehat{\chi}\,.
\end{aligned}
\label{1p1p2_eqs:Commutation_relations_scalars}
\end{equation}

In the case of the covariant derivatives of a 1-form field $\chi$ defined on the sheet, that is $\chi_{\alpha}=N_{\alpha}{}^{\mu}\chi_{\mu}$,
 the commutation relations take the form:
\begin{equation}
\begin{aligned}N_{\alpha}{}^{\mu}\widehat{\dot{\chi}_{\mu}}-N_{\alpha}{}^{\mu}\left(\widehat{\chi}{}_{\mu}\right)^{\cdot} & =\left(\frac{1}{3}\theta+\Sigma\right)N_{\alpha}{}^{\mu}\widehat{\chi}_{\mu}-\mathcal{A}N_{\alpha}{}^{\mu}\dot{\chi}{}_{\mu}+\chi_{\mu}\left(\Sigma^{\mu}+\varepsilon^{\mu\nu}\Omega_{\nu}\right)\mathcal{A}_{\alpha}\\
 & +\left(\Sigma^{\mu}+\varepsilon^{\mu\nu}\Omega_{\nu}-\alpha^{\mu}\right)\left(\delta_{\mu}\chi_{\alpha}\right)+\varepsilon_{\alpha}{}^{\mu}\chi_{\mu}\mathcal{H}\,,
\end{aligned}
\end{equation}
\begin{equation}
\begin{aligned}\delta_{\alpha}\dot{\chi}_{\beta}-N_{\alpha}{}^{\mu}N_{\beta}{}^{\nu}\left(\delta_{\mu}\chi{}_{\nu}\right)^{\cdot} & =\left(\Sigma_{\alpha}+\alpha_{\alpha}-\varepsilon_{\alpha\lambda}\Omega^{\lambda}\right)N_{\beta}{}^{\mu}\widehat{\chi}_{\mu}-\mathcal{A}_{\alpha}N_{\beta}{}^{\mu}\dot{\chi}{}_{\mu}-\frac{1}{2}N_{\alpha\beta}Q^{\mu}\chi_{\mu}+\mathcal{H}_{\alpha}\varepsilon_{\beta}{}^{\mu}\chi_{\mu}\\
 & +\left[N_{\alpha}{}^{\mu}\left(\frac{1}{3}\theta-\frac{1}{2}\Sigma\right)+\Sigma_{\alpha}{}^{\mu}+\varepsilon_{\alpha}{}^{\mu}\Omega\right]\left(\delta_{\mu}\chi_{\beta}\right)-\chi_{\mu}\left(\zeta_{\alpha}{}^{\mu}+\varepsilon_{\alpha}{}^{\mu}\xi\right)\alpha_{\beta}\\
 & +\chi_{\alpha}\left[\mathcal{A}_{\beta}\left(\frac{1}{3}\theta-\frac{1}{2}\Sigma\right)+\frac{1}{2}Q_{\beta}-\frac{1}{2}\phi\alpha_{\beta}\right]+\chi_{\mu}\left(\Sigma_{\alpha}{}^{\mu}+\varepsilon_{\alpha}{}^{\mu}\Omega\right)\mathcal{A}_{\beta}\,,
\end{aligned}
\end{equation}
\begin{equation}
	\begin{aligned}\delta_{\alpha}\widehat{\chi}_{\beta}-N_{\alpha}{}^{\mu}N_{\beta}{}^{\nu}\left(\widehat{\delta_{\mu}\chi_{\nu}}\right) & =a_{\alpha}N_{\beta}{}^{\mu}\widehat{\chi}_{\mu}-2\varepsilon_{\alpha\mu}\Omega^{\mu}N_{\beta}{}^{\nu}\dot{\chi}_{\nu}+\left(\frac{1}{2}N_{\alpha}{}^{\mu}\phi+\zeta_{\alpha}{}^{\mu}+\varepsilon_{\alpha}{}^{\mu}\xi\right)\delta_{\mu}\chi_{\beta}\\
	 & +\chi_{\alpha}\left[\left(\Sigma_{\beta}+\varepsilon_{\beta\mu}\Omega^{\mu}\right)\left(\frac{1}{3}\theta-\frac{1}{2}\Sigma\right)-\mathcal{E}_{\beta}-\frac{1}{2}\Pi_{\beta}-\frac{1}{2}\phi a_{\beta}\right]\\
	 & +\chi_{\mu}\left[\left(\Sigma_{\alpha}{}^{\mu}+\varepsilon_{\alpha}{}^{\mu}\Omega\right)\left(\Sigma_{\beta}+\varepsilon_{\beta\lambda}\Omega^{\lambda}\right)-\left(\zeta_{\alpha}{}^{\mu}+\varepsilon_{\alpha}{}^{\mu}\xi\right)a_{\beta}\right]\\
	 & +N_{\alpha\beta}\chi_{\mu}\left[\mathcal{E}^{\mu}+\frac{1}{2}\Pi^{\mu}-\left(\frac{1}{3}\theta-\frac{1}{2}\Sigma\right)\left(\Sigma^{\mu}+\varepsilon^{\mu\nu}\Omega_{\nu}\right)\right]\\
	 & -\chi_{\mu}\left(\Sigma_{\alpha\beta}+\varepsilon_{\alpha\beta}\Omega\right)\left(\Sigma^{\mu}+\varepsilon^{\mu\nu}\Omega_{\nu}\right)\,,
	\end{aligned}
\end{equation}
\begin{equation}
\begin{aligned}\delta_{\alpha}\delta_{\beta}\chi_{\gamma}-\delta_{\beta}\delta_{\alpha}\chi_{\gamma} & =2\varepsilon_{\alpha\beta}N_{\gamma}{}^{\mu}\left(\dot{\chi}_{\mu}\Omega-\widehat{\chi}_{\mu}\xi\right)+2\chi_{\left[\alpha\right.}N_{\left.\beta\right]\gamma}\left[\left(\frac{1}{3}\theta-\frac{1}{2}\Sigma\right)^{2}-\frac{1}{4}\phi^{2}+\mathcal{E}+\frac{1}{2}\Pi-\frac{1}{3}\left(\mu+\Lambda\right)\right]\\
 & +2\left[\left(\Sigma_{\left[\alpha\right|}{}^{\mu}+\varepsilon_{\left[\alpha\right|}{}^{\mu}\Omega\right)\left(\Sigma_{\left|\beta\right]\gamma}+\varepsilon_{\left|\beta\right]\gamma}\Omega\right)-\left(\zeta_{\left[\alpha\right|}{}^{\mu}+\varepsilon_{\left[\alpha\right|}{}^{\mu}\xi\right)\left(\zeta_{\left|\beta\right]\gamma}+\varepsilon_{\left|\beta\right]\gamma}\xi\right)\right]\chi_{\mu}\\
 & +2\left[\left(\frac{1}{3}\theta-\frac{1}{2}\Sigma\right)\left(\Sigma_{\left[\alpha\right|}{}^{\mu}+\varepsilon_{\left[\alpha\right|}{}^{\mu}\Omega\right)-\frac{1}{2}\phi\left(\zeta_{\left[\alpha\right|}{}^{\mu}+\varepsilon_{\left[\alpha\right|}{}^{\mu}\xi\right)-\frac{1}{2}\Pi_{\left[\alpha\right|}{}^{\mu}\right]N_{\left|\beta\right]\gamma}\chi_{\mu}\\
 & +2\chi_{\left[\alpha\right|}\left[\left(\frac{1}{3}\theta-\frac{1}{2}\Sigma\right)\left(\Sigma_{\left|\beta\right]\gamma}+\varepsilon_{\left|\beta\right]\gamma}\Omega\right)-\frac{1}{2}\phi\left(\zeta_{\left|\beta\right]\gamma}+\varepsilon_{\left|\beta\right]\gamma}\xi\right)-\frac{1}{2}\Pi_{\left|\beta\right]\gamma}\right]\,.
\end{aligned}
\end{equation}

\subsection{The set of 1+1+2 equations}

The 1+1+2 variables introduced in the previous subsections completely
describe the geometry of the spacetime manifold and the properties of the matter fields
that permeate it. Using the Ricci and Bianchi identities and the Einstein
field equations for the theory of General Relativity,
\begin{equation}
R_{\alpha\beta}-\frac{1}{2}g_{\alpha\beta}R+g_{\alpha\beta}\Lambda=T_{\alpha\beta}\,,
\end{equation}
where $\Lambda$ represents the cosmological constant, we can find
a set of evolution, propagation, and constraint equations for the 1+1+2
variables. In Appendix~\ref{Appendix:General_1p1p2_eqs} we give the general set of
these equations. Using the freedom of choice of frame to describe
the setup and providing an equation of state for the matter fields,
those equations form a closed system, describing the geometry of the
manifold and the dynamics of the permeating matter fields.

\section{Linearized equations for the perturbed spacetime}\label{sec:Linearized_equations}

\subsection{Background spacetime\label{subsec:Background_spacetime_static_LRSII_general}}

Having properly introduced the 1+1+2 covariant formalism,
we are now in a position to derive a set of covariant and gauge invariant
equations to describe linear order perturbations of a spacetime assumed
to be static, LRS II, and permeated by a general matter fluid. As is customary in relativistic perturbation theory,
throughout the article, we will refer to the unperturbed spacetime
as the ``background spacetime''.

For a proper choice of frame, in an LRS II spacetime, all covariantly
defined vector and tensor quantities of the 1+1+2 decomposition can
be made to vanish identically. If, in addition, the spacetime is static,
we can align the $u$ vector field with the timelike, hypersurface
orthogonal Killing vector field, such that all dot-derivatives of
the covariantly defined quantities are zero and the scalars $\left\{ \theta,\Sigma,\Omega,\xi,\mathcal{H},Q\right\} $
also vanish. Hence, a static LRS II background spacetime can be completely
characterized by the quantities $\left\{ \phi_{0},\mathcal{A}_{0},\mathcal{E}_{0},\mu_{0},p_{0},\Pi_{0},\Lambda\right\} $, which satisfy  the following equations
\begin{equation}
\begin{aligned}\mathcal{\widehat{A}}_{0} & =\frac{1}{2}\left(\mu_{0}+3p_{0}\right)-\Lambda-\mathcal{A}_{0}\left(\mathcal{A}_{0}+\phi_{0}\right)\,,\\
\widehat{\phi}_{0} & =-\frac{1}{2}\phi_{0}^{2}-\frac{2}{3}\left(\mu_{0}+\Lambda\right)-\frac{1}{2}\Pi_{0}-\mathcal{E}_{0}\,,\\
\widehat{p}_{0}+\widehat{\Pi}_{0} & =-\left(\frac{3}{2}\phi_{0}+\mathcal{A}_{0}\right)\Pi_{0}-\left(\mu_{0}+p_{0}\right)\mathcal{A}_{0}\,,\\
\widehat{\mathcal{E}}_{0}-\frac{1}{3}\widehat{\mu}_{0}+\frac{1}{2}\widehat{\Pi}_{0} & =-\frac{3}{2}\phi_{0}\left(\mathcal{E}_{0}+\frac{1}{2}\Pi_{0}\right)\,,
\end{aligned}
\label{eq:LRSII_back_general_eqs_1}
\end{equation}
and the constraint
\begin{align}
\mu_{0}+3p_{0}-2\Lambda-3\mathcal{A}_{0}\phi_{0} & =-\frac{3}{2}\Pi_{0}+3\mathcal{E}_{0}\,,\label{eq:LRSII_back_general_eqs_2}
\end{align}
where the subscript ``$0$'' from hereon will be used to refer to
the quantities that characterize the equilibrium configuration. The system is closed by providing an equation of state that relates the pressure components, $p_{0}$ and $\Pi_{0}$, with the energy density, $\mu_{0}$,
or some functional dependencies for these matter variables are imposed.

\subsection{Gauge invariant variables}

Following the Stewart-Walker lemma \citep{Stewart_Walker_1974}, to
write a set of equations for the linear perturbations that are identification
gauge invariant, we will consider perturbation variables that
either vanish in the background spacetime or are scalars with the same constant value in the background and the perturbed spacetimes (e.g., the cosmological constant). Since the background spacetime is assumed to be static LRS II, provided the choice of frame adopted in the previous subsection, all vector and tensor quantities can be used as identification gauge invariant perturbation quantities. Moreover, for such spacetimes, the scalars $\left\{ \theta,\Sigma,\Omega,\xi,\mathcal{H},Q\right\}$ are identically zero, hence are identification gauge invariant quantities. Then, we only have to find a set of variables
that vanish in the background spacetime, and that can be used to characterize
the linear perturbations of the remaining scalars, i.e. $\left\{ \mathcal{A},\phi,\mathcal{E},\mu,p,\Pi\right\} $.
 Given the symmetries of the background spacetime, two natural sets of variables can be adopted. The first one is given by
\begin{equation}
\begin{aligned}\mathbb{A}_{\alpha} & :=\delta_{\alpha}\mathcal{A}\,, & \mathbb{F}_{\alpha} & :=\delta_{\alpha}\phi\,, & \mathbb{E}_{\alpha} & :=\delta_{\alpha}\mathcal{E}\,,\\
\mathfrak{m}_{\alpha} & :=\delta_{\alpha}\mu\,, & \mathfrak{p}_{\alpha} & :=\delta_{\alpha}p\,, & \mathbb{P}_{a} & :=\delta_{\alpha}\Pi\,,
\end{aligned}
\label{eq:GI_angular_gradients_definition}
\end{equation}
which represent the gradients on the sheets of the various scalar quantities
that describe the background spacetime. The second one instead contains
\begin{equation}
\begin{aligned}\mathsf{A} & :=\dot{\mathcal{A}}\,, & \mathsf{F} & :=\dot{\phi}\,, & \mathsf{E} & :=\dot{\mathcal{E}}\,,\\
\mathsf{m} & :=\dot{\mu}\,, & \mathsf{p} & :=\dot{p}\,, & \mathcal{P} & :=\dot{\Pi}\,,
\end{aligned}
\label{eq:GI_dot_derivatives_definition}
\end{equation}
which represent the dot-derivatives of the various scalar quantities that describe
the spacetime.
In a static LRS II background, both the $\delta$-gradients
in Eq.~(\ref{eq:GI_angular_gradients_definition}), which are vectors,  and the quantities
in Eq.~(\ref{eq:GI_dot_derivatives_definition}), which are related to the proper-time variation, vanish. Therefore, the quantities of the sets above can be used
as gauge invariant variables that characterize the linear perturbations of a static LRS II spacetime.

In principle, we can use either set among  Eq.~\eqref{eq:GI_angular_gradients_definition} and \eqref{eq:GI_dot_derivatives_definition}.
Moreover, as shown in Eq.~\eqref{Var_Sets_Conn}, the above perturbation variables are not independent. Then, in general, we can also construct a set of variables containing elements of both Eq.~\eqref{eq:GI_angular_gradients_definition} and \eqref{eq:GI_dot_derivatives_definition}, provided that they form a closed system.
Note, however, that dependence does not imply equivalence. For example, the $\delta$-gradients variables
in Eq.~(\ref{eq:GI_angular_gradients_definition}) preserve the information
regarding the degrees of freedom on the sheet, however, if we only
perturb the background spacetime in directions parallel to the $u$
and $e$ vector fields, these variables by themselves are not suitable
to fully describe the perturbed spacetime, as they will remain identically zero. Conversely, the variables in Eq.~(\ref{eq:GI_dot_derivatives_definition}) can be used to fully characterize linear perturbations
of $\left\{ \mathcal{A},\phi,\mathcal{E},\mu,p,\Pi\right\} $, only
losing information in the case of constant-in-time perturbations.  Thus, this second set of variables can appear more convenient to analyze the dynamics of the perturbations. The drawback is that some of the linearized
1+1+2 equations for the variables~(\ref{eq:GI_dot_derivatives_definition})
are second order in time (cf.~Appendix~\ref{Appendix:Linear_eqs_dot-derivatives_variables}), introducing extra complexity to the problem. Then, in what follows, we will opt for using the $\delta$-gradients in Eq.~\eqref{eq:GI_angular_gradients_definition} only to describe vector and tensor perturbations, since those will be characterized by first-order equations. Conversely, we will use the variables in Eq.~\eqref{eq:GI_dot_derivatives_definition} to describe the scalar modes.

\subsection{The linearization procedure}

The procedure to find the set of gauge invariant structure equations for linear perturbations of static, LRS II spacetimes is relatively straightforward, although rather laborious. After choosing the gauge-invariant quantities from those in Eqs.~\eqref{eq:GI_angular_gradients_definition} or \eqref{eq:GI_dot_derivatives_definition}, we deduce the equations for these quantities by applying the projected derivative operators ``dot" and $\delta_\alpha$ to the  1+1+2 equations in Appendix~\ref{Appendix:General_1p1p2_eqs}. Then, using the commutation relations in subsection~\ref{subsec:Commutation-relations}, we express all gauge-dependent terms as a combination of gauge-independent quantities. Successively, the terms containing the products of two or more first-order quantities are discarded as higher order. We present in Appendices~\ref{Appendix:Linear_eqs_angular_variables} and \ref{Appendix:Linear_eqs_dot-derivatives_variables} the general set of gauge independent field equations, valid at linear level, for the perturbations for both the $\delta$-gradients and the dot-derivatives variables, respectively. As explained above, we will opt to use a mix of these equations to describe general linear perturbations. However, the set of equations in each appendix mentioned above is completely general and can be used by themselves.

\subsection{Harmonic decomposition\label{subsec:Harmonic-decomposition}}

The sets of the linearized equations in Appendices \ref{Appendix:Linear_eqs_angular_variables}
and \ref{Appendix:Linear_eqs_dot-derivatives_variables} form two
systems of partial differential equations. The appearance of $\delta$-derivatives makes the system particularly complicated
to integrate, such that finding solutions, even in relatively simple
setups, an intractable problem. Following Refs.~\citep{Clarkson_Barrett_2003,Betschart_Clarkson_2004,Clarkson_2007},
to transform this system into a system of ordinary differential equations
(ODEs) valid at linear level, we can use the fact that the background
spacetime is static and LRS II.

Using the eigenfunctions of the projected covariant Laplace-Beltrami
operator, $\delta^{2}\equiv\delta_{\mu}\delta^{\mu}$, locally defined on
the sheets of the LRSII background spacetime, we can make a harmonic
decomposition and write the various quantities that characterize the
perturbed spacetime as a linear combination of these eigenfunctions.
In Appendix~\ref{Appendix:Harmonics_properties}, we list the definitions
and various useful properties of the scalar, vector, and tensor harmonics
of the $\delta^{2}$ operator. Then, given some scalar, 1-tensor, or  symmetric,
traceless 2-tensor quantity defined on a sheet of the perturbed
spacetime, say $\chi$, $\chi_{\alpha}$ or $\chi_{\alpha\beta}$, at linear level
we can formally write these as the following infinite sum
\begin{align}\label{Harm_Dec_Sp}
\chi & =\Psi_{\chi}^{\left(k,S\right)}\mathcal{Q}^{\left(k\right)}\,,\nonumber \\
\chi_{\alpha} & =\Psi_{\chi}^{\left(k,V\right)}\mathcal{Q}_{\alpha}^{\left(k\right)}+\overline{\Psi}_{\chi}^{\left(k,V\right)}\bar{\mathcal{Q}}_{\alpha}^{\left(k\right)}\,,\\
\chi_{\alpha\beta} & =\Psi_{\chi}^{\left(k,T\right)}\mathcal{Q}_{\alpha\beta}^{\left(k\right)}+\overline{\Psi}_{\chi}^{\left(k,T\right)}\bar{\mathcal{Q}}_{\alpha\beta}^{\left(k\right)}\,,\nonumber 
\end{align}
where summation or integration in the Laplace-Beltrami eigenvalue $k$ (the so-called ``modes'' ) is assumed, depending on the
geometry of the sheets. We will use the compact notation $\Psi_{\chi}^{\left(k,S\right)}$
to refer to the harmonic coefficients associated with the eigenfunctions
$\mathcal{Q}^{\left(k\right)}$ of the scalar quantity $\chi$, $\Psi_{\chi}^{\left(k,V\right)}$
and $\overline{\Psi}_{\chi}^{\left(k,V\right)}$ to refer to the coefficients
associated with the eigenfunctions $\mathcal{Q}_{\alpha}^{\left(k\right)}$
or $\bar{\mathcal{Q}}_{\alpha}^{\left(k\right)}$ of the 1-tensor
quantity $\chi_{\alpha}$,
and $\Psi_{\chi}^{\left(k,T\right)}$ and $\overline{\Psi}_{\chi}^{\left(k,T\right)}$
to refer to the coefficients associated with the eigenfunctions $\mathcal{Q}_{\alpha\beta}^{\left(k\right)}$
or $\bar{\mathcal{Q}}_{\alpha\beta}^{\left(k\right)}$ of the 2-tensor
quantity $\chi_{\alpha\beta}$.

Additionally, since the background spacetime is assumed to be static, we can Fourier transform the harmonic coefficients of Eq.~\eqref{Harm_Dec_Sp}, explicitly factorizing their time dependence, writing them as a linear combination of the eigenfunctions of the Laplace operator in $\mathbb{R}$ (or some subset of it) for the appropriate boundary conditions. That is, we can write all coefficients for linear perturbations: $\Psi_{\chi}^{\left(k,I\right)}$, where $I=\left\{S, V, T \right\}$, as linear combinations of the eigenfunctions  $e^{i\upsilon\tau}$, where $\tau$ represents the proper time of an observer with 4-velocity $u$ and $\upsilon$ represent the eigenvalues of the Laplace operator in that manifold. The scalar $\upsilon$ takes discrete or continuous values depending on the chosen boundary conditions. To make this idea precise, consider the time-harmonic functions $T^{\left(\upsilon\right)}$ with the following properties
\begin{equation}
\begin{aligned}\dot{T}^{\left(\upsilon\right)} & =i\upsilon T^{\left(\upsilon\right)}\,,\\
\widehat{T}^{\left(\upsilon\right)}=\delta_{\alpha}T^{\left(\upsilon\right)} & =0\,,\\
\dot{\upsilon}=\delta_{\alpha}\upsilon & =0\,,
\end{aligned}
\end{equation}
where $i$ represents the imaginary unit. These properties imply that
$\widehat{\dot{T}}^{\left(\upsilon\right)}=-\mathcal{A}_{0}\dot{T}^{\left(\upsilon\right)}$,
from which we find 
\begin{equation}
\widehat{\upsilon}=-\mathcal{A}_{0}\upsilon\,.\label{eq:upsilon_hat_A_relation}
\end{equation}
Knowing the function $\mathcal{A}_{0}=\mathcal{A}_{0}\left(x^{\alpha}\right)$
in the background, where $\left\{ x^{\alpha}\right\} $ represent
some local coordinate system, Eq.~(\ref{eq:upsilon_hat_A_relation})
allow us to relate the eigenfrequencies $\upsilon$ defined with respect
to  $\tau$ and
the eigenfrequencies defined with respect to the $x^{0}$ time coordinate. 

Gathering these results, formally, we can expand any scalar, 1-tensor
or a symmetric, traceless 2-tensor characterizing first-order quantities
as
\begin{equation}
\begin{aligned}\chi & =\Psi_{\chi}^{\left(\upsilon,k,S\right)}\mathcal{Q}^{\left(k\right)}T^{\left(\upsilon\right)}\,,\\
\chi_{\alpha} & =\left(\Psi_{\chi}^{\left(\upsilon,k,V\right)}\mathcal{Q}_{\alpha}^{\left(k\right)}+\overline{\Psi}_{\chi}^{\left(\upsilon,k,V\right)}\bar{\mathcal{Q}}_{\alpha}^{\left(k\right)}\right)T^{\left(\upsilon\right)}\,,\\
\chi_{\alpha\beta} & =\left(\Psi_{\chi}^{\left(\upsilon,k,T\right)}\mathcal{Q}_{\alpha\beta}^{\left(k\right)}+\overline{\Psi}_{\chi}^{\left(\upsilon,k,T\right)}\bar{\mathcal{Q}}_{\alpha\beta}^{\left(k\right)}\right)T^{\left(\upsilon\right)}\,,
\end{aligned}
\end{equation}
where either discrete sums or multiple integrals in $\upsilon$ and
$k$ are assumed, once again, depending on the boundary conditions of the
problem. 

Using the harmonic decomposition just described, we can turn the sets of
partial differential equations in Appendices \ref{Appendix:Linear_eqs_angular_variables}
and \ref{Appendix:Linear_eqs_dot-derivatives_variables}, in a set
of ODEs. We remark, though, that since the different harmonics are not defined for all values of $k$,  we have to distinguish between the
modes where the vector and tensor harmonics are not identically zero
and those where these harmonics are zero. Then, we have the following
set of gauge invariant equations for the linear perturbations that
characterize the perturbed spacetime.

\subsection{Even sector\label{subsec:General_eqs_coefficients_Even}}

\subsubsection{$k$-Modes where $\left\{ \mathcal{Q}_{\alpha\beta}^{\left(k\right)},\mathcal{\bar{Q}}_{\alpha\beta}^{\left(k\right)}\right\} \protect\neq0$}

\paragraph{Equations for the kinematical quantities associated with the timelike
congruence\protect \\
}
\begin{itemize}
\item Evolution and propagation equations for the harmonic coefficients
of gradients of scalar quantities:
\begin{equation}
\begin{aligned}\frac{1}{r}i\upsilon\Psi_{\mathbb{\theta}}^{\left(\upsilon,k,S\right)}-\widehat{\Psi}_{\mathbb{A}}^{\left(\upsilon,k,V\right)} & =\mathcal{\widehat{A}}_{0}\Psi_{a}^{\left(\upsilon,k,V\right)}-\frac{1}{2}\left(\Psi_{\mathfrak{m}}^{\left(\upsilon,k,V\right)}+3\Psi_{\mathfrak{p}}^{\left(\upsilon,k,V\right)}\right)-\frac{k^{2}}{r^{2}}\Psi_{\mathcal{A}}^{\left(\upsilon,k,V\right)}\\
 & +\mathcal{A}_{0}\Psi_{\mathbb{F}}^{\left(\upsilon,k,V\right)}+\left(\frac{3}{2}\phi_{0}+2\mathcal{A}_{0}\right)\Psi_{\mathbb{A}}^{\left(\upsilon,k,V\right)}\,,
\end{aligned}
\end{equation}
\begin{equation}
\begin{aligned}\frac{2}{3r}i\upsilon\Psi_{\mathbb{\theta}}^{\left(\upsilon,k,S\right)}-\frac{1}{r}i\upsilon\Psi_{\Sigma}^{\left(\upsilon,k,S\right)} & =\phi_{0}\Psi_{\mathbb{A}}^{\left(\upsilon,k,V\right)}+\mathcal{A}_{0}\Psi_{\mathbb{F}}^{\left(\upsilon,k,V\right)}-\frac{k^{2}}{r^{2}}\Psi_{\mathcal{A}}^{\left(\upsilon,k,V\right)}\\
 & -\frac{1}{3}\Psi_{\mathfrak{m}}^{\left(\upsilon,k,V\right)}-\Psi_{\mathfrak{p}}^{\left(\upsilon,k,V\right)}-\frac{1}{2}\Psi_{\mathbb{P}}^{\left(\upsilon,k,V\right)}+\Psi_{\mathbb{E}}^{\left(\upsilon,k,V\right)}\,,
\end{aligned}
\end{equation}
\begin{equation}
\frac{3}{2}\widehat{\Psi}_{\Sigma}^{\left(\upsilon,k,S\right)}-\widehat{\Psi}_{\mathbb{\theta}}^{\left(\upsilon,k,S\right)}=\frac{3k^{2}}{2r}\left(\Psi_{\Sigma}^{\left(\upsilon,k,V\right)}-\overline{\Psi}_{\Omega}^{\left(\upsilon,k,V\right)}\right)-\frac{3}{2}\Psi_{Q}^{\left(\upsilon,k,S\right)}-\frac{9}{4}\phi_{0}\Psi_{\Sigma}^{\left(\upsilon,k,S\right)}\,;
\end{equation}

\item Evolution and propagation equations for the harmonic coefficients
of vector quantities:
\begin{equation}
2i\upsilon\overline{\Psi}_{\Omega}^{\left(\upsilon,k,V\right)}+\widehat{\Psi}_{\mathcal{A}}^{\left(\upsilon,k,V\right)}=\Psi_{\mathbb{A}}^{\left(\upsilon,k,V\right)}-\frac{1}{2}\phi_{0}\Psi_{\mathcal{A}}^{\left(\upsilon,k,V\right)}-\mathcal{A}_{0}\Psi_{a}^{\left(\upsilon,k,V\right)}\,,
\end{equation}
\begin{equation}
i\upsilon\left(\Psi_{\Sigma}^{\left(\upsilon,k,V\right)}+\overline{\Psi}_{\Omega}^{\left(\upsilon,k,V\right)}\right)=\left(\mathcal{A}_{0}-\frac{1}{2}\phi_{0}\right)\Psi_{\mathcal{A}}^{\left(\upsilon,k,V\right)}+\Psi_{\mathbb{A}}^{\left(\upsilon,k,V\right)}-\Psi_{\mathcal{E}}^{\left(\upsilon,k,V\right)}+\frac{1}{2}\Psi_{\Pi}^{\left(\upsilon,k,V\right)}\,,
\end{equation}

\begin{equation}
\begin{aligned}\widehat{\Psi}_{\Sigma}^{\left(\upsilon,k,V\right)}+\widehat{\overline{\Psi}}_{\Omega}^{\left(\upsilon,k,V\right)} & =\frac{1}{2r}\Psi_{\Sigma}^{\left(\upsilon,k,S\right)}+\frac{2}{3r}\Psi_{\theta}^{\left(\upsilon,k,S\right)}-\frac{3}{2}\phi_{0}\Psi_{\Sigma}^{\left(\upsilon,k,V\right)}-\left(\frac{1}{2}\phi_{0}+2\mathcal{A}_{0}\right)\overline{\Psi}_{\Omega}^{\left(\upsilon,k,V\right)}\\
 & -\frac{2-k^{2}}{2r}\Psi_{\Sigma}^{\left(\upsilon,k,T\right)}-\Psi_{Q}^{\left(\upsilon,k,V\right)}\,;
\end{aligned}
\end{equation}

\item Evolution and propagation equations for the harmonic coefficients
of tensor quantities:

\begin{equation}
i\upsilon\Psi_{\Sigma}^{\left(\upsilon,k,T\right)}=\frac{1}{r}\Psi_{\mathcal{A}}^{\left(\upsilon,k,V\right)}+\mathcal{A}_{0}\Psi_{\zeta}^{\left(\upsilon,k,T\right)}-\Psi_{\mathcal{E}}^{\left(\upsilon,k,T\right)}+\frac{1}{2}\Psi_{\Pi}^{\left(\upsilon,k,T\right)}\,,
\end{equation}
\begin{equation}
\widehat{\Psi}_{\Sigma}^{\left(\upsilon,k,T\right)}=\frac{1}{r}\left(\Psi_{\Sigma}^{\left(\upsilon,k,V\right)}-\overline{\Psi}_{\Omega}^{\left(\upsilon,k,V\right)}\right)-\frac{1}{2}\phi_{0}\Psi_{\Sigma}^{\left(\upsilon,k,T\right)}+\overline{\Psi}_{\mathcal{H}}^{\left(\upsilon,k,T\right)}\,;
\end{equation}

\item Constraint equations for the harmonic coefficients:
\begin{equation}
\phi_{0}\left(\Psi_{\Sigma}^{\left(\upsilon,k,V\right)}+\overline{\Psi}_{\Omega}^{\left(\upsilon,k,V\right)}\right)+\frac{1}{r}\Psi_{\Sigma}^{\left(\upsilon,k,S\right)}-\frac{2}{3r}\Psi_{\theta}^{\left(\upsilon,k,S\right)}+\Psi_{Q}^{\left(\upsilon,k,V\right)}-2\overline{\Psi}_{\mathcal{H}}^{\left(\upsilon,k,V\right)}+\frac{2-k^{2}}{r}\Psi_{\Sigma}^{\left(\upsilon,k,T\right)}=0\,.
\end{equation}

\end{itemize} \hfill\hfill

\paragraph{Equations for the kinematical quantities associated with the spacelike
congruence\protect \\
}
\begin{itemize}

\item Evolution and propagation equations for the harmonic coefficients
of gradients of scalar quantities:
\begin{equation}
\begin{aligned}i\upsilon\Psi_{\mathbb{F}}^{\left(\upsilon,k,V\right)}= & -\widehat{\phi}_{0}\left(\Psi_{\Sigma}^{\left(\upsilon,k,V\right)}+\Psi_{\alpha}^{\left(\upsilon,k,V\right)}+\overline{\Psi}_{\Omega}^{\left(\upsilon,k,V\right)}\right)+\frac{1}{r}\Psi_{Q}^{\left(\upsilon,k,S\right)}\\
 & +\left(\mathcal{A}_{0}-\frac{1}{2}\phi_{0}\right)\left(\frac{2}{3r}\Psi_{\theta}^{\left(\upsilon,k,S\right)}-\frac{1}{r}\Psi_{\Sigma}^{\left(\upsilon,k,S\right)}\right)-\frac{k^{2}}{r^{2}}\Psi_{\alpha}^{\left(\upsilon,k,V\right)}\,,
\end{aligned}
\end{equation}
\begin{equation}
\widehat{\Psi}_{\mathbb{F}}^{\left(\upsilon,k,V\right)}=-\left(\widehat{\phi}_{0}+\frac{k^{2}}{r^{2}}\right)\Psi_{a}^{\left(\upsilon,k,V\right)}-\frac{3}{2}\phi_{0}\Psi_{\mathbb{F}}^{\left(\upsilon,k,V\right)}-\frac{2}{3}\Psi_{\mathfrak{m}}^{\left(\upsilon,k,V\right)}-\frac{1}{2}\Psi_{\mathbb{P}}^{\left(\upsilon,k,V\right)}-\Psi_{\mathbb{E}}^{\left(\upsilon,k,V\right)}\,;
\end{equation}

\item Evolution and propagation equations for the harmonic coefficients
of vector quantities:
\begin{equation}
\begin{aligned}\widehat{\Psi}_{\alpha}^{\left(\upsilon,k,V\right)}-i\upsilon\Psi_{a}^{\left(\upsilon,k,V\right)} & =\overline{\Psi}_{\mathcal{H}}^{\left(\upsilon,k,V\right)}-\left(\mathcal{A}_{0}+\frac{1}{2}\phi_{0}\right)\Psi_{\alpha}^{\left(\upsilon,k,V\right)}+\frac{1}{2}\Psi_{Q}^{\left(\upsilon,k,V\right)}\\
 & +\left(\frac{1}{2}\phi_{0}-\mathcal{A}_{0}\right)\left(\Psi_{\Sigma}^{\left(\upsilon,k,V\right)}-\overline{\Psi}_{\Omega}^{\left(\upsilon,k,V\right)}\right)\,;
\end{aligned}
\end{equation}

\item Evolution and propagation equations for the harmonic coefficients
of tensor quantities:

\begin{equation}
i\upsilon\Psi_{\zeta}^{\left(\upsilon,k,T\right)}=\left(\mathcal{A}_{0}-\frac{1}{2}\phi_{0}\right)\Psi_{\Sigma}^{\left(\upsilon,k,T\right)}+\frac{1}{r}\Psi_{\alpha}^{\left(\upsilon,k,V\right)}+\overline{\Psi}_{\mathcal{H}}^{\left(\upsilon,k,T\right)}\,,
\end{equation}
\begin{equation}
\widehat{\Psi}_{\zeta}^{\left(\upsilon,k,T\right)}=\frac{1}{r}\Psi_{a}^{\left(\upsilon,k,V\right)}-\phi_{0}\Psi_{\zeta}^{\left(\upsilon,k,T\right)}-\Psi_{\mathcal{E}}^{\left(\upsilon,k,T\right)}-\frac{1}{2}\Psi_{\Pi}^{\left(\upsilon,k,T\right)}\,;
\end{equation}

\item Constraint equations for the harmonic coefficients:
\begin{equation}
\frac{2-k^{2}}{r}\Psi_{\zeta}^{\left(\upsilon,k,T\right)}-\Psi_{\mathbb{F}}^{\left(\upsilon,k,V\right)}-2\Psi_{\mathcal{E}}^{\left(\upsilon,k,V\right)}-\Psi_{\Pi}^{\left(\upsilon,k,V\right)}=0\,.
\end{equation}

\end{itemize} \hfill\hfill

\paragraph{Equations for the Weyl tensor components and the matter variables\protect \\
}
\begin{itemize}

\item Evolution and propagation equations for the harmonic coefficients
of gradients of scalar quantities:
\begin{equation}
\begin{aligned}\frac{1}{r}\widehat{\Psi}_{Q}^{\left(\upsilon,k,S\right)}+i\upsilon\Psi_{\mathfrak{m}}^{\left(\upsilon,k,V\right)}= & -\widehat{\mu}_{0}\left(\overline{\Psi}_{\Omega}^{\left(\upsilon,k,V\right)}+\Psi_{\Sigma}^{\left(\upsilon,k,V\right)}+\Psi_{\alpha}^{\left(\upsilon,k,V\right)}\right)-\frac{1}{r}\left(\mu_{0}+p_{0}\right)\Psi_{\theta}^{\left(\upsilon,k,S\right)}\\
 & -\frac{3}{2r}\Pi_{0}\Psi_{\Sigma}^{\left(\upsilon,k,S\right)}-\frac{1}{r}\left(\phi_{0}+2\mathcal{A}_{0}\right)\Psi_{Q}^{\left(\upsilon,k,S\right)}+\frac{k^{2}}{r^{2}}\Psi_{Q}^{\left(\upsilon,k,V\right)}\,,
\end{aligned}
\end{equation}
\begin{equation}
\begin{aligned}\frac{1}{3}i\upsilon\Psi_{\mathfrak{m}}^{\left(\upsilon,k,V\right)}-\frac{1}{2}i\upsilon\Psi_{\mathbb{P}}^{\left(\upsilon,k,V\right)}-i\upsilon\Psi_{\mathbb{E}}^{\left(\upsilon,k,V\right)} & =\left(\mu_{0}+p_{0}-\frac{1}{2}\Pi_{0}-3\mathcal{E}_{0}\right)\left(\frac{1}{2r}\Psi_{\Sigma}^{\left(\upsilon,k,S\right)}-\frac{1}{3r}\Psi_{\theta}^{\left(\upsilon,k,S\right)}\right)\\
 & -\frac{3}{2}\phi_{0}\left(\mathcal{E}_{0}+\frac{1}{2}\Pi_{0}\right)\left(\overline{\Psi}_{\Omega}^{\left(\upsilon,k,V\right)}+\Psi_{\Sigma}^{\left(\upsilon,k,V\right)}+\Psi_{\alpha}^{\left(\upsilon,k,V\right)}\right)\\
 & +\frac{k^{2}}{r^{2}}\left(\frac{1}{2}\Psi_{Q}^{\left(\upsilon,k,V\right)}-\overline{\Psi}_{\mathcal{H}}^{\left(\upsilon,k,V\right)}\right)-\frac{1}{2r}\phi_{0}\Psi_{Q}^{\left(\upsilon,k,S\right)}\,,
\end{aligned}
\end{equation}

\begin{equation}
\begin{aligned}\frac{1}{r}i\upsilon\Psi_{Q}^{\left(\upsilon,k,S\right)}+\widehat{\Psi}_{\mathbb{P}}^{\left(\upsilon,k,V\right)}+\widehat{\Psi}_{\mathfrak{p}}^{\left(\upsilon,k,V\right)} & =\left(\mu_{0}+p_{0}\right)\left(\mathcal{A}_{0}\Psi_{a}^{\left(\upsilon,k,V\right)}-\Psi_{\mathbb{A}}^{\left(\upsilon,k,V\right)}\right)-\mathcal{A}_{0}\left(\Psi_{\mathfrak{m}}^{\left(\upsilon,k,V\right)}+\Psi_{\mathfrak{p}}^{\left(\upsilon,k,V\right)}+\Psi_{\mathbb{P}}^{\left(\upsilon,k,V\right)}\right)\\
 & -\frac{1}{2}\phi_{0}\left(\Psi_{\mathfrak{p}}^{\left(\upsilon,k,V\right)}+4\Psi_{\mathbb{P}}^{\left(\upsilon,k,V\right)}\right)+\left(\frac{3}{2}\phi_{0}+\mathcal{A}_{0}\right)\Pi_{0}\Psi_{a}^{\left(\upsilon,k,V\right)}\\
 & -\Pi_{0}\left(\frac{3}{2}\Psi_{\mathbb{F}}^{\left(\upsilon,k,V\right)}+\Psi_{\mathbb{A}}^{\left(\upsilon,k,V\right)}\right)+\frac{k^{2}}{r^{2}}\Psi_{\Pi}^{\left(\upsilon,k,V\right)}\,,
\end{aligned}
\end{equation}

\begin{equation}
\begin{aligned}\widehat{\Psi}_{\mathbb{E}}^{\left(\upsilon,k,V\right)}+\frac{1}{2}\widehat{\Psi}_{\mathbb{P}}^{\left(\upsilon,k,V\right)}-\frac{1}{3}\widehat{\Psi}_{\mathfrak{m}}^{\left(\upsilon,k,V\right)}= & -2\phi_{0}\left(\Psi_{\mathbb{E}}^{\left(\upsilon,k,V\right)}+\frac{1}{2}\Psi_{\mathbb{P}}^{\left(\upsilon,k,V\right)}-\frac{1}{12}\Psi_{\mathfrak{m}}^{\left(\upsilon,k,V\right)}\right)\\
 & -\frac{3}{2}\left(\mathcal{E}_{0}+\frac{1}{2}\Pi_{0}\right)\left(\Psi_{\mathbb{F}}^{\left(\upsilon,k,V\right)}-\phi_{0}\Psi_{a}^{\left(\upsilon,k,V\right)}\right)\\
 & +\frac{k^{2}}{r^{2}}\left(\Psi_{\mathcal{E}}^{\left(\upsilon,k,V\right)}+\frac{1}{2}\Psi_{\Pi}^{\left(\upsilon,k,V\right)}\right)\,;
\end{aligned}
\end{equation}

\item Evolution and propagation equations for the harmonic coefficients
of vector quantities:

\begin{equation}
\begin{aligned}i\upsilon\left(\Psi_{\mathcal{E}}^{\left(\upsilon,k,V\right)}+\frac{1}{2}\Psi_{\Pi}^{\left(\upsilon,k,V\right)}\right) & =\left(\frac{1}{2}\phi_{0}-\mathcal{A}_{0}\right)\left(\frac{1}{2}\Psi_{Q}^{\left(\upsilon,k,V\right)}-\overline{\Psi}_{\mathcal{H}}^{\left(\upsilon,k,V\right)}\right)-\frac{3}{2}\left(\mathcal{E}_{0}+\frac{1}{2}\Pi_{0}\right)\Psi_{\alpha}^{\left(\upsilon,k,V\right)}\\
 & -\frac{1}{2}\left(\mu_{0}+p_{0}+\Pi_{0}\right)\left(\Psi_{\Sigma}^{\left(\upsilon,k,V\right)}+\overline{\Psi}_{\Omega}^{\left(\upsilon,k,V\right)}\right)\\
 & -\frac{1}{2r}\Psi_{Q}^{\left(\upsilon,k,S\right)}+\frac{2-k^{2}}{2r}\overline{\Psi}_{\mathcal{H}}^{\left(\upsilon,k,T\right)}\,,
\end{aligned}
\end{equation}
\begin{equation}
\begin{aligned}\widehat{\Psi}_{\mathcal{E}}^{\left(\upsilon,k,V\right)}+\frac{1}{2}\widehat{\Psi}_{\Pi}^{\left(\upsilon,k,V\right)}= & -\frac{3}{2}\phi_{0}\left(\Psi_{\mathcal{E}}^{\left(\upsilon,k,V\right)}+\frac{1}{2}\Psi_{\Pi}^{\left(\upsilon,k,V\right)}\right)-\frac{3}{2}\left(\mathcal{E}_{0}+\frac{1}{2}\Pi_{0}\right)\Psi_{a}^{\left(\upsilon,k,V\right)}\\
 & +\frac{1}{2}\Psi_{\mathbb{E}}^{\left(\upsilon,k,V\right)}+\frac{1}{3}\Psi_{\mathfrak{m}}^{\left(\upsilon,k,V\right)}+\frac{1}{4}\Psi_{\mathbb{P}}^{\left(\upsilon,k,V\right)}\\
 & -\frac{2-k^{2}}{2r}\left(\Psi_{\mathcal{E}}^{\left(\upsilon,k,T\right)}+\frac{1}{2}\Psi_{\Pi}^{\left(\upsilon,k,T\right)}\right)\,,
\end{aligned}
\end{equation}
\begin{equation}
\begin{aligned}\frac{1}{2}\widehat{\Psi}_{\mathcal{E}}^{\left(\upsilon,k,V\right)}-i\upsilon\overline{\Psi}_{\mathcal{H}}^{\left(\upsilon,k,V\right)}-\frac{1}{4}\widehat{\Psi}_{\Pi}^{\left(\upsilon,k,V\right)} & =\frac{3}{2}\mathcal{E}_{0}\Psi_{\mathcal{A}}^{\left(\upsilon,k,V\right)}+\frac{3}{4}\left(\Psi_{\mathbb{E}}^{\left(\upsilon,k,V\right)}-\frac{1}{2}\Psi_{\mathbb{P}}^{\left(\upsilon,k,V\right)}\right)-\frac{3}{4}\left(\mathcal{E}_{0}-\frac{1}{2}\Pi_{0}\right)\Psi_{a}^{\left(\upsilon,k,V\right)}\\
 & -\frac{1}{4}\phi_{0}\left(\Psi_{\mathcal{E}}^{\left(\upsilon,k,V\right)}-\frac{1}{2}\Psi_{\Pi}^{\left(\upsilon,k,V\right)}\right)-\mathcal{A}_{0}\Psi_{\mathcal{E}}^{\left(\upsilon,k,V\right)}\\
 & +\frac{2-k^{2}}{4r}\left(\Psi_{\mathcal{E}}^{\left(\upsilon,k,T\right)}-\frac{1}{2}\Psi_{\Pi}^{\left(\upsilon,k,T\right)}\right)\,,
\end{aligned}
\end{equation}
\begin{equation}
\begin{aligned}\widehat{\overline{\Psi}}_{\mathcal{H}}^{\left(\upsilon,k,V\right)}-\frac{1}{2}\widehat{\Psi}_{Q}^{\left(\upsilon,k,V\right)} & =\frac{2-k^{2}}{2r}\overline{\Psi}_{\mathcal{H}}^{\left(\upsilon,k,T\right)}-\frac{1}{2r}\Psi_{Q}^{\left(\upsilon,k,S\right)}-\frac{3}{2}\left(\mathcal{E}_{0}+\frac{1}{2}\Pi_{0}\right)\Psi_{\Sigma}^{\left(\upsilon,k,V\right)}\\
 & -\left(\mu_{0}+p_{0}+\frac{1}{4}\Pi_{0}-\frac{3}{2}\mathcal{E}_{0}\right)\overline{\Psi}_{\Omega}^{\left(\upsilon,k,V\right)}-\frac{3}{2}\phi_{0}\left(\overline{\Psi}_{\mathcal{H}}^{\left(\upsilon,k,V\right)}-\frac{1}{6}\Psi_{Q}^{\left(\upsilon,k,V\right)}\right)\,,
\end{aligned}
\end{equation}
\begin{equation}
\begin{aligned}i\upsilon\Psi_{Q}^{\left(\upsilon,k,V\right)}+\widehat{\Psi}_{\Pi}^{\left(\upsilon,k,V\right)} & =\frac{1}{2}\Psi_{\mathbb{P}}^{\left(\upsilon,k,V\right)}-\Psi_{\mathfrak{p}}^{\left(\upsilon,k,V\right)}-\frac{3}{2}\Pi_{0}\Psi_{a}^{\left(\upsilon,k,V\right)}-\frac{2-k^{2}}{2r}\Psi_{\Pi}^{\left(\upsilon,k,T\right)}\\
 & -\left(\frac{3}{2}\phi_{0}+\mathcal{A}_{0}\right)\Psi_{\Pi}^{\left(\upsilon,k,V\right)}-\left(\mu_{0}+p_{0}-\frac{1}{2}\Pi_{0}\right)\Psi_{\mathcal{A}}^{\left(\upsilon,k,V\right)}\,;
\end{aligned}
\end{equation}

\item Evolution and propagation equations for the harmonic coefficients
of tensor quantities:
\begin{equation}
\begin{aligned}i\upsilon\left(\Psi_{\mathcal{E}}^{\left(\upsilon,k,T\right)}+\frac{1}{2}\Psi_{\Pi}^{\left(\upsilon,k,T\right)}\right)+\widehat{\overline{\Psi}}_{\mathcal{H}}^{\left(\upsilon,k,T\right)}= & -\frac{1}{r}\left(\overline{\Psi}_{\mathcal{H}}^{\left(\upsilon,k,V\right)}+\frac{1}{2}\Psi_{Q}^{\left(\upsilon,k,V\right)}\right)-\left(\frac{1}{2}\phi_{0}+2\mathcal{A}_{0}\right)\overline{\Psi}_{\mathcal{H}}^{\left(\upsilon,k,T\right)}\\
 & -\frac{1}{2}\left(\mu_{0}+p_{0}+3\mathcal{E}_{0}-\frac{1}{2}\Pi_{0}\right)\Psi_{\Sigma}^{\left(\upsilon,k,T\right)}\,,
\end{aligned}
\end{equation}
\begin{equation}
\begin{aligned}\frac{1}{2}\widehat{\Psi}_{\Pi}^{\left(\upsilon,k,T\right)}-\widehat{\Psi}_{\mathcal{E}}^{\left(\upsilon,k,T\right)}-i\upsilon\overline{\Psi}_{\mathcal{H}}^{\left(\upsilon,k,T\right)} & =\frac{1}{r}\left(\frac{1}{2}\Psi_{\Pi}^{\left(\upsilon,k,V\right)}-\Psi_{\mathcal{E}}^{\left(\upsilon,k,V\right)}\right)-\frac{1}{4}\phi_{0}\Psi_{\Pi}^{\left(\upsilon,k,T\right)}\\
 & -\frac{3}{2}\left(\mathcal{E}_{0}-\frac{1}{2}\Pi_{0}\right)\Psi_{\zeta}^{\left(\upsilon,k,T\right)}+\left(\frac{1}{2}\phi_{0}+2\mathcal{A}_{0}\right)\Psi_{\mathcal{E}}^{\left(\upsilon,k,T\right)}\,.
\end{aligned}
\end{equation}

\end{itemize} 

\subsubsection{$k$-Modes where $\left\{ \mathcal{Q}_{\alpha}^{\left(k\right)},\bar{\mathcal{Q}}_{\alpha}^{\left(k\right)}\right\} \protect\neq0\wedge\left\{ \mathcal{Q}_{\alpha\beta}^{\left(k\right)},\mathcal{\bar{Q}}_{\alpha\beta}^{\left(k\right)}\right\} =0$:
$k^{2}=2$}

For the $k^2=2$ modes, the tensor harmonics are not defined, but in general, the  vector and scalar harmonics do not vanish. For those modes, we find the following equations for the harmonic coefficients.
\\

\paragraph{Equations for the kinematical quantities associated with the timelike
congruence\protect \\
}
\begin{itemize}

\item Evolution and propagation equations for the harmonic coefficients
of gradients of scalar quantities:
\begin{equation}
\begin{aligned}\frac{1}{r}i\upsilon\Psi_{\mathbb{\theta}}^{\left(\upsilon,k,S\right)}-\widehat{\Psi}_{\mathbb{A}}^{\left(\upsilon,k,V\right)} & =\mathcal{\widehat{A}}_{0}\Psi_{a}^{\left(\upsilon,k,V\right)}-\frac{2}{r^{2}}\Psi_{\mathcal{A}}^{\left(\upsilon,k,V\right)}-\frac{1}{2}\left(\Psi_{\mathfrak{m}}^{\left(\upsilon,k,V\right)}+3\Psi_{\mathfrak{p}}^{\left(\upsilon,k,V\right)}\right)\\
 & +\mathcal{A}_{0}\Psi_{\mathbb{F}}^{\left(\upsilon,k,V\right)}+\left(\frac{3}{2}\phi_{0}+2\mathcal{A}_{0}\right)\Psi_{\mathbb{A}}^{\left(\upsilon,k,V\right)}\,,
\end{aligned}
\end{equation}
\begin{equation}
\begin{aligned}\frac{2}{3r}i\upsilon\Psi_{\mathbb{\theta}}^{\left(\upsilon,k,S\right)}-\frac{1}{r}i\upsilon\Psi_{\Sigma}^{\left(\upsilon,k,S\right)} & =\phi_{0}\Psi_{\mathbb{A}}^{\left(\upsilon,k,V\right)}+\mathcal{A}_{0}\Psi_{\mathbb{F}}^{\left(\upsilon,k,V\right)}-\frac{2}{r^{2}}\Psi_{\mathcal{A}}^{\left(\upsilon,k,V\right)}\\
 & -\frac{1}{3}\Psi_{\mathfrak{m}}^{\left(\upsilon,k,V\right)}-\Psi_{\mathfrak{p}}^{\left(\upsilon,k,V\right)}-\frac{1}{2}\Psi_{\mathbb{P}}^{\left(\upsilon,k,V\right)}+\Psi_{\mathbb{E}}^{\left(\upsilon,k,V\right)}\,,
\end{aligned}
\end{equation}
\begin{equation}
\frac{3}{2}\widehat{\Psi}_{\Sigma}^{\left(\upsilon,k,S\right)}-\widehat{\Psi}_{\mathbb{\theta}}^{\left(\upsilon,k,S\right)}=\frac{3}{r}\left(\Psi_{\Sigma}^{\left(\upsilon,k,V\right)}-\overline{\Psi}_{\Omega}^{\left(\upsilon,k,V\right)}\right)-\frac{3}{2}\Psi_{Q}^{\left(\upsilon,k,S\right)}-\frac{9}{4}\phi_{0}\Psi_{\Sigma}^{\left(\upsilon,k,S\right)}\,,
\end{equation}

\item Evolution and propagation equations for the harmonic coefficients
of vector quantities:
\begin{equation}
2i\upsilon\overline{\Psi}_{\Omega}^{\left(\upsilon,k,V\right)}+\widehat{\Psi}_{\mathcal{A}}^{\left(\upsilon,k,V\right)}=\Psi_{\mathbb{A}}^{\left(\upsilon,k,V\right)}-\frac{1}{2}\phi_{0}\Psi_{\mathcal{A}}^{\left(\upsilon,k,V\right)}-\mathcal{A}_{0}\Psi_{a}^{\left(\upsilon,k,V\right)}\,,
\end{equation}
\begin{equation}
i\upsilon\left(\Psi_{\Sigma}^{\left(\upsilon,k,V\right)}+\overline{\Psi}_{\Omega}^{\left(\upsilon,k,V\right)}\right)=\left(\mathcal{A}_{0}-\frac{1}{2}\phi_{0}\right)\Psi_{\mathcal{A}}^{\left(\upsilon,k,V\right)}+\Psi_{\mathbb{A}}^{\left(\upsilon,k,V\right)}-\Psi_{\mathcal{E}}^{\left(\upsilon,k,V\right)}+\frac{1}{2}\Psi_{\Pi}^{\left(\upsilon,k,V\right)}\,,
\end{equation}

\begin{equation}
\widehat{\Psi}_{\Sigma}^{\left(\upsilon,k,V\right)}+\widehat{\overline{\Psi}}_{\Omega}^{\left(\upsilon,k,V\right)}=
\frac{1}{2r}\Psi_{\Sigma}^{\left(\upsilon,k,S\right)}+
\frac{2}{3r}\Psi_{\theta}^{\left(\upsilon,k,S\right)}-
\frac{3}{2}\phi_{0}\Psi_{\Sigma}^{\left(\upsilon,k,V\right)}
-\left(\frac{1}{2}\phi_{0}+2\mathcal{A}_{0}\right)\overline{\Psi}_{\Omega}^{\left(\upsilon,k,V\right)}-\Psi_{Q}^{\left(\upsilon,k,V\right)}\,;
\end{equation}

\item Constraint equations for the harmonic coefficients:
\begin{equation}
\phi_{0}\left(\Psi_{\Sigma}^{\left(\upsilon,k,V\right)}+\overline{\Psi}_{\Omega}^{\left(\upsilon,k,V\right)}\right)+\frac{1}{r}\Psi_{\Sigma}^{\left(\upsilon,k,S\right)}-\frac{2}{3r}\Psi_{\theta}^{\left(\upsilon,k,S\right)}+\Psi_{Q}^{\left(\upsilon,k,V\right)}-2\overline{\Psi}_{\mathcal{H}}^{\left(\upsilon,k,V\right)}=0\,.
\end{equation}
 
\end{itemize} \hfill\hfill

\paragraph{Equations for the kinematical quantities associated with the spacelike
congruence\protect \\
}

\begin{itemize}
\item Evolution and propagation equations for the harmonic coefficients
of gradients of scalar quantities:
\begin{equation}
\begin{aligned}i\upsilon\Psi_{\mathbb{F}}^{\left(\upsilon,k,V\right)}= & -\widehat{\phi}_{0}\left(\Psi_{\Sigma}^{\left(\upsilon,k,V\right)}+\Psi_{\alpha}^{\left(\upsilon,k,V\right)}+\overline{\Psi}_{\Omega}^{\left(\upsilon,k,V\right)}\right)+\frac{1}{r}\Psi_{Q}^{\left(\upsilon,k,S\right)}\\
 & +\left(\mathcal{A}_{0}-\frac{1}{2}\phi_{0}\right)\left(\frac{2}{3r}\Psi_{\theta}^{\left(\upsilon,k,S\right)}-\frac{1}{r}\Psi_{\Sigma}^{\left(\upsilon,k,S\right)}\right)-\frac{2}{r^{2}}\Psi_{\alpha}^{\left(\upsilon,k,V\right)}\,,
\end{aligned}
\end{equation}
\begin{equation}
\widehat{\Psi}_{\mathbb{F}}^{\left(\upsilon,k,V\right)}= -\left(\widehat{\phi}_{0}+\frac{2}{r^{2}}\right)\Psi_{a}^{\left(\upsilon,k,V\right)}-\frac{3}{2}\phi_{0}\Psi_{\mathbb{F}}^{\left(\upsilon,k,V\right)}
-\frac{2}{3}\Psi_{\mathfrak{m}}^{\left(\upsilon,k,V\right)}-\frac{1}{2}\Psi_{\mathbb{P}}^{\left(\upsilon,k,V\right)}-\Psi_{\mathbb{E}}^{\left(\upsilon,k,V\right)}\,;
\end{equation}

\item Evolution and propagation equations for the harmonic coefficients
of vector quantities:
\begin{equation}
\begin{aligned}\widehat{\Psi}_{\alpha}^{\left(\upsilon,k,V\right)}-i\upsilon\Psi_{a}^{\left(\upsilon,k,V\right)} & =\overline{\Psi}_{\mathcal{H}}^{\left(\upsilon,k,V\right)}-\left(\mathcal{A}_{0}+\frac{1}{2}\phi_{0}\right)\Psi_{\alpha}^{\left(\upsilon,k,V\right)}+\frac{1}{2}\Psi_{Q}^{\left(\upsilon,k,V\right)}\\
 & +\left(\frac{1}{2}\phi_{0}-\mathcal{A}_{0}\right)\left(\Psi_{\Sigma}^{\left(\upsilon,k,V\right)}-\overline{\Psi}_{\Omega}^{\left(\upsilon,k,V\right)}\right)\,;
\end{aligned}
\end{equation}

\item Constraint equations for the harmonic coefficients:
\begin{equation}
\Psi_{\mathbb{F}}^{\left(\upsilon,k,V\right)}+2\Psi_{\mathcal{E}}^{\left(\upsilon,k,V\right)}+\Psi_{\Pi}^{\left(\upsilon,k,V\right)}=0\,.
\end{equation}

\end{itemize} \hfill\hfill

\paragraph{Equations for the Weyl tensor components and the matter variables\protect \\
}
\begin{itemize}

\item Evolution and propagation equations for the harmonic coefficients
of gradients of scalar quantities:
\begin{equation}
\begin{aligned}\frac{1}{r}\widehat{\Psi}_{Q}^{\left(\upsilon,k,S\right)}+i\upsilon\Psi_{\mathfrak{m}}^{\left(\upsilon,k,V\right)}= & -\widehat{\mu}_{0}\left(\overline{\Psi}_{\Omega}^{\left(\upsilon,k,V\right)}+\Psi_{\Sigma}^{\left(\upsilon,k,V\right)}+\Psi_{\alpha}^{\left(\upsilon,k,V\right)}\right)-\frac{1}{r}\left(\mu_{0}+p_{0}\right)\Psi_{\theta}^{\left(\upsilon,k,S\right)}\\
 & -\frac{3}{2r}\Pi_{0}\Psi_{\Sigma}^{\left(\upsilon,k,S\right)}-\frac{1}{r}\left(\phi_{0}+2\mathcal{A}_{0}\right)\Psi_{Q}^{\left(\upsilon,k,S\right)}+\frac{2}{r^{2}}\Psi_{Q}^{\left(\upsilon,k,V\right)}\,,
\end{aligned}
\end{equation}
\begin{equation}
\begin{aligned}\frac{1}{3}i\upsilon\Psi_{\mathfrak{m}}^{\left(\upsilon,k,V\right)}-\frac{1}{2}i\upsilon\Psi_{\mathbb{P}}^{\left(\upsilon,k,V\right)}-i\upsilon\Psi_{\mathbb{E}}^{\left(\upsilon,k,V\right)} & =\left(\mu_{0}+p_{0}-\frac{1}{2}\Pi_{0}-3\mathcal{E}_{0}\right)\left(\frac{1}{2r}\Psi_{\Sigma}^{\left(\upsilon,k,S\right)}-\frac{1}{3r}\Psi_{\theta}^{\left(\upsilon,k,S\right)}\right)\\
 & -\frac{3}{2}\phi_{0}\left(\mathcal{E}_{0}+\frac{1}{2}\Pi_{0}\right)\left(\overline{\Psi}_{\Omega}^{\left(\upsilon,k,V\right)}+\Psi_{\Sigma}^{\left(\upsilon,k,V\right)}+\Psi_{\alpha}^{\left(\upsilon,k,V\right)}\right)\\
 & +\frac{2}{r^{2}}\left(\frac{1}{2}\Psi_{Q}^{\left(\upsilon,k,V\right)}-\overline{\Psi}_{\mathcal{H}}^{\left(\upsilon,k,V\right)}\right)-\frac{1}{2r}\phi_{0}\Psi_{Q}^{\left(\upsilon,k,S\right)}\,,
\end{aligned}
\end{equation}

\begin{equation}
\begin{aligned}\frac{1}{r}i\upsilon\Psi_{Q}^{\left(\upsilon,k,S\right)}+\widehat{\Psi}_{\mathbb{P}}^{\left(\upsilon,k,V\right)}+\widehat{\Psi}_{\mathfrak{p}}^{\left(\upsilon,k,V\right)} & =\left(\mu_{0}+p_{0}\right)\left(\mathcal{A}_{0}\Psi_{a}^{\left(\upsilon,k,V\right)}-\Psi_{\mathbb{A}}^{\left(\upsilon,k,V\right)}\right)-\mathcal{A}_{0}\left(\Psi_{\mathfrak{m}}^{\left(\upsilon,k,V\right)}+\Psi_{\mathfrak{p}}^{\left(\upsilon,k,V\right)}+\Psi_{\mathbb{P}}^{\left(\upsilon,k,V\right)}\right)\\
 & -\frac{1}{2}\phi_{0}\left(\Psi_{\mathfrak{p}}^{\left(\upsilon,k,V\right)}+4\Psi_{\mathbb{P}}^{\left(\upsilon,k,V\right)}\right)+\left(\frac{3}{2}\phi_{0}+\mathcal{A}_{0}\right)\Pi_{0}\Psi_{a}^{\left(\upsilon,k,V\right)}\\
 & -\Pi_{0}\left(\frac{3}{2}\Psi_{\mathbb{F}}^{\left(\upsilon,k,V\right)}+\Psi_{\mathbb{A}}^{\left(\upsilon,k,V\right)}\right)+\frac{2}{r^{2}}\Psi_{\Pi}^{\left(\upsilon,k,V\right)}\,,
\end{aligned}
\end{equation}

\begin{equation}
\begin{aligned}\widehat{\Psi}_{\mathbb{E}}^{\left(\upsilon,k,V\right)}+\frac{1}{2}\widehat{\Psi}_{\mathbb{P}}^{\left(\upsilon,k,V\right)}-\frac{1}{3}\widehat{\Psi}_{\mathfrak{m}}^{\left(\upsilon,k,V\right)}= & -2\phi_{0}\left(\Psi_{\mathbb{E}}^{\left(\upsilon,k,V\right)}+\frac{1}{2}\Psi_{\mathbb{P}}^{\left(\upsilon,k,V\right)}-\frac{1}{12}\Psi_{\mathfrak{m}}^{\left(\upsilon,k,V\right)}\right)\\
 & -\frac{3}{2}\left(\mathcal{E}_{0}+\frac{1}{2}\Pi_{0}\right)\left(\Psi_{\mathbb{F}}^{\left(\upsilon,k,V\right)}-\phi_{0}\Psi_{a}^{\left(\upsilon,k,V\right)}\right)\\
 & +\frac{2}{r^{2}}\left(\Psi_{\mathcal{E}}^{\left(\upsilon,k,V\right)}+\frac{1}{2}\Psi_{\Pi}^{\left(\upsilon,k,V\right)}\right)\,;
\end{aligned}
\end{equation}

\item Evolution and propagation equations for the harmonic coefficients
of vector quantities:

\begin{equation}
\begin{aligned}i\upsilon\left(\Psi_{\mathcal{E}}^{\left(\upsilon,k,V\right)}+\frac{1}{2}\Psi_{\Pi}^{\left(\upsilon,k,V\right)}\right) & =\left(\frac{1}{2}\phi_{0}-\mathcal{A}_{0}\right)\left(\frac{1}{2}\Psi_{Q}^{\left(\upsilon,k,V\right)}-\overline{\Psi}_{\mathcal{H}}^{\left(\upsilon,k,V\right)}\right)-\frac{3}{2}\left(\mathcal{E}_{0}+\frac{1}{2}\Pi_{0}\right)\Psi_{\alpha}^{\left(\upsilon,k,V\right)}\\
 & -\frac{1}{2}\left(\mu_{0}+p_{0}+\Pi_{0}\right)\left(\Psi_{\Sigma}^{\left(\upsilon,k,V\right)}+\overline{\Psi}_{\Omega}^{\left(\upsilon,k,V\right)}\right)-\frac{1}{2r}\Psi_{Q}^{\left(\upsilon,k,S\right)}\,,
\end{aligned}
\end{equation}
\begin{equation}
\begin{aligned}\widehat{\Psi}_{\mathcal{E}}^{\left(\upsilon,k,V\right)}+\frac{1}{2}\widehat{\Psi}_{\Pi}^{\left(\upsilon,k,V\right)}= & -\frac{3}{2}\phi_{0}\left(\Psi_{\mathcal{E}}^{\left(\upsilon,k,V\right)}+\frac{1}{2}\Psi_{\Pi}^{\left(\upsilon,k,V\right)}\right)-\frac{3}{2}\left(\mathcal{E}_{0}+\frac{1}{2}\Pi_{0}\right)\Psi_{a}^{\left(\upsilon,k,V\right)}\\
 & +\frac{1}{2}\Psi_{\mathbb{E}}^{\left(\upsilon,k,V\right)}+\frac{1}{3}\Psi_{\mathfrak{m}}^{\left(\upsilon,k,V\right)}+\frac{1}{4}\Psi_{\mathbb{P}}^{\left(\upsilon,k,V\right)}\,,
\end{aligned}
\end{equation}
\begin{equation}
\begin{aligned}\frac{1}{2}\widehat{\Psi}_{\mathcal{E}}^{\left(\upsilon,k,V\right)}-i\upsilon\overline{\Psi}_{\mathcal{H}}^{\left(\upsilon,k,V\right)}-\frac{1}{4}\widehat{\Psi}_{\Pi}^{\left(\upsilon,k,V\right)} & =\frac{3}{2}\mathcal{E}_{0}\Psi_{\mathcal{A}}^{\left(\upsilon,k,V\right)}+\frac{3}{4}\left(\Psi_{\mathbb{E}}^{\left(\upsilon,k,V\right)}-\frac{1}{2}\Psi_{\mathbb{P}}^{\left(\upsilon,k,V\right)}\right)-\frac{3}{4}\left(\mathcal{E}_{0}-\frac{1}{2}\Pi_{0}\right)\Psi_{a}^{\left(\upsilon,k,V\right)}\\
 & -\frac{1}{4}\phi_{0}\left(\Psi_{\mathcal{E}}^{\left(\upsilon,k,V\right)}-\frac{1}{2}\Psi_{\Pi}^{\left(\upsilon,k,V\right)}\right)-\mathcal{A}_{0}\Psi_{\mathcal{E}}^{\left(\upsilon,k,V\right)}\,,
\end{aligned}
\end{equation}
\begin{equation}
\begin{aligned}\widehat{\overline{\Psi}}_{\mathcal{H}}^{\left(\upsilon,k,V\right)}-\frac{1}{2}\widehat{\Psi}_{Q}^{\left(\upsilon,k,V\right)}= & -\frac{1}{2r}\Psi_{Q}^{\left(\upsilon,k,S\right)}-\left(\mu_{0}+p_{0}+\frac{1}{4}\Pi_{0}-\frac{3}{2}\mathcal{E}_{0}\right)\overline{\Psi}_{\Omega}^{\left(\upsilon,k,V\right)}\\
 & -\frac{3}{2}\phi_{0}\left(\overline{\Psi}_{\mathcal{H}}^{\left(\upsilon,k,V\right)}-\frac{1}{6}\Psi_{Q}^{\left(\upsilon,k,V\right)}\right)-\frac{3}{2}\left(\mathcal{E}_{0}+\frac{1}{2}\Pi_{0}\right)\Psi_{\Sigma}^{\left(\upsilon,k,V\right)}\,,
\end{aligned}
\end{equation}
\begin{equation}
\begin{aligned}i\upsilon\Psi_{Q}^{\left(\upsilon,k,V\right)}+\widehat{\Psi}_{\Pi}^{\left(\upsilon,k,V\right)} & =\frac{1}{2}\Psi_{\mathbb{P}}^{\left(\upsilon,k,V\right)}-\Psi_{\mathfrak{p}}^{\left(\upsilon,k,V\right)}-\left(\mu_{0}+p_{0}-\frac{1}{2}\Pi_{0}\right)\Psi_{\mathcal{A}}^{\left(\upsilon,k,V\right)}\\
 & -\frac{3}{2}\Pi_{0}\Psi_{a}^{\left(\upsilon,k,V\right)}-\left(\frac{3}{2}\phi_{0}+\mathcal{A}_{0}\right)\Psi_{\Pi}^{\left(\upsilon,k,V\right)}\,.
\end{aligned}
\end{equation}

\end{itemize} 

\subsubsection{\label{subsubsec:General_eqs_coefficients_Even_k_0}$k$-Modes where
$\mathcal{Q}^{\left(k\right)}\protect\neq0\wedge\left\{ \mathcal{Q}_{\alpha}^{\left(k\right)},\bar{\mathcal{Q}}_{\alpha}^{\left(k\right)}\right\} =0\wedge\left\{ \mathcal{Q}_{\alpha\beta}^{\left(k\right)},\mathcal{\bar{Q}}_{\alpha\beta}^{\left(k\right)}\right\} =0$:
$k=0$}

As mentioned above, for the $k$-modes where the vector and tensor harmonics are not defined,
we cannot use the equations in Appendix~\ref{Appendix:Linear_eqs_angular_variables},
since those only relate to vector and tensor quantities, that is, those
equations are trivially verified and do not contain information regarding linear perturbations
that are only along the $u$ and the $e$ directions. Then, to describe
the dynamics of the perturbations that are not along the directions
on the sheets, we will use, instead, the relations in Appendix~\ref{Appendix:Linear_eqs_dot-derivatives_variables}. The equations for the harmonic coefficients for the $k=0$ mode are
\begin{equation}
\begin{aligned}\widehat{\Psi}_{\mathsf{A}}^{\left(\upsilon,0,S\right)}+\upsilon^{2}\Psi_{\theta}^{\left(\upsilon,0,S\right)} & =\frac{1}{2}\left(\Psi_{\mathsf{m}}^{\left(\upsilon,0,S\right)}+3\Psi_{\mathsf{p}}^{\left(\upsilon,0,S\right)}\right)+\widehat{\mathcal{A}}_{0}\left(\frac{1}{3}\Psi_{\theta}^{\left(\upsilon,0,S\right)}+\Psi_{\Sigma}^{\left(\upsilon,0,S\right)}\right)\\
 & -\left(3\mathcal{A}_{0}+\phi_{0}\right)\Psi_{\mathsf{A}}^{\left(\upsilon,0,S\right)}-\mathcal{A}_{0}\Psi_{\mathsf{F}}^{\left(\upsilon,0,S\right)}\,,
\end{aligned}
\end{equation}
\begin{equation}
\begin{aligned}\upsilon^{2}\left(\frac{2}{3}\Psi_{\theta}^{\left(\upsilon,0,S\right)}-\Psi_{\Sigma}^{\left(\upsilon,0,S\right)}\right) & =\frac{1}{3}\left(\Psi_{\mathsf{m}}^{\left(\upsilon,0,S\right)}+3\Psi_{\mathsf{p}}^{\left(\upsilon,0,S\right)}\right)+\frac{1}{2}\Psi_{\mathcal{P}}^{\left(\upsilon,0,S\right)}-\Psi_{\mathsf{E}}^{\left(\upsilon,0,S\right)}\\
 & -\mathcal{A}_{0}\Psi_{\mathsf{F}}^{\left(\upsilon,0,S\right)}-\phi_{0}\Psi_{\mathsf{A}}^{\left(\upsilon,0,S\right)}\,,
\end{aligned}
\end{equation}

\begin{equation}
\frac{2}{3}\widehat{\Psi}_{\theta}^{\left(\upsilon,0,S\right)}-\widehat{\Psi}_{\Sigma}^{\left(\upsilon,0,S\right)}=\Psi_{Q}^{\left(\upsilon,0,S\right)}+\frac{3}{2}\phi_{0}\Psi_{\Sigma}^{\left(\upsilon,0,S\right)}\,,
\end{equation}

\begin{equation}
\begin{aligned}-\upsilon^{2}\Psi_{Q}^{\left(\upsilon,0,S\right)}+\widehat{\Psi}_{\mathsf{p}}^{\left(\upsilon,0,S\right)}+\widehat{\Psi}_{\mathcal{P}}^{\left(\upsilon,0,S\right)} & =\left(\widehat{p}_{0}+\widehat{\Pi}_{0}\right)\left(\frac{1}{3}\Psi_{\theta}^{\left(\upsilon,0,S\right)}+\Psi_{\Sigma}^{\left(\upsilon,0,S\right)}\right)-\left(\mu_{0}+p_{0}\right)\Psi_{\mathsf{A}}^{\left(\upsilon,0,S\right)}\\
 & -\Pi_{0}\left(\frac{3}{2}\Psi_{\mathsf{F}}^{\left(\upsilon,0,S\right)}+\Psi_{\mathsf{A}}^{\left(\upsilon,0,S\right)}\right)-\left(\frac{3}{2}\phi_{0}+2\mathcal{A}_{0}\right)\Psi_{\mathcal{P}}^{\left(\upsilon,0,S\right)}\\
 & -\mathcal{A}_{0}\left(\Psi_{\mathsf{m}}^{\left(\upsilon,0,S\right)}+2\Psi_{\mathsf{p}}^{\left(\upsilon,0,S\right)}\right)\,,
\end{aligned}
\end{equation}

\begin{equation}
\Psi_{\mathsf{m}}^{\left(\upsilon,0,S\right)}+\widehat{\Psi}_{Q}^{\left(\upsilon,0,S\right)}=-\left(\phi_{0}+2\mathcal{A}_{0}\right)\Psi_{Q}^{\left(\upsilon,0,S\right)}-\frac{3}{2}\Pi_{0}\Psi_{\Sigma}^{\left(\upsilon,0,S\right)}-\left(\mu_{0}+p_{0}\right)\Psi_{\theta}^{\left(\upsilon,0,S\right)}\,;
\end{equation}
and the constraints

\begin{equation}
\Psi_{\mathsf{F}}^{\left(\upsilon,0,S\right)}=\Psi_{Q}^{\left(\upsilon,0,S\right)}+\left(2\mathcal{A}_{0}-\phi_{0}\right)\left(\frac{1}{3}\Psi_{\theta}^{\left(\upsilon,0,S\right)}-\frac{1}{2}\Psi_{\Sigma}^{\left(\upsilon,0,S\right)}\right)\,,
\end{equation}

\begin{equation}
\Psi_{\mathsf{E}}^{\left(\upsilon,0,S\right)}+\frac{1}{2}\Psi_{\mathcal{P}}^{\left(\upsilon,0,S\right)}-\frac{1}{3}\Psi_{\mathsf{m}}^{\left(\upsilon,0,S\right)}=\left(\mu_{0}+p_{0}-\frac{1}{2}\Pi_{0}-3\mathcal{E}_{0}\right)\left(\frac{1}{3}\Psi_{\theta}^{\left(\upsilon,0,S\right)}-\frac{1}{2}\Psi_{\Sigma}^{\left(\upsilon,0,S\right)}\right)+\frac{1}{2}\phi_{0}\Psi_{Q}^{\left(\upsilon,0,S\right)}\,.
\end{equation}

\subsection{Odd sector\label{subsec:General_eqs_coefficients_Odd}}

\subsubsection{$k$-Modes where $\left\{ \mathcal{Q}_{\alpha\beta}^{\left(k\right)},\mathcal{\bar{Q}}_{\alpha\beta}^{\left(k\right)}\right\} \protect\neq0$}

\paragraph{Equations for the kinematical quantities associated with the timelike
congruence\protect \\
}
\begin{itemize}

\item Evolution and propagation equations for the harmonic coefficients
of scalars and gradients of scalar quantities:
\begin{equation}
\begin{aligned}\widehat{\overline{\Psi}}_{\mathbb{A}}^{\left(\upsilon,k,V\right)}= & -\mathcal{\widehat{A}}_{0}\overline{\Psi}_{a}^{\left(\upsilon,k,V\right)}-\left(\frac{3}{2}\phi_{0}+2\mathcal{A}_{0}\right)\overline{\Psi}_{\mathbb{A}}^{\left(\upsilon,k,V\right)}\\
 & -\mathcal{A}_{0}\overline{\Psi}_{\mathbb{F}}^{\left(\upsilon,k,V\right)}+\frac{1}{2}\left(\overline{\Psi}_{\mathfrak{m}}^{\left(\upsilon,k,V\right)}+3\overline{\Psi}_{\mathfrak{p}}^{\left(\upsilon,k,V\right)}\right)\,,
\end{aligned}
\end{equation}
\begin{equation}
i\upsilon\Psi_{\Omega}^{\left(\upsilon,k,S\right)}=\frac{k^{2}}{2r}\overline{\Psi}_{\mathcal{A}}^{\left(\upsilon,k,V\right)}+\mathcal{A}_{0}\Psi_{\xi}^{\left(\upsilon,k,S\right)}\,,
\end{equation}
\begin{equation}
\widehat{\Psi}_{\Omega}^{\left(\upsilon,k,S\right)}=\left(\mathcal{A}_{0}-\phi_{0}\right)\Psi_{\Omega}^{\left(\upsilon,k,S\right)}+\frac{k^{2}}{r}\Psi_{\Omega}^{\left(\upsilon,k,V\right)}\,;
\end{equation}

\item Evolution and propagation equations for the harmonic coefficients
of vector quantities:
\begin{equation}
i\upsilon\Psi_{\Omega}^{\left(\upsilon,k,V\right)}-\frac{1}{2}\widehat{\overline{\Psi}}_{\mathcal{A}}^{\left(\upsilon,k,V\right)}=\frac{1}{2}\left(\frac{1}{2}\phi_{0}\overline{\Psi}_{\mathcal{A}}^{\left(\upsilon,k,V\right)}+\mathcal{A}_{0}\overline{\Psi}_{a}^{\left(\upsilon,k,V\right)}-\overline{\Psi}_{\mathbb{A}}^{\left(\upsilon,k,V\right)}\right)\,,
\end{equation}

\begin{equation}
i\upsilon\left(\overline{\Psi}_{\Sigma}^{\left(\upsilon,k,V\right)}-\Psi_{\Omega}^{\left(\upsilon,k,V\right)}\right)=\left(\mathcal{A}_{0}-\frac{1}{2}\phi_{0}\right)\overline{\Psi}_{\mathcal{A}}^{\left(\upsilon,k,V\right)}+\overline{\Psi}_{\mathbb{A}}^{\left(\upsilon,k,V\right)}-\overline{\Psi}_{\mathcal{E}}^{\left(\upsilon,k,V\right)}+\frac{1}{2}\overline{\Psi}_{\Pi}^{\left(\upsilon,k,V\right)}\,,
\end{equation}

\begin{equation}
\begin{aligned}\widehat{\overline{\Psi}}_{\Sigma}^{\left(\upsilon,k,V\right)}-\widehat{\Psi}_{\Omega}^{\left(\upsilon,k,V\right)}= & -\frac{1}{r}\Psi_{\Omega}^{\left(\upsilon,k,S\right)}-\frac{3}{2}\phi_{0}\overline{\Psi}_{\Sigma}^{\left(\upsilon,k,V\right)}-\overline{\Psi}_{Q}^{\left(\upsilon,k,V\right)}\\
 & +\left(\frac{1}{2}\phi_{0}+2\mathcal{A}_{0}\right)\Psi_{\Omega}^{\left(\upsilon,k,V\right)}+\frac{2-k^{2}}{2r}\overline{\Psi}_{\Sigma}^{\left(\upsilon,k,T\right)}\,;
\end{aligned}
\end{equation}

\item Evolution and propagation equations for the harmonic coefficients
of tensor quantities:

\begin{equation}
i\upsilon\overline{\Psi}_{\Sigma}^{\left(\upsilon,k,T\right)}=-\frac{1}{r}\overline{\Psi}_{\mathcal{A}}^{\left(\upsilon,k,V\right)}+\mathcal{A}_{0}\overline{\Psi}_{\zeta}^{\left(\upsilon,k,T\right)}-\overline{\Psi}_{\mathcal{E}}^{\left(\upsilon,k,T\right)}+\frac{1}{2}\overline{\Psi}_{\Pi}^{\left(\upsilon,k,T\right)}\,,
\end{equation}
\begin{equation}
\widehat{\overline{\Psi}}_{\Sigma}^{\left(\upsilon,k,T\right)}=-\frac{1}{r}\left(\overline{\Psi}_{\Sigma}^{\left(\upsilon,k,V\right)}+\Psi_{\Omega}^{\left(\upsilon,k,V\right)}\right)-\frac{1}{2}\phi_{0}\overline{\Psi}_{\Sigma}^{\left(\upsilon,k,T\right)}-\Psi_{\mathcal{H}}^{\left(\upsilon,k,T\right)}\,;
\end{equation}

\item Constraint equations for the harmonic coefficients:

\begin{equation}
\frac{k^{2}}{2r}\overline{\Psi}_{\mathbb{A}}^{\left(\upsilon,k,V\right)}+\hat{\mathcal{A}}_{0}\Psi_{\xi}^{\left(\upsilon,k,S\right)}=0\,,
\end{equation}
\begin{equation}
-\Psi_{\mathcal{H}}^{\left(\upsilon,k,S\right)}+\left(\phi_{0}-2\mathcal{A}_{0}\right)\Psi_{\Omega}^{\left(\upsilon,k,S\right)}+\frac{k^{2}}{r}\left(\overline{\Psi}_{\Sigma}^{\left(\upsilon,k,V\right)}-\Psi_{\Omega}^{\left(\upsilon,k,V\right)}\right)=0\,,
\end{equation}

\begin{equation}
\phi_{0}\left(\overline{\Psi}_{\Sigma}^{\left(\upsilon,k,V\right)}-\Psi_{\Omega}^{\left(\upsilon,k,V\right)}\right)+\overline{\Psi}_{Q}^{\left(\upsilon,k,V\right)}+2\Psi_{\mathcal{H}}^{\left(\upsilon,k,V\right)}+\frac{2}{r}\Psi_{\Omega}^{\left(\upsilon,k,S\right)}-\frac{2-k^{2}}{r}\overline{\Psi}_{\Sigma}^{\left(\upsilon,k,T\right)}=0\,,
\end{equation}
\begin{equation}
\phi_{0}\overline{\Psi}_{\mathbb{A}}^{\left(\upsilon,k,V\right)}+\mathcal{A}_{0}\overline{\Psi}_{\mathbb{F}}^{\left(\upsilon,k,V\right)}-\frac{1}{3}\overline{\Psi}_{\mathfrak{m}}^{\left(\upsilon,k,V\right)}-\overline{\Psi}_{\mathfrak{p}}^{\left(\upsilon,k,V\right)}-\frac{1}{2}\overline{\Psi}_{\mathbb{P}}^{\left(\upsilon,k,V\right)}+\overline{\Psi}_{\mathbb{E}}^{\left(\upsilon,k,V\right)}=0\,.
\end{equation}

\end{itemize} \hfill\hfill 
\hfill \hfill

\paragraph{Equations for the kinematical quantities associated with the spacelike
congruence\protect \\
}

\begin{itemize}
\item Evolution and propagation equations for the harmonic coefficients
of scalars and gradients of scalar quantities:

\begin{equation}
i\upsilon\overline{\Psi}_{\mathbb{F}}^{\left(\upsilon,k,V\right)}=-\widehat{\phi}_{0}\left(\overline{\Psi}_{\Sigma}^{\left(\upsilon,k,V\right)}+\overline{\Psi}_{\alpha}^{\left(\upsilon,k,V\right)}-\Psi_{\Omega}^{\left(\upsilon,k,V\right)}\right)\,,
\end{equation}
\begin{equation}
\widehat{\overline{\Psi}}_{\mathbb{F}}^{\left(\upsilon,k,V\right)}=-\widehat{\phi}_{0}\overline{\Psi}_{a}^{\left(\upsilon,k,V\right)}-\frac{2}{3}\overline{\Psi}_{\mathfrak{m}}^{\left(\upsilon,k,V\right)}-\frac{1}{2}\overline{\Psi}_{\mathbb{P}}^{\left(\upsilon,k,V\right)}-\overline{\Psi}_{\mathbb{E}}^{\left(\upsilon,k,V\right)}-\frac{3}{2}\phi_{0}\overline{\Psi}_{\mathbb{F}}^{\left(\upsilon,k,V\right)}\,,
\end{equation}
\begin{equation}
i\upsilon\Psi_{\xi}^{\left(\upsilon,k,S\right)}=\frac{1}{2}\Psi_{\mathcal{H}}^{\left(\upsilon,k,S\right)}+\left(\mathcal{A}_{0}-\frac{1}{2}\phi_{0}\right)\Psi_{\Omega}^{\left(\upsilon,k,S\right)}+\frac{k^{2}}{2r}\overline{\Psi}_{\alpha}^{\left(\upsilon,k,V\right)}\,,
\end{equation}
\begin{equation}
\widehat{\Psi}_{\xi}^{\left(\upsilon,k,S\right)}=-\phi_{0}\Psi_{\xi}^{\left(\upsilon,k,S\right)}+\frac{k^{2}}{2r}\overline{\Psi}_{a}^{\left(\upsilon,k,V\right)}\,;
\end{equation}

\item Evolution and propagation equations for the harmonic coefficients
of vector quantities:
\begin{equation}
\begin{aligned}\widehat{\overline{\Psi}}_{\alpha}^{\left(\upsilon,k,V\right)}-i\upsilon\overline{\Psi}_{a}^{\left(\upsilon,k,V\right)}= & -\Psi_{\mathcal{H}}^{\left(\upsilon,k,V\right)}-\left(\mathcal{A}_{0}+\frac{1}{2}\phi_{0}\right)\overline{\Psi}_{\alpha}^{\left(\upsilon,k,V\right)}+\frac{1}{2}\overline{\Psi}_{Q}^{\left(\upsilon,k,V\right)}\\
 & +\left(\frac{1}{2}\phi_{0}-\mathcal{A}_{0}\right)\left(\overline{\Psi}_{\Sigma}^{\left(\upsilon,k,V\right)}+\Psi_{\Omega}^{\left(\upsilon,k,V\right)}\right)\,;
\end{aligned}
\end{equation}

\item Evolution and propagation equations for the harmonic coefficients
of tensor quantities:

\begin{equation}
i\upsilon\overline{\Psi}_{\zeta}^{\left(\upsilon,k,T\right)}=\left(\mathcal{A}_{0}-\frac{1}{2}\phi_{0}\right)\overline{\Psi}_{\Sigma}^{\left(\upsilon,k,T\right)}-\frac{1}{r}\overline{\Psi}_{\alpha}^{\left(\upsilon,k,V\right)}-\Psi_{\mathcal{H}}^{\left(\upsilon,k,T\right)}\,,
\end{equation}

\begin{equation}
\widehat{\overline{\Psi}}_{\zeta}^{\left(\upsilon,k,T\right)}=-\frac{1}{r}\overline{\Psi}_{a}^{\left(\upsilon,k,V\right)}-\phi_{0}\overline{\Psi}_{\zeta}^{\left(\upsilon,k,T\right)}-\overline{\Psi}_{\mathcal{E}}^{\left(\upsilon,k,T\right)}-\frac{1}{2}\overline{\Psi}_{\Pi}^{\left(\upsilon,k,T\right)}\,;
\end{equation}

\item Constraint equations for the harmonic coefficients:

\begin{equation}
\frac{k^{2}}{2r}\overline{\Psi}_{\mathbb{F}}^{\left(\upsilon,k,V\right)}=-\widehat{\phi}_{0}\Psi_{\xi}^{\left(\upsilon,k,S\right)}\,,
\end{equation}
\begin{equation}
\frac{2}{r}\Psi_{\xi}^{\left(\upsilon,k,S\right)}-\frac{2-k^{2}}{r}\overline{\Psi}_{\zeta}^{\left(\upsilon,k,T\right)}-\overline{\Psi}_{\mathbb{F}}^{\left(\upsilon,k,V\right)}-2\overline{\Psi}_{\mathcal{E}}^{\left(\upsilon,k,V\right)}-\overline{\Psi}_{\Pi}^{\left(\upsilon,k,V\right)}=0\,.
\end{equation}

\end{itemize} \hfill\hfill

\paragraph{Equations for the Weyl tensor components and the matter variables\protect \\
}

\begin{itemize}
\item Evolution and propagation equations for the harmonic coefficients
of scalars and gradients of scalar quantities:

\begin{equation}
\begin{aligned}i\upsilon\overline{\Psi}_{\mathfrak{m}}^{\left(\upsilon,k,V\right)} & =\hat{\mu}_{0}\left(\Psi_{\Omega}^{\left(\upsilon,k,V\right)}-\overline{\Psi}_{\Sigma}^{\left(\upsilon,k,V\right)}-\overline{\Psi}_{\alpha}^{\left(\upsilon,k,V\right)}\right)\,,\end{aligned}
\end{equation}

\begin{equation}
i\upsilon\overline{\Psi}_{\mathbb{E}}^{\left(\upsilon,k,V\right)}+\frac{1}{2}i\upsilon\overline{\Psi}_{\mathbb{P}}^{\left(\upsilon,k,V\right)}=\left(\widehat{\mathcal{E}}_{0}+\frac{1}{2}\widehat{\Pi}_{0}\right)\left(\Psi_{\Omega}^{\left(\upsilon,k,V\right)}-\overline{\Psi}_{\Sigma}^{\left(\upsilon,k,V\right)}-\overline{\Psi}_{\alpha}^{\left(\upsilon,k,V\right)}\right)\,,
\end{equation}

\begin{equation}
\begin{aligned}\widehat{\overline{\Psi}}_{\mathbb{P}}^{\left(\upsilon,k,V\right)}+\widehat{\overline{\Psi}}_{\mathfrak{p}}^{\left(\upsilon,k,V\right)} & =\left(\mu_{0}+p_{0}\right)\left(\mathcal{A}_{0}\overline{\Psi}_{a}^{\left(\upsilon,k,V\right)}-\overline{\Psi}_{\mathbb{A}}^{\left(\upsilon,k,V\right)}\right)-\mathcal{A}_{0}\left(\overline{\Psi}_{\mathfrak{m}}^{\left(\upsilon,k,V\right)}+\overline{\Psi}_{\mathfrak{p}}^{\left(\upsilon,k,V\right)}+\overline{\Psi}_{\mathbb{P}}^{\left(\upsilon,k,V\right)}\right)\\
 & -\frac{1}{2}\phi_{0}\left(\overline{\Psi}_{\mathfrak{p}}^{\left(\upsilon,k,V\right)}+4\overline{\Psi}_{\mathbb{P}}^{\left(\upsilon,k,V\right)}\right)+\left(\frac{3}{2}\phi_{0}+\mathcal{A}_{0}\right)\Pi_{0}\overline{\Psi}_{a}^{\left(\upsilon,k,V\right)}\\
 & -\Pi_{0}\left(\frac{3}{2}\overline{\Psi}_{\mathbb{F}}^{\left(\upsilon,k,V\right)}+\overline{\Psi}_{\mathbb{A}}^{\left(\upsilon,k,V\right)}\right)\,,
\end{aligned}
\end{equation}

\begin{equation}
\begin{aligned}\widehat{\overline{\Psi}}_{\mathbb{E}}^{\left(\upsilon,k,V\right)}+\frac{1}{2}\widehat{\overline{\Psi}}_{\mathbb{P}}^{\left(\upsilon,k,V\right)}-\frac{1}{3}\widehat{\overline{\Psi}}_{\mathfrak{m}}^{\left(\upsilon,k,V\right)}= & -2\phi_{0}\left(\overline{\Psi}_{\mathbb{E}}^{\left(\upsilon,k,V\right)}+\frac{1}{2}\overline{\Psi}_{\mathbb{P}}^{\left(\upsilon,k,V\right)}-\frac{1}{12}\overline{\Psi}_{\mathfrak{m}}^{\left(\upsilon,k,V\right)}\right)\\
 & -\frac{3}{2}\left(\mathcal{E}_{0}+\frac{1}{2}\Pi_{0}\right)\left(\overline{\Psi}_{\mathbb{F}}^{\left(\upsilon,k,V\right)}-\phi_{0}\overline{\Psi}_{a}^{\left(\upsilon,k,V\right)}\right)\,,
\end{aligned}
\end{equation}

\begin{equation}
i\upsilon\Psi_{\mathcal{H}}^{\left(\upsilon,k,S\right)}=\frac{k^{2}}{r}\left(\frac{1}{2}\overline{\Psi}_{\Pi}^{\left(\upsilon,k,V\right)}-\overline{\Psi}_{\mathcal{E}}^{\left(\upsilon,k,V\right)}\right)+3\left(\frac{1}{2}\Pi_{0}-\mathcal{E}_{0}\right)\Psi_{\xi}^{\left(\upsilon,k,S\right)}\,,
\end{equation}
\begin{equation}
\widehat{\Psi}_{\mathcal{H}}^{\left(\upsilon,k,S\right)}=\frac{k^{2}}{r}\left(\Psi_{\mathcal{H}}^{\left(\upsilon,k,V\right)}-\frac{1}{2}\overline{\Psi}_{Q}^{\left(\upsilon,k,V\right)}\right)-\frac{3}{2}\phi_{0}\Psi_{\mathcal{H}}^{\left(\upsilon,k,S\right)}-\left(3\mathcal{E}_{0}+\mu_{0}+p_{0}-\frac{1}{2}\Pi_{0}\right)\Psi_{\Omega}^{\left(\upsilon,k,S\right)}\,;
\end{equation}

\item Evolution and propagation equations for the harmonic coefficients
of vector quantities:

\begin{equation}
\begin{aligned}i\upsilon\left(\overline{\Psi}_{\mathcal{E}}^{\left(\upsilon,k,V\right)}+\frac{1}{2}\overline{\Psi}_{\Pi}^{\left(\upsilon,k,V\right)}\right) & =\left(\frac{1}{2}\phi_{0}-\mathcal{A}_{0}\right)\left(\frac{1}{2}\overline{\Psi}_{Q}^{\left(\upsilon,k,V\right)}+\Psi_{\mathcal{H}}^{\left(\upsilon,k,V\right)}\right)-\frac{3}{2}\left(\mathcal{E}_{0}+\frac{1}{2}\Pi_{0}\right)\overline{\Psi}_{\alpha}^{\left(\upsilon,k,V\right)}\\
 & -\frac{1}{2}\left(\mu_{0}+p_{0}+\Pi_{0}\right)\left(\overline{\Psi}_{\Sigma}^{\left(\upsilon,k,V\right)}-\Psi_{\Omega}^{\left(\upsilon,k,V\right)}\right)\\
 & +\frac{1}{2r}\Psi_{\mathcal{H}}^{\left(\upsilon,k,S\right)}+\frac{2-k^{2}}{2r}\Psi_{\mathcal{H}}^{\left(\upsilon,k,T\right)}\,,
\end{aligned}
\end{equation}
\begin{equation}
\begin{aligned}\widehat{\overline{\Psi}}_{\mathcal{E}}^{\left(\upsilon,k,V\right)}+\frac{1}{2}\widehat{\overline{\Psi}}_{\Pi}^{\left(\upsilon,k,V\right)}= & -\frac{3}{2}\phi_{0}\left(\overline{\Psi}_{\mathcal{E}}^{\left(\upsilon,k,V\right)}+\frac{1}{2}\overline{\Psi}_{\Pi}^{\left(\upsilon,k,V\right)}\right)-\frac{3}{2}\left(\mathcal{E}_{0}+\frac{1}{2}\Pi_{0}\right)\overline{\Psi}_{a}^{\left(\upsilon,k,V\right)}\\
 & +\frac{1}{2}\overline{\Psi}_{\mathbb{E}}^{\left(\upsilon,k,V\right)}+\frac{1}{3}\overline{\Psi}_{\mathfrak{m}}^{\left(\upsilon,k,V\right)}+\frac{1}{4}\overline{\Psi}_{\mathbb{P}}^{\left(\upsilon,k,V\right)}+\frac{2-k^{2}}{2r}\left(\overline{\Psi}_{\mathcal{E}}^{\left(\upsilon,k,T\right)}+\frac{1}{2}\overline{\Psi}_{\Pi}^{\left(\upsilon,k,T\right)}\right)\,,
\end{aligned}
\end{equation}
\begin{equation}
\begin{aligned}\frac{1}{4}\widehat{\overline{\Psi}}_{\Pi}^{\left(\upsilon,k,V\right)}-\frac{1}{2}\widehat{\overline{\Psi}}_{\mathcal{E}}^{\left(\upsilon,k,V\right)}-i\upsilon\Psi_{\mathcal{H}}^{\left(\upsilon,k,V\right)}= & -\frac{3}{4}\left(\overline{\Psi}_{\mathbb{E}}^{\left(\upsilon,k,V\right)}-\frac{1}{2}\overline{\Psi}_{\mathbb{P}}^{\left(\upsilon,k,V\right)}\right)+\frac{2-k^{2}}{4r}\left(\overline{\Psi}_{\mathcal{E}}^{\left(\upsilon,k,T\right)}-\frac{1}{2}\overline{\Psi}_{\Pi}^{\left(\upsilon,k,T\right)}\right)\\
 & -\frac{3}{2}\mathcal{E}_{0}\overline{\Psi}_{\mathcal{A}}^{\left(\upsilon,k,V\right)}+\frac{1}{4}\phi_{0}\left(\overline{\Psi}_{\mathcal{E}}^{\left(\upsilon,k,V\right)}-\frac{1}{2}\overline{\Psi}_{\Pi}^{\left(\upsilon,k,V\right)}\right)\\
 & +\frac{3}{4}\left(\mathcal{E}_{0}-\frac{1}{2}\Pi_{0}\right)\overline{\Psi}_{a}^{\left(\upsilon,k,V\right)}+\mathcal{A}_{0}\overline{\Psi}_{\mathcal{E}}^{\left(\upsilon,k,V\right)}\,,
\end{aligned}
\end{equation}
\begin{equation}
\begin{aligned}\widehat{\Psi}_{\mathcal{H}}^{\left(\upsilon,k,V\right)}+\frac{1}{2}\widehat{\overline{\Psi}}_{Q}^{\left(\upsilon,k,V\right)} & =\frac{1}{2r}\Psi_{\mathcal{H}}^{\left(\upsilon,k,S\right)}-\frac{2-k^{2}}{2r}\Psi_{\mathcal{H}}^{\left(\upsilon,k,T\right)}+\frac{3}{2}\left(\mathcal{E}_{0}+\frac{1}{2}\Pi_{0}\right)\overline{\Psi}_{\Sigma}^{\left(\upsilon,k,V\right)}\\
 & -\frac{3}{2}\phi_{0}\left(\Psi_{\mathcal{H}}^{\left(\upsilon,k,V\right)}+\frac{1}{6}\overline{\Psi}_{Q}^{\left(\upsilon,k,V\right)}\right)-\left(\mu_{0}+p_{0}+\frac{1}{4}\Pi_{0}-\frac{3}{2}\mathcal{E}_{0}\right)\Psi_{\Omega}^{\left(\upsilon,k,V\right)}\,,
\end{aligned}
\end{equation}

\begin{equation}
\begin{aligned}i\upsilon\overline{\Psi}_{Q}^{\left(\upsilon,k,V\right)}+\widehat{\overline{\Psi}}_{\Pi}^{\left(\upsilon,k,V\right)} & =\frac{1}{2}\overline{\Psi}_{\mathbb{P}}^{\left(\upsilon,k,V\right)}-\overline{\Psi}_{\mathfrak{p}}^{\left(\upsilon,k,V\right)}-\frac{3}{2}\Pi_{0}\overline{\Psi}_{a}^{\left(\upsilon,k,V\right)}+\frac{2-k^{2}}{2r}\overline{\Psi}_{\Pi}^{\left(\upsilon,k,T\right)}\\
 & -\left(\frac{3}{2}\phi_{0}+\mathcal{A}_{0}\right)\overline{\Psi}_{\Pi}^{\left(\upsilon,k,V\right)}-\left(\mu_{0}+p_{0}-\frac{1}{2}\Pi_{0}\right)\overline{\Psi}_{\mathcal{A}}^{\left(\upsilon,k,V\right)}\,;
\end{aligned}
\end{equation}

\item Evolution and propagation equations for the harmonic coefficients
of tensor quantities:

\begin{equation}
\begin{aligned}i\upsilon\overline{\Psi}_{\mathcal{E}}^{\left(\upsilon,k,T\right)}+\frac{1}{2}i\upsilon\overline{\Psi}_{\Pi}^{\left(\upsilon,k,T\right)}-\widehat{\Psi}_{\mathcal{H}}^{\left(\upsilon,k,T\right)}= & -\frac{1}{r}\left(\Psi_{\mathcal{H}}^{\left(\upsilon,k,V\right)}-\frac{1}{2}\overline{\Psi}_{Q}^{\left(\upsilon,k,V\right)}\right)+\left(\frac{1}{2}\phi_{0}+2\mathcal{A}_{0}\right)\Psi_{\mathcal{H}}^{\left(\upsilon,k,T\right)}\\
 & -\frac{1}{2}\left(\mu_{0}+p_{0}+3\mathcal{E}_{0}-\frac{1}{2}\Pi_{0}\right)\overline{\Psi}_{\Sigma}^{\left(\upsilon,k,T\right)}\,,
\end{aligned}
\end{equation}

\begin{equation}
\begin{aligned}\widehat{\overline{\Psi}}_{\mathcal{E}}^{\left(\upsilon,k,T\right)}-\frac{1}{2}\widehat{\overline{\Psi}}_{\Pi}^{\left(\upsilon,k,T\right)}-i\upsilon\Psi_{\mathcal{H}}^{\left(\upsilon,k,T\right)} & =\frac{1}{r}\left(\frac{1}{2}\overline{\Psi}_{\Pi}^{\left(\upsilon,k,V\right)}-\overline{\Psi}_{\mathcal{E}}^{\left(\upsilon,k,V\right)}\right)+\frac{1}{4}\phi_{0}\overline{\Psi}_{\Pi}^{\left(\upsilon,k,T\right)}\\
 & +\frac{3}{2}\left(\mathcal{E}_{0}-\frac{1}{2}\Pi_{0}\right)\overline{\Psi}_{\zeta}^{\left(\upsilon,k,T\right)}-\left(\frac{1}{2}\phi_{0}+2\mathcal{A}_{0}\right)\overline{\Psi}_{\mathcal{E}}^{\left(\upsilon,k,T\right)}\,;
\end{aligned}
\end{equation}

\item Constraint equations for the harmonic coefficients:

\begin{equation}
\begin{aligned}\frac{k^{2}}{r}\left(\frac{1}{3}\overline{\Psi}_{\mathfrak{m}}^{\left(\upsilon,k,V\right)}-\overline{\Psi}_{\mathbb{E}}^{\left(\upsilon,k,V\right)}-\frac{1}{2}\overline{\Psi}_{\mathbb{P}}^{\left(\upsilon,k,V\right)}\right) & =-3\phi_{0}\left(\mathcal{E}_{0}+\frac{1}{2}\Pi_{0}\right)\Psi_{\xi}^{\left(\upsilon,k,S\right)}\,,\end{aligned}
\end{equation}

\begin{equation}
\frac{k^{2}}{2r}\overline{\Psi}_{\mathfrak{p}}^{\left(\upsilon,k,V\right)}+\widehat{p}_{0}\Psi_{\xi}^{\left(\upsilon,k,S\right)}=0\,.
\end{equation}

\end{itemize} \hfill\hfill

\subsubsection{$k$-Modes where $\left\{ \mathcal{Q}_{\alpha}^{\left(k\right)},\bar{\mathcal{Q}}_{\alpha}^{\left(k\right)}\right\} \protect\neq0\wedge\left\{ \mathcal{Q}_{\alpha\beta}^{\left(k\right)},\mathcal{\bar{Q}}_{\alpha\beta}^{\left(k\right)}\right\} =0$:
$k^{2}=2$}

As was discussed in the even sector, for the $k^2=2$ modes, the tensor harmonics are not defined, but the vector and scalar harmonics do not vanish necessarily. For those modes, we find the following equations for the harmonic coefficients.
\\

\paragraph{Equations for the kinematical quantities associated with the timelike
congruence\protect \\
}
 \begin{itemize}
 
\item Evolution and propagation equations for the harmonic coefficients
of scalars and gradients of scalar quantities:
\begin{equation}
\widehat{\overline{\Psi}}_{\mathbb{A}}^{\left(\upsilon,k,V\right)}=
-\mathcal{\widehat{A}}_{0}\overline{\Psi}_{a}^{\left(\upsilon,k,V\right)}-
\mathcal{A}_{0}\overline{\Psi}_{\mathbb{F}}^{\left(\upsilon,k,V\right)}-
\left(\frac{3}{2}\phi_{0}+
2\mathcal{A}_{0}\right)\overline{\Psi}_{\mathbb{A}}^{\left(\upsilon,k,V\right)}
+\frac{1}{2}\left(\overline{\Psi}_{\mathfrak{m}}^{\left(\upsilon,k,V\right)}+
3\overline{\Psi}_{\mathfrak{p}}^{\left(\upsilon,k,V\right)}\right)\,,
\end{equation}

\begin{equation}
i\upsilon\Psi_{\Omega}^{\left(\upsilon,k,S\right)}=\frac{1}{r}\overline{\Psi}_{\mathcal{A}}^{\left(\upsilon,k,V\right)}+\mathcal{A}_{0}\Psi_{\xi}^{\left(\upsilon,k,S\right)}\,,
\end{equation}
\begin{equation}
\widehat{\Psi}_{\Omega}^{\left(\upsilon,k,S\right)}=\left(\mathcal{A}_{0}-\phi_{0}\right)\Psi_{\Omega}^{\left(\upsilon,k,S\right)}+\frac{2}{r}\Psi_{\Omega}^{\left(\upsilon,k,V\right)}\,;
\end{equation}

\item Evolution and propagation equations for the harmonic coefficients
of vector quantities:
\begin{equation}
2i\upsilon\Psi_{\Omega}^{\left(\upsilon,k,V\right)}-\widehat{\overline{\Psi}}_{\mathcal{A}}^{\left(\upsilon,k,V\right)}=\frac{1}{2}\phi_{0}\overline{\Psi}_{\mathcal{A}}^{\left(\upsilon,k,V\right)}+\mathcal{A}_{0}\overline{\Psi}_{a}^{\left(\upsilon,k,V\right)}-\overline{\Psi}_{\mathbb{A}}^{\left(\upsilon,k,V\right)}\,,
\end{equation}

\begin{equation}
i\upsilon\left(\overline{\Psi}_{\Sigma}^{\left(\upsilon,k,V\right)}-\Psi_{\Omega}^{\left(\upsilon,k,V\right)}\right)=\left(\mathcal{A}_{0}-\frac{1}{2}\phi_{0}\right)\overline{\Psi}_{\mathcal{A}}^{\left(\upsilon,k,V\right)}+\overline{\Psi}_{\mathbb{A}}^{\left(\upsilon,k,V\right)}-\overline{\Psi}_{\mathcal{E}}^{\left(\upsilon,k,V\right)}+\frac{1}{2}\overline{\Psi}_{\Pi}^{\left(\upsilon,k,V\right)}\,,
\end{equation}

\begin{equation}
\widehat{\overline{\Psi}}_{\Sigma}^{\left(\upsilon,k,V\right)}-\widehat{\Psi}_{\Omega}^{\left(\upsilon,k,V\right)}= -\frac{1}{r}\Psi_{\Omega}^{\left(\upsilon,k,S\right)}-
\frac{3}{2}\phi_{0}\overline{\Psi}_{\Sigma}^{\left(\upsilon,k,V\right)}-
\overline{\Psi}_{Q}^{\left(\upsilon,k,V\right)}
+\left(\frac{1}{2}\phi_{0}+
2\mathcal{A}_{0}\right)\Psi_{\Omega}^{\left(\upsilon,k,V\right)}\,;
\end{equation}

\item Constraint equations for the harmonic coefficients:

\begin{equation}
\frac{1}{r}\overline{\Psi}_{\mathbb{A}}^{\left(\upsilon,k,V\right)}+\widehat{\mathcal{A}}_{0}\Psi_{\xi}^{\left(\upsilon,k,S\right)}=0\,,
\end{equation}
\begin{equation}
-\Psi_{\mathcal{H}}^{\left(\upsilon,k,S\right)}+\left(\phi_{0}-2\mathcal{A}_{0}\right)\Psi_{\Omega}^{\left(\upsilon,k,S\right)}+\frac{2}{r}\left(\overline{\Psi}_{\Sigma}^{\left(\upsilon,k,V\right)}-\Psi_{\Omega}^{\left(\upsilon,k,V\right)}\right)=0\,,
\end{equation}

\begin{equation}
\phi_{0}\left(\overline{\Psi}_{\Sigma}^{\left(\upsilon,k,V\right)}-\Psi_{\Omega}^{\left(\upsilon,k,V\right)}\right)+\overline{\Psi}_{Q}^{\left(\upsilon,k,V\right)}+2\Psi_{\mathcal{H}}^{\left(\upsilon,k,V\right)}+\frac{2}{r}\Psi_{\Omega}^{\left(\upsilon,k,S\right)}=0\,,
\end{equation}
\begin{equation}
\phi_{0}\overline{\Psi}_{\mathbb{A}}^{\left(\upsilon,k,V\right)}+\mathcal{A}_{0}\overline{\Psi}_{\mathbb{F}}^{\left(\upsilon,k,V\right)}-\frac{1}{3}\overline{\Psi}_{\mathfrak{m}}^{\left(\upsilon,k,V\right)}-\overline{\Psi}_{\mathfrak{p}}^{\left(\upsilon,k,V\right)}-\frac{1}{2}\overline{\Psi}_{\mathbb{P}}^{\left(\upsilon,k,V\right)}+\overline{\Psi}_{\mathbb{E}}^{\left(\upsilon,k,V\right)}=0\,.
\end{equation}

 \end{itemize} \hfill\hfill
 
\paragraph{Equations for the kinematical quantities associated with the spacelike
congruence\protect \\
}
\begin{itemize}

\item Evolution and propagation equations for the harmonic coefficients
of scalars and gradients of scalar quantities:

\begin{equation}
i\upsilon\overline{\Psi}_{\mathbb{F}}^{\left(\upsilon,k,V\right)}=-\widehat{\phi}_{0}\left(\overline{\Psi}_{\Sigma}^{\left(\upsilon,k,V\right)}+\overline{\Psi}_{\alpha}^{\left(\upsilon,k,V\right)}-\Psi_{\Omega}^{\left(\upsilon,k,V\right)}\right)\,,
\end{equation}
\begin{equation}
\widehat{\overline{\Psi}}_{\mathbb{F}}^{\left(\upsilon,k,V\right)}=-\widehat{\phi}_{0}\overline{\Psi}_{a}^{\left(\upsilon,k,V\right)}-\frac{2}{3}\overline{\Psi}_{\mathfrak{m}}^{\left(\upsilon,k,V\right)}-\frac{1}{2}\overline{\Psi}_{\mathbb{P}}^{\left(\upsilon,k,V\right)}-\overline{\Psi}_{\mathbb{E}}^{\left(\upsilon,k,V\right)}-\frac{3}{2}\phi_{0}\overline{\Psi}_{\mathbb{F}}^{\left(\upsilon,k,V\right)}\,,
\end{equation}
\begin{equation}
i\upsilon\Psi_{\xi}^{\left(\upsilon,k,S\right)}=\frac{1}{2}\Psi_{\mathcal{H}}^{\left(\upsilon,k,S\right)}+\left(\mathcal{A}_{0}-\frac{1}{2}\phi_{0}\right)\Psi_{\Omega}^{\left(\upsilon,k,S\right)}+\frac{1}{r}\overline{\Psi}_{\alpha}^{\left(\upsilon,k,V\right)}\,,
\end{equation}
\begin{equation}
\widehat{\Psi}_{\xi}^{\left(\upsilon,k,S\right)}=-\phi_{0}\Psi_{\xi}^{\left(\upsilon,k,S\right)}+\frac{1}{r}\overline{\Psi}_{a}^{\left(\upsilon,k,V\right)}\,;
\end{equation}

\item Evolution and propagation equations for the harmonic coefficients
of vector quantities:
\begin{equation}
\begin{aligned}\widehat{\overline{\Psi}}_{\alpha}^{\left(\upsilon,k,V\right)}-i\upsilon\overline{\Psi}_{a}^{\left(\upsilon,k,V\right)}= & -\Psi_{\mathcal{H}}^{\left(\upsilon,k,V\right)}-\left(\mathcal{A}_{0}+\frac{1}{2}\phi_{0}\right)\overline{\Psi}_{\alpha}^{\left(\upsilon,k,V\right)}+\frac{1}{2}\overline{\Psi}_{Q}^{\left(\upsilon,k,V\right)}\\
 & +\left(\frac{1}{2}\phi_{0}-\mathcal{A}_{0}\right)\left(\overline{\Psi}_{\Sigma}^{\left(\upsilon,k,V\right)}+\Psi_{\Omega}^{\left(\upsilon,k,V\right)}\right)\,;
\end{aligned}
\end{equation}

\item Constraint equations for the harmonic coefficients:

\begin{equation}
\frac{1}{r}\overline{\Psi}_{\mathbb{F}}^{\left(\upsilon,k,V\right)}=-\widehat{\phi}_{0}\Psi_{\xi}^{\left(\upsilon,k,S\right)}\,,
\end{equation}

\begin{equation}
\frac{2}{r}\Psi_{\xi}^{\left(\upsilon,k,S\right)}-\overline{\Psi}_{\mathbb{F}}^{\left(\upsilon,k,V\right)}-2\overline{\Psi}_{\mathcal{E}}^{\left(\upsilon,k,V\right)}-\overline{\Psi}_{\Pi}^{\left(\upsilon,k,V\right)}=0\,.
\end{equation}

\end{itemize} \hfill\hfill

\paragraph{Equations for the Weyl tensor components and the matter variables\protect \\
}

\begin{itemize}

\item Evolution and propagation equations for the harmonic coefficients
of scalars and gradients of scalar quantities:

\begin{equation}
\begin{aligned}i\upsilon\overline{\Psi}_{\mathfrak{m}}^{\left(\upsilon,k,V\right)} & =\widehat{\mu}_{0}\left(\Psi_{\Omega}^{\left(\upsilon,k,V\right)}-\overline{\Psi}_{\Sigma}^{\left(\upsilon,k,V\right)}-\overline{\Psi}_{\alpha}^{\left(\upsilon,k,V\right)}\right)\,,\end{aligned}
\end{equation}

\begin{equation}
i\upsilon\overline{\Psi}_{\mathbb{E}}^{\left(\upsilon,k,V\right)}+\frac{1}{2}i\upsilon\overline{\Psi}_{\mathbb{P}}^{\left(\upsilon,k,V\right)}=\left(\widehat{\mathcal{E}}_{0}+\frac{1}{2}\widehat{\Pi}_{0}\right)\left(\Psi_{\Omega}^{\left(\upsilon,k,V\right)}-\overline{\Psi}_{\Sigma}^{\left(\upsilon,k,V\right)}-\overline{\Psi}_{\alpha}^{\left(\upsilon,k,V\right)}\right)\,,
\end{equation}

\begin{equation}
\begin{aligned}\widehat{\overline{\Psi}}_{\mathbb{P}}^{\left(\upsilon,k,V\right)}+\widehat{\overline{\Psi}}_{\mathfrak{p}}^{\left(\upsilon,k,V\right)} & =\left(\mu_{0}+p_{0}\right)\left(\mathcal{A}_{0}\overline{\Psi}_{a}^{\left(\upsilon,k,V\right)}-\overline{\Psi}_{\mathbb{A}}^{\left(\upsilon,k,V\right)}\right)-\mathcal{A}_{0}\left(\overline{\Psi}_{\mathfrak{m}}^{\left(\upsilon,k,V\right)}+\overline{\Psi}_{\mathfrak{p}}^{\left(\upsilon,k,V\right)}+\overline{\Psi}_{\mathbb{P}}^{\left(\upsilon,k,V\right)}\right)\\
 & -\frac{1}{2}\phi_{0}\left(\overline{\Psi}_{\mathfrak{p}}^{\left(\upsilon,k,V\right)}+4\overline{\Psi}_{\mathbb{P}}^{\left(\upsilon,k,V\right)}\right)+\left(\frac{3}{2}\phi_{0}+\mathcal{A}_{0}\right)\Pi_{0}\overline{\Psi}_{a}^{\left(\upsilon,k,V\right)}\\
 & -\Pi_{0}\left(\frac{3}{2}\overline{\Psi}_{\mathbb{F}}^{\left(\upsilon,k,V\right)}+\overline{\Psi}_{\mathbb{A}}^{\left(\upsilon,k,V\right)}\right)\,,
\end{aligned}
\end{equation}

\begin{equation}
\begin{aligned}\widehat{\overline{\Psi}}_{\mathbb{E}}^{\left(\upsilon,k,V\right)}+\frac{1}{2}\widehat{\overline{\Psi}}_{\mathbb{P}}^{\left(\upsilon,k,V\right)}-\frac{1}{3}\widehat{\overline{\Psi}}_{\mathfrak{m}}^{\left(\upsilon,k,V\right)}= & -2\phi_{0}\left(\overline{\Psi}_{\mathbb{E}}^{\left(\upsilon,k,V\right)}+\frac{1}{2}\overline{\Psi}_{\mathbb{P}}^{\left(\upsilon,k,V\right)}-\frac{1}{12}\overline{\Psi}_{\mathfrak{m}}^{\left(\upsilon,k,V\right)}\right)\\
 & -\frac{3}{2}\left(\mathcal{E}_{0}+\frac{1}{2}\Pi_{0}\right)\left(\overline{\Psi}_{\mathbb{F}}^{\left(\upsilon,k,V\right)}-\phi_{0}\overline{\Psi}_{a}^{\left(\upsilon,k,V\right)}\right)\,,
\end{aligned}
\end{equation}

\begin{equation}
i\upsilon\Psi_{\mathcal{H}}^{\left(\upsilon,k,S\right)}=\frac{2}{r}\left(\frac{1}{2}\overline{\Psi}_{\Pi}^{\left(\upsilon,k,V\right)}-\overline{\Psi}_{\mathcal{E}}^{\left(\upsilon,k,V\right)}\right)+3\left(\frac{1}{2}\Pi_{0}-\mathcal{E}_{0}\right)\Psi_{\xi}^{\left(\upsilon,k,S\right)}\,,
\end{equation}
\begin{equation}
\widehat{\Psi}_{\mathcal{H}}^{\left(\upsilon,k,S\right)}=\frac{2}{r}\left(\Psi_{\mathcal{H}}^{\left(\upsilon,k,V\right)}-\frac{1}{2}\overline{\Psi}_{Q}^{\left(\upsilon,k,V\right)}\right)-\frac{3}{2}\phi_{0}\Psi_{\mathcal{H}}^{\left(\upsilon,k,S\right)}-\left(3\mathcal{E}_{0}+\mu_{0}+p_{0}-\frac{1}{2}\Pi_{0}\right)\Psi_{\Omega}^{\left(\upsilon,k,S\right)}\,;
\end{equation}

\item Evolution and propagation equations for the harmonic coefficients
of vector quantities:

\begin{equation}
\begin{aligned}i\upsilon\left(\overline{\Psi}_{\mathcal{E}}^{\left(\upsilon,k,V\right)}+\frac{1}{2}\overline{\Psi}_{\Pi}^{\left(\upsilon,k,V\right)}\right) & =\left(\frac{1}{2}\phi_{0}-\mathcal{A}_{0}\right)\left(\frac{1}{2}\overline{\Psi}_{Q}^{\left(\upsilon,k,V\right)}+\Psi_{\mathcal{H}}^{\left(\upsilon,k,V\right)}\right)-\frac{3}{2}\left(\mathcal{E}_{0}+\frac{1}{2}\Pi_{0}\right)\overline{\Psi}_{\alpha}^{\left(\upsilon,k,V\right)}\\
 & -\frac{1}{2}\left(\mu_{0}+p_{0}+\Pi_{0}\right)\left(\overline{\Psi}_{\Sigma}^{\left(\upsilon,k,V\right)}-\Psi_{\Omega}^{\left(\upsilon,k,V\right)}\right)+\frac{1}{2r}\Psi_{\mathcal{H}}^{\left(\upsilon,k,S\right)}\,,
\end{aligned}
\end{equation}
\begin{equation}
\begin{aligned}\widehat{\overline{\Psi}}_{\mathcal{E}}^{\left(\upsilon,k,V\right)}+\frac{1}{2}\widehat{\overline{\Psi}}_{\Pi}^{\left(\upsilon,k,V\right)}= & -\frac{3}{2}\phi_{0}\left(\overline{\Psi}_{\mathcal{E}}^{\left(\upsilon,k,V\right)}+\frac{1}{2}\overline{\Psi}_{\Pi}^{\left(\upsilon,k,V\right)}\right)-\frac{3}{2}\left(\mathcal{E}_{0}+\frac{1}{2}\Pi_{0}\right)\overline{\Psi}_{a}^{\left(\upsilon,k,V\right)}\\
 & +\frac{1}{2}\overline{\Psi}_{\mathbb{E}}^{\left(\upsilon,k,V\right)}+\frac{1}{3}\overline{\Psi}_{\mathfrak{m}}^{\left(\upsilon,k,V\right)}+\frac{1}{4}\overline{\Psi}_{\mathbb{P}}^{\left(\upsilon,k,V\right)}\,,
\end{aligned}
\end{equation}
\begin{equation}
\begin{aligned}\frac{1}{4}\widehat{\overline{\Psi}}_{\Pi}^{\left(\upsilon,k,V\right)}-\frac{1}{2}\widehat{\overline{\Psi}}_{\mathcal{E}}^{\left(\upsilon,k,V\right)}-i\upsilon\Psi_{\mathcal{H}}^{\left(\upsilon,k,V\right)}= & -\frac{3}{4}\left(\overline{\Psi}_{\mathbb{E}}^{\left(\upsilon,k,V\right)}-\frac{1}{2}\overline{\Psi}_{\mathbb{P}}^{\left(\upsilon,k,V\right)}\right)+\frac{3}{4}\left(\mathcal{E}_{0}-\frac{1}{2}\Pi_{0}\right)\overline{\Psi}_{a}^{\left(\upsilon,k,V\right)}\\
 & -\frac{3}{2}\mathcal{E}_{0}\overline{\Psi}_{\mathcal{A}}^{\left(\upsilon,k,V\right)}+\frac{1}{4}\phi_{0}\left(\overline{\Psi}_{\mathcal{E}}^{\left(\upsilon,k,V\right)}-\frac{1}{2}\overline{\Psi}_{\Pi}^{\left(\upsilon,k,V\right)}\right)\\
 & +\mathcal{A}_{0}\overline{\Psi}_{\mathcal{E}}^{\left(\upsilon,k,V\right)}\,,
\end{aligned}
\end{equation}
\begin{equation}
\begin{aligned}\widehat{\Psi}_{\mathcal{H}}^{\left(\upsilon,k,V\right)}+\frac{1}{2}\widehat{\overline{\Psi}}_{Q}^{\left(\upsilon,k,V\right)} & =\frac{1}{2r}\Psi_{\mathcal{H}}^{\left(\upsilon,k,S\right)}-\frac{3}{2}\phi_{0}\left(\Psi_{\mathcal{H}}^{\left(\upsilon,k,V\right)}+\frac{1}{6}\overline{\Psi}_{Q}^{\left(\upsilon,k,V\right)}\right)\\
 & +\frac{3}{2}\left(\mathcal{E}_{0}+\frac{1}{2}\Pi_{0}\right)\overline{\Psi}_{\Sigma}^{\left(\upsilon,k,V\right)}-\left(\mu_{0}+p_{0}+\frac{1}{4}\Pi_{0}-\frac{3}{2}\mathcal{E}_{0}\right)\Psi_{\Omega}^{\left(\upsilon,k,V\right)}\,,
\end{aligned}
\end{equation}

\begin{equation}
\begin{aligned}i\upsilon\overline{\Psi}_{Q}^{\left(\upsilon,k,V\right)}+\widehat{\overline{\Psi}}_{\Pi}^{\left(\upsilon,k,V\right)} & =\frac{1}{2}\overline{\Psi}_{\mathbb{P}}^{\left(\upsilon,k,V\right)}-\overline{\Psi}_{\mathfrak{p}}^{\left(\upsilon,k,V\right)}-\left(\frac{3}{2}\phi_{0}+\mathcal{A}_{0}\right)\overline{\Psi}_{\Pi}^{\left(\upsilon,k,V\right)}\\
 & -\frac{3}{2}\Pi_{0}\overline{\Psi}_{a}^{\left(\upsilon,k,V\right)}-\left(\mu_{0}+p_{0}-\frac{1}{2}\Pi_{0}\right)\overline{\Psi}_{\mathcal{A}}^{\left(\upsilon,k,V\right)}\,;
\end{aligned}
\end{equation}

\item Constraint equations for the harmonic coefficients:

\begin{equation}
\begin{aligned}\frac{2}{r}\left(\frac{1}{3}\overline{\Psi}_{\mathfrak{m}}^{\left(\upsilon,k,V\right)}-\overline{\Psi}_{\mathbb{E}}^{\left(\upsilon,k,V\right)}-\frac{1}{2}\overline{\Psi}_{\mathbb{P}}^{\left(\upsilon,k,V\right)}\right) & =-3\phi_{0}\left(\mathcal{E}_{0}+\frac{1}{2}\Pi_{0}\right)\Psi_{\xi}^{\left(\upsilon,k,S\right)}\,,\end{aligned}
\end{equation}

\begin{equation}
\frac{1}{r}\overline{\Psi}_{\mathfrak{p}}^{\left(\upsilon,k,V\right)}+\widehat{p}_{0}\Psi_{\xi}^{\left(\upsilon,k,S\right)}=0\,.
\end{equation}

\end{itemize} \hfill\hfill

\subsubsection{\label{subsubsec:General_eqs_coefficients_Odd_k_0}$k$-Modes where
$\mathcal{Q}^{\left(k\right)}\protect\neq0\wedge\left\{ \mathcal{Q}_{\alpha}^{\left(k\right)},\bar{\mathcal{Q}}_{\alpha}^{\left(k\right)}\right\} =0\wedge\left\{ \mathcal{Q}_{\alpha\beta}^{\left(k\right)},\mathcal{\bar{Q}}_{\alpha\beta}^{\left(k\right)}\right\} =0$:
$k=0$}

Once again, as was discussed in the even sector, for $k$-modes where the vector and tensor harmonics are not defined, to describe the dynamics of the perturbations that are not along the directions on the sheets, we will use the relations in Appendix~\ref{Appendix:Linear_eqs_dot-derivatives_variables}. The equations for the harmonic coefficients for the $k=0$ mode are:

\begin{equation}
i\upsilon\Psi_{\Omega}^{\left(\upsilon,0,S\right)}=\mathcal{A}_{0}\Psi_{\xi}^{\left(\upsilon,0,S\right)}\,,
\end{equation}
\begin{equation}
\widehat{\Psi}_{\Omega}^{\left(\upsilon,0,S\right)}=\left(\mathcal{A}_{0}-\phi_{0}\right)\Psi_{\Omega}^{\left(\upsilon,0,S\right)}\,,
\end{equation}
\begin{equation}
i\upsilon\Psi_{\xi}^{\left(\upsilon,0,S\right)}=0\,,
\end{equation}
\begin{equation}
\widehat{\Psi}_{\xi}^{\left(\upsilon,0,S\right)}=-\phi_{0}\Psi_{\xi}^{\left(\upsilon,0,S\right)}\,,
\end{equation}
\begin{equation}
i\upsilon\Psi_{\mathcal{H}}^{\left(\upsilon,0,S\right)}=-3\left(\mathcal{E}_{0}-\frac{1}{2}\Pi_{0}\right)\Psi_{\xi}^{\left(\upsilon,0,S\right)}\,,
\end{equation}
\begin{equation}
\widehat{\Psi}_{\mathcal{H}}^{\left(\upsilon,0,S\right)}=-\frac{3}{2}\phi_{0}\Psi_{\mathcal{H}}^{\left(\upsilon,0,S\right)}-\left(3\mathcal{E}_{0}+\mu_{0}+p_{0}-\frac{1}{2}\Pi_{0}\right)\Psi_{\Omega}^{\left(\upsilon,0,S\right)}\,;
\end{equation}
and the constraint
\begin{equation}
\left(2\mathcal{A}_{0}-\phi_{0}\right)\Psi_{\Omega}^{\left(\upsilon,0,S\right)}+\Psi_{\mathcal{H}}^{\left(\upsilon,0,S\right)}=0\,.
\end{equation}

\section{\label{sec:Adiabatic-isotropic-perturbations}Adiabatic isotropic
perturbations}

As a first application of the system of covariant, gauge invariant
equations presented in the previous section, in the remainder of the article
we will consider linear isotropic perturbations of a static, spherically
symmetric spacetime permeated by a perfect fluid, which we will call, for brevity and its immediate physical application, ``star''. Due to the scope of this article, here we will analyze the
general properties of the system of differential equations and relegate
finding and discussing the solutions for specific background
spacetimes to another article.

\subsection{The equilibrium spacetime\label{subsec:Background_spacetime_static_LRSII_perfect_fluid}}

In the language of the 1+1+2 decomposition, static, spherically symmetric
spacetimes with a perfect fluid source can be completely characterized
by the six scalars $\left\{ \mathcal{A}_{0},\phi_{0},\mathcal{E}_{0},\mu_{0},p_{0},\Lambda\right\} $,
such that all their ``dot derivatives'' vanish. For simplicity, we will
set the cosmological constant to zero: $\Lambda=0$. Then, the covariant
defined scalars verify (cf.~Eqs.~(\ref{eq:LRSII_back_general_eqs_1})
and (\ref{eq:LRSII_back_general_eqs_2})):
\begin{align}
\widehat{p}_{0} & =-\left(\mu_{0}+p_{0}\right)\mathcal{A}_{0}\,,\label{eq:Radial_Adiabatic_p_hat_background}\\
\widehat{\phi}_{0} & =-\frac{1}{2}\phi_{0}^{2}-\frac{2}{3}\mu_{0}-\mathcal{E}_{0}\,,\label{eq:Radial_Adiabatic_phi_hat_background}\\
\widehat{\mathcal{A}}_{0} & =\frac{3}{2}\mathcal{E}_{0}+\left(\frac{1}{2}\phi_{0}-\mathcal{A}_{0}\right)\mathcal{A}_{0}\,,\\
\widehat{\mathcal{E}}_{0}+\frac{3}{2}\phi_{0}\mathcal{E}_{0} & =\frac{1}{3}\widehat{\mu}_{0}\,,
\end{align}
and the constraint
\begin{align}
\frac{1}{3}\left(\mu_{0}+3p_{0}\right)-\mathcal{A}_{0}\phi_{0} & =\mathcal{E}_{0}\,.\label{eq:Radial_Adiabatic_constraint_background}
\end{align}

We will consider the setup where two solutions of the Einstein field equations are smoothly matched at a common timelike hypersurface. The interior of the star is described by a static, spatially compact solution with a perfect fluid source, while the exterior spacetime is described by an asymptotically flat branch of the vacuum Schwarzschild solution with no event horizons.

\subsection{Choice of frame in the perturbed spacetime}

The sheets of the background spacetime have spherical symmetry. Hence, by
choosing the $e$ tangent vector field of the background spacetime
to be aligned with the gradient of the circumferential radius, $r$, we can particularize
the harmonics $\mathcal{Q}^{\left(k\right)}$ to be the spherical
harmonics: $Y_{l m}$, where the eigenvalues $k$ verify $k^{2}=l\left(l+1\right)$,
with $l\geq0$, and $-l\le m\leq l$. Moreover, this choice
of frame implies that,
\begin{equation}
\mathcal{Q}_{\alpha}^{\left(0\right)}=0, \quad \bar{\mathcal{Q}}_{\alpha}^{\left(0\right)}=0, \quad \mathcal{Q}_{\alpha\beta}^{\left(0\right)}=\mathcal{Q}_{\alpha\beta}^{\left(1\right)}=0, \quad \bar{\mathcal{Q}}_{\alpha\beta}^{\left(0\right)}=\bar{\mathcal{Q}}_{\alpha\beta}^{\left(1\right)}=0.
\end{equation}

The perturbed spacetime is assumed to maintain spherical symmetry,
therefore, it is useful also to consider the spacelike vector
field $e$ to not have angular components
and dependencies in the perturbed spacetime. Notice that, at this point, we are free to choose
any smooth mapping between the background and the perturbed spacetimes
given that covariantly defined tensors, vectors, and the quantities in
Eqs.~(\ref{eq:GI_angular_gradients_definition}) and (\ref{eq:GI_dot_derivatives_definition})
are identification gauge invariant. Considering the equations in subsections~\ref{subsec:General_eqs_coefficients_Even}
and \ref{subsec:General_eqs_coefficients_Odd}, since the perturbations
are spherically symmetric, all coefficients, in both even and odd
sectors, with $l\geq1$ vanish identically. Then, only the coefficients
for the $l=0$ mode, the monopole, may not be trivial.

In addition to the imposition that the perturbation maintains the
spherical symmetry of the spacetime, for simplicity, we will further
assume that the perturbation does not generate anisotropic pressure
and is adiabatic, that is, in the comoving matter frame, the perturbation
will not give rise to heat flows within the fluid. Moreover, we will
also impose that the vorticity of the timelike congruence is identically
zero in the perturbed spacetime. These extra conditions set the coefficients
\begin{equation}
\Psi_{\mathcal{P}}^{\left(\upsilon,0,S\right)},\ \Psi_{\Omega}^{\left(\upsilon,0,S\right)},\ \Psi_{\xi}^{\left(\upsilon,0,S\right)},\
\Psi_{\mathcal{H}}^{\left(\upsilon,0,S\right)}=0\,,
\end{equation}
and in the comoving frame 
\begin{equation}
\Psi_{Q}^{\left(\upsilon,0,S\right)}=0\,.
\end{equation}

Now, the choice of frame is not completely determined by the choice
of frame in the background spacetime and the choice that the $e$
vector field does not have angular components and dependencies in
the perturbed spacetime. Since the background spacetime is assumed
to be static, comoving observers with the fluid in equilibrium are
also static. Therefore, to describe the perturbed fluid, we can choose
either to consider a congruence that describes the world-lines of
observers locally comoving with the elements of the fluid or a congruence
that describes the world-lines of static, $\dot{r}=0$, observers.
Both choices are valid, and both have advantages. For the former choice,
the matter field is described by the source fluid in its rest frame, i.e., a perfect fluid model. Hence, it can
be characterized simply by its energy density and pressure. Moreover,
all quantities that characterize the observer directly characterize
the elements of the fluid, allowing for a clear interpretation of
the quantities and how their dependencies give rise to the
various physical effects. In the latter case, we have the constraint
$\frac{2}{3}\theta=\Sigma$ between the expansion and shear kinematical
variables, Eq.~(\ref{Harmonics_eq:r_definitions}). However, for
static observers, the fluid will no longer be perfect: such observers
will measure momentum flows, and we have to consider a heat flow
term in the perturbed stress-energy tensor: $Q\neq0$ (cf. Appendix~\ref{Appendix:isotropic frame transformations}). Nonetheless,
the description of the problem in this frame is greatly simplified,
allowing us to find some general properties of the solutions easily.
Below, we show the system of equations for both frames. 

In order to lighten the notation, here and in the following sections, we will indicate the perturbation-coefficients as 
\begin{equation}
{\Psi}_\chi^{\left(\upsilon,0,S\right)}={\Psi}_\chi\,.
\end{equation}
where $\chi$ is a generic perturbation variable. This will not compromise clarity as only the monopole perturbations mode is nontrivial for the considered setup.

\subsection{Comoving observers}\label{Com_Obs}

In the comoving frame, the perturbed fluid can be modeled by a perfect
fluid. Hence, it can be fully characterized by its energy density and pressure.
To close the system, we have to provide a matter model for the perturbed
fluid. As a simplifying assumption, we will consider that the perturbed
matter fluid still verifies a barotropic equation of state such that
\begin{equation}
p=f\left(\mu\right)\,,
\end{equation}
where $f$ is assumed to be non-vanishing and of class $\mathcal{C}^1$ in some neighborhood containing
$\mu_{0}$. Then, at linear order 
\begin{equation}
\mathsf{p}\approx f'\left(\mu_{0}\right)\mathsf{m}\,,
\label{barotropic_approx}
\end{equation}
where prime represents derivative with respect to the function's parameter,
so that $f'\left(\mu_{0}\right)$ represents the square of the adiabatic speed
of sound, to be assumed non-vanishing in the interior of the perturbed
star.
Notice that the function $f$ does not have to be equal to the equation
of state of the background configuration, in particular,
the equilibrium fluid is not even required to verify a barotropic equation of state.

Imposing the previous conditions in the equations of subsections~\ref{subsubsec:General_eqs_coefficients_Even_k_0}
and \ref{subsubsec:General_eqs_coefficients_Odd_k_0}, we are left
with the following system for the non-trivial coefficients of the
perturbation variables~

\begin{equation}
\widehat{\Psi}_{\mathsf{p}} +2\mathcal{A}_{0}\left(1+\frac{1}{3f'\left(\mu_{0}\right)}\right)\Psi_{\mathsf{p}} =-\left(\mu_{0}+p_{0}\right)\left(\Psi_{\mathsf{A}} +\mathcal{A}_{0}\Psi_{\Sigma} \right)\,,\label{eq:Comoving_Radial_Adiabatic_pdot_hat}
\end{equation}
\begin{equation}
\widehat{\Psi}_{\mathsf{A}} +\left(3\mathcal{A}_{0}-\frac{1}{2}\phi_{0}\right)\Psi_{\mathsf{A}} =\frac{\mathcal{E}_{0}}{\left(\mu_{0}+p_{0}\right)f'\left(\mu_{0}\right)}\Psi_{\mathsf{p}} -\frac{3}{2}\left(\upsilon^{2}+\mathcal{A}_{0}^{2}+\frac{1}{3}\mu_{0}-2\mathcal{E}_{0}\right)\Psi_{\Sigma} \,,\label{eq:Comoving_Radial_Adiabatic_Adot_hat}
\end{equation}
\begin{equation}
\begin{aligned}\widehat{\Psi}_{\Sigma} +\left(\frac{3}{2}\phi_{0}-\frac{2\mathcal{A}_{0}}{3f'\left(\mu_{0}\right)}\right)\Psi_{\Sigma}  & =\frac{2}{3\left(\mu_{0}+p_{0}\right)f'\left(\mu_{0}\right)}\left[\frac{3f''\left(\mu_{0}\right)\widehat{\mu}_{0}+2\mathcal{A}_{0}}{3f'\left(\mu_{0}\right)}+\frac{\widehat{\mu}_{0}}{\mu_{0}+p_{0}}+\mathcal{A}_{0}\right]\Psi_{\mathsf{p}} \\
 & +\frac{2}{3f'\left(\mu_{0}\right)}\Psi_{\mathsf{A}} \,,
\end{aligned}
\label{eq:Comoving_Radial_Adiabatic_sigma_hat}
\end{equation}
and the constraints~

\begin{equation}
\left(\upsilon^{2}+\mathcal{A}_{0}\phi_{0}+\mathcal{A}_{0}^{2}-p_{0}\right)\left(\frac{2}{3}\Psi_{\theta} -\Psi_{\Sigma} \right)=\Psi_{\mathsf{p}} -\phi_{0}\Psi_{\mathsf{A}} \,,\label{eq:Comoving_Radial_Adiabatic_vsquare_constraint}
\end{equation}

\begin{equation}
\Psi_{\mathsf{E}} =\mathcal{E}_{0}\left(\frac{3}{2}\Psi_{\Sigma} +\frac{\Psi_{\mathsf{p}} }{f'\left(\mu_{0}\right)\left(\mu_{0}+p_{0}\right)}\right)-\frac{1}{2}\left(\mu_{0}+p_{0}\right)\Psi_{\Sigma} \,,
\end{equation}

\begin{equation}
\Psi_{\mathsf{F}} =\left(\frac{1}{2}\phi_{0}-\mathcal{A}_{0}\right)\left(\frac{2\Psi_{\mathsf{p}} }{3f'\left(\mu_{0}\right)\left(\mu_{0}+p_{0}\right)}+\Psi_{\Sigma} \right)\,,
\end{equation}
\begin{equation}
\Psi_{\mathsf{m}} =-\left(\mu_{0}+p_{0}\right)\Psi_{\theta} \,,\label{eq:Comoving_Radial_Adiabatic_mudot_constraint}
\end{equation}
\begin{equation}
\Psi_{\mathsf{p}} =f'\left(\mu_{0}\right)\Psi_{\mathsf{m}} \,,\label{eq:Comoving_Radial_Adiabatic_mudot_pdot_eos}
\end{equation}
where Eq.~\eqref{eq:Comoving_Radial_Adiabatic_mudot_pdot_eos} follows from Eq.~\eqref{barotropic_approx}.
It can be readily shown that Eqs.~(\ref{eq:Comoving_Radial_Adiabatic_pdot_hat})--(\ref{eq:Comoving_Radial_Adiabatic_vsquare_constraint})
are consistent with the relation~(\ref{eq:upsilon_hat_A_relation}).

To select the physically acceptable solutions and formalize the boundary
value problem, we impose the following boundary conditions:
\begin{enumerate}[label=(\roman*)]
\item \label{enu:general_boundary_condition_1}the energy density and the
pressure perturbations at the center of the star must be finite in
a neighborhood of the initial instant;
\item \label{enu:general_boundary_condition_2} the spacetime interior to
the perturbed star can be smoothly matched to an exterior vacuum Schwarzschild
spacetime at a timelike hypersurface $\mathfrak B$, which will represent the ``surface of the star''.
\end{enumerate}
From the point of view of the comoving observer, the boundary condition \ref{enu:general_boundary_condition_2}
implies that the pressure of the perturbed fluid is identically zero
at all times at the surface of the star and thus, $\mathsf{p}$, hence $\Psi_{\mathsf{p}} $,
are also identically zero at the boundary $\mathfrak B$.

\subsection{Static observers}

An advantage of the adopted covariant formalism is the ability to
readily change frames without deriving a new set of equations
for the perturbations. Hence, an alternative way to describe isotropic
adiabatic perturbations is to consider the frame of static observers.

As was discussed previously, static observers will not describe the
perturbed fluid as a perfect fluid since, in general, these will measure
a local net momentum flow. Therefore, the relation between the energy density,
pressure and heat flow terms of the perturbed stress-energy tensor
is more complex than that of the comoving frame above. It is shown in Appendix~\ref{Appendix:isotropic frame transformations} 
that
a barotropic equation of state in the comoving frame
leads to the following relation between the matter variables measured
in a static frame
\begin{equation}
	\mathsf{m}=\frac{1}{f'\left(\mu_{0}\right)}\mathsf{p}-
	\frac{1}{\mu_{0}+p_{0}}\left(\widehat{\mu}_{0}-
	\frac{\widehat{p}_{0}}{f'\left(\mu_{0}\right)}\right)Q\,,
	\label{eq:Static_Radial_Adiabatic_corrected_eos}
\end{equation}
where $f'\left(\mu_{0}\right)$ represents the square of the speed
of sound measured in the comoving frame. We highlight that although
we have kept the nomenclature of the previous subsection, $\mathsf{m}$
and $\mathsf{p}$ refer here, respectively, to the dot derivatives
of the energy density and pressure, and $Q$ to the scalar heat flow
density, all three measured by a radially static observer.
In this form, it is immediate to see that if the equilibrium and the perturbed fluids verify the same equation of state, the second term in the right-hand side of Eq.~\eqref{eq:Static_Radial_Adiabatic_corrected_eos} vanishes. 
Therefore, the extra term accounts for the phase change between the two fluids. This is a very interesting result, showing that in the static frame the correction to the equation of state has to be taken into account already at linear level.

Imposing condition $\Psi_{\Sigma} =\frac{2}{3}\Psi_{\theta} $,
Eq.~(\ref{Harmonics_eq:r_definitions}), and using Eq.~(\ref{eq:Static_Radial_Adiabatic_corrected_eos})
in the equations of subsections~\ref{subsubsec:General_eqs_coefficients_Even_k_0}
and \ref{subsubsec:General_eqs_coefficients_Odd_k_0}, we find the
following system for the non-trivial coefficients of the perturbation
variables in the static frame

\begin{equation}
\widehat{\Psi}_{\mathsf{p}} +\left[\frac{\mu_{0}+p_{0}}{\phi_{0}}+\left(2+\frac{1}{f'\left(\mu_{0}\right)}\right)\mathcal{A}_{0}\right]\Psi_{\mathsf{p}} =\left[\frac{\mu_{0}+p_{0}}{\phi_{0}}\left(\frac{1}{2}\phi_{0}+2\mathcal{A}_{0}\right)+\frac{\mathcal{A}_{0}^{2}}{f'\left(\mu_{0}\right)}+\frac{\mathcal{A}_{0}\widehat{\mu}_{0}}{\mu_{0}+p_{0}}+\upsilon^{2}\right]\Psi_{Q} \,,\label{eq:Static_Radial_Adiabatic_pdot_hat}
\end{equation}

\begin{equation}
\widehat{\Psi}_{Q} +\left(\phi_{0}+2\mathcal{A}_{0}-\frac{\mathcal{A}_{0}}{f'\left(\mu_{0}\right)}-\frac{\widehat{\mu}_{0}}{\mu_{0}+p_{0}}-\frac{\mu_{0}+p_{0}}{\phi_{0}}\right)\Psi_{Q} =-\frac{1}{f'\left(\mu_{0}\right)}\Psi_{\mathsf{p}} \,;
\label{eq:Static_Radial_Adiabatic_Q_hat}
\end{equation}
and the constraints
\begin{equation}
\begin{aligned}\Psi_{\mathsf{m}}  & =\frac{1}{f'\left(\mu_{0}\right)}\Psi_{\mathsf{p}} -\left(\frac{\mathcal{A}_{0}}{f'\left(\mu_{0}\right)}+\frac{\widehat{\mu}_{0}}{\mu_{0}+p_{0}}\right)\Psi_{Q} \,,\\
\Psi_{\mathsf{A}}  & =\frac{1}{\phi_{0}}\left[\Psi_{\mathsf{p}} -\left(\frac{1}{2}\phi_{0}+\mathcal{A}_{0}\right)\Psi_{Q} \right]\,,\\
\Psi_{\mathsf{E}}  & =\frac{1}{2}\phi_{0}\Psi_{Q} +\frac{1}{3f'\left(\mu_{0}\right)}\Psi_{\mathsf{p}} \,,\\
\Psi_{\theta}  & =-\frac{1}{\phi_{0}}\Psi_{Q} \,,\\
\Psi_{\Sigma}  & =\frac{2}{3}\Psi_{\theta} \,,\\
\Psi_{\mathsf{F}}  & =\Psi_{Q} \,.
\label{eq:Static_Radial_Adiabatic_constraints}
\end{aligned}
\end{equation}

To select the physically relevant solutions, we impose the boundary
conditions \ref{enu:general_boundary_condition_1} and \ref{enu:general_boundary_condition_2}.
In particular, for the static observer, condition~\ref{enu:general_boundary_condition_2}
implies that
at the surface of the star, at all times, we must have (cf, Eq.~(\ref{Isotropic_frame_transformations_eq:relation_static_comoving_matter_variables_final}))
\begin{equation}
\left.\mathsf{p}-\mathcal{A}_{0}Q\right|_{\mathfrak B}=0\,,
\end{equation}
hence
\begin{equation}
\left.\Psi_{\mathsf{p}} -\mathcal{A}_{0}\Psi_{Q} \right|_{\mathfrak B}=0\,.\label{eq:Static_Radial_Adiabatic_boundary_condition_coefficients}
\end{equation}

\section{General solutions, Sturm-Liouville problem and properties of eigenfrequencies}

The system~(\ref{eq:Comoving_Radial_Adiabatic_pdot_hat})--(\ref{eq:Comoving_Radial_Adiabatic_mudot_pdot_eos})
found for the comoving frame and the one found for the static frame,
Eqs.~(\ref{eq:Static_Radial_Adiabatic_pdot_hat})--(\ref{eq:Static_Radial_Adiabatic_constraints}),
both describe the same setup. Nonetheless, comparing the equations
of each system, it is manifest that there is a trade-off when considering
one picture over the other: in the comoving frame, the characterization
of the matter fluid is simpler, but the dynamic description of the
perturbed star is more complex, whereas from the point of view of
static observers the description of the dynamics of the perturbed
star is simpler, but the characterization of the matter fluid is more
complex. The solutions of each system with appropriate boundary
conditions are exactly the same, however when analyzing the properties
and computing the solutions, from a technical point of view, it is easier
to work in the static frame.

One key result that can be readily proven from Eqs.~(\ref{eq:Static_Radial_Adiabatic_pdot_hat})
and (\ref{eq:Static_Radial_Adiabatic_Q_hat}) is that if the background
spacetime is a solution of the Tolman-Oppenheimer-Volkoff equations
for energy density and pressure functions, $\mu_{0}$ and $p_{0}$,
that are real analytic in the interior and boundary of the star, we
can find real analytic solutions for the perturbations. Moreover, in those conditions the coefficients $\Psi_{Q} $
verify a Sturm-Liouville eigenvalue problem with a limit-point-non-oscillating
(LPNO) endpoint and a regular endpoint, with separated self-adjoint
boundary conditions. This fact, in particular, allows us to relate
the properties of the eigenfrequencies $\upsilon$ with the properties
of the eigenvalues of the Sturm-Liouville problem. This section focuses on the derivation of these results.

\subsection{General solutions for the perturbations}

To find explicit solutions for the system~(\ref{eq:Static_Radial_Adiabatic_pdot_hat})
and (\ref{eq:Static_Radial_Adiabatic_Q_hat}),  we must break covariance. In that regard, consider the parameter $r$ defined
in Eq.~(\ref{Harmonics_eq:r_definitions}). Following Ref.~\citep{Betschart_Clarkson_2004},
if the sheets are isometric to 2-spheres, $r^{-2}$ is, up to a multiplicative
constant, equal to the Gauss curvature of the sheets. Therefore, the
parameter $r$ represents the circumferential radius function both
in the background and the perturbed spacetime. Consequently, let $r=0$
represents the center $\mathfrak C$ of the star, and we will consider that the background
quantities $\left\{ \mathcal{A}_{0},\phi_{0},\mathcal{E}_{0},\mu_{0},p_{0}\right\} $
and the coefficients $\Psi_{\mathsf{p}} $
and $\Psi_{Q} $ are functions of $r$.

Now, we will impose that the following regularity constraints hold:
\begin{itemize}
\item \label{enu:regularity_conditions}the equilibrium fluid verifies the
weak energy condition;
\item the background spacetime is a solution of the Tolman-Oppenheimer-Volkoff
equation for real analytic, non-trivial energy density and pressure
functions for the whole range within the equilibrium star;
\item the square of the speed of sound of the perturbed fluid, $f'$, is
positive and real analytic in the interior and at the boundary of
the star.
\end{itemize}
Requiring real analytical background solutions is a rather strong
constraint; nonetheless, to our knowledge, all classical exact solutions
for compact astrophysical objects, verify this hypothesis in some open
neighborhood of the center, at $r=0$. Therefore, the following results
are appropriate for treating perturbations of physically relevant setups.
The radius of convergence of the power series, of course, may or may not
be greater than the radius of the equilibrium star, in which case
further treatment has to be carried out, nonetheless, for the following
results, it suffices that the radius of convergence of the power series
around the center is non-zero.

Imposing these conditions, Eqs.~(\ref{eq:Radial_Adiabatic_phi_hat_background}),
(\ref{eq:Radial_Adiabatic_constraint_background}) and (\ref{Harmonics_eq:r_definitions_algebraic}) imply
\begin{equation}
\begin{aligned}\phi_{0} & =\frac{2}{r}\sqrt{1-\frac{2M\left(r\right)}{r}}\,,\\
\mathcal{E}_{0} & =\frac{1}{3}\mu_{0}-\frac{2M\left(r\right)}{r^{3}}\,,\\
\mathcal{A}_{0}\phi_{0} & =p_{0}+\frac{2M\left(r\right)}{r^{3}}\,,
\end{aligned}
\end{equation}
where 
\begin{equation}
M\left(r\right):=\frac{1}{2}\int_{0}^{r}\mu_{0}x^{2}dx\,,
\end{equation} 
is usually dubbed the mass function, and Eq.~(\ref{eq:upsilon_hat_A_relation})
yields
\begin{equation}
\upsilon\left(r\right)=\lambda e^{-\int\frac{2\mathcal{A}_{0}}{r\phi_{0}}dr}\,,\label{eq:Radial_Adiabatic_eigenfrequencies_v-sigma_relation}
\end{equation}
relating the eigenfrequencies
measured by an observer comoving with the fluid, $\upsilon$, with
the constant eigenfrequencies $\lambda$, measured by a free-falling observer
at spatial infinity. 
If the functions $\mu_{0}$ and $p_{0}$ verify the weak
energy condition and are real analytic within the star, so are the
functions $\mathcal{A}_{0}$ and $\mathcal{E}_{0}$, and $\phi_{0}$
has a simple pole at the center $\mathfrak C$, but is otherwise real analytic
in the interior and boundary of the star.

Hence, in the considered
setup, Eqs.~(\ref{eq:Static_Radial_Adiabatic_pdot_hat}) and (\ref{eq:Static_Radial_Adiabatic_Q_hat})
form a system of ordinary differential equations with real analytic
coefficients in a neighborhood of $\mathfrak C$, with a simple pole at $r=0$, and solutions
can be found around the singular point.

Using Eq.~\eqref{Harmonics_eq:r_definitions} to relate the hat derivatives
with the derivatives with respect to $r$ of first order
quantities, the system of ODEs~(\ref{eq:Static_Radial_Adiabatic_pdot_hat})
and (\ref{eq:Static_Radial_Adiabatic_Q_hat}) is given by

\begin{equation}
\frac{r\phi_{0}}{2}\frac{d\Psi_{\mathsf{p}}}{dr}+\left[\frac{\mu_{0}+p_{0}}{\phi_{0}}+\left(2+\frac{1}{f'\left(\mu_{0}\right)}\right)\mathcal{A}_{0}\right]\Psi_{\mathsf{p}}=\left[\frac{\mu_{0}+p_{0}}{\phi_{0}}\left(\frac{1}{2}\phi_{0}+2\mathcal{A}_{0}\right)+\frac{\mathcal{A}_{0}^{2}}{f'\left(\mu_{0}\right)}+\frac{r\mathcal{A}_{0}\phi_{0}}{2\left(\mu_{0}+p_{0}\right)}\frac{d\mu_{0}}{dr}+\upsilon^{2}\right]\Psi_{Q}\,,
\end{equation}

\begin{equation}
\frac{r\phi_{0}}{2}\frac{d\Psi_{Q}}{dr}+\left(\phi_{0}+2\mathcal{A}_{0}-\frac{\mathcal{A}_{0}}{f'\left(\mu_{0}\right)}-\frac{r\phi_{0}}{2\left(\mu_{0}+p_{0}\right)}\frac{d\mu_{0}}{dr}-\frac{\mu_{0}+p_{0}}{\phi_{0}}\right)\Psi_{Q}=-\frac{1}{f'\left(\mu_{0}\right)}\Psi_{\mathsf{p}}\,.
\end{equation}
To find solutions for this system, it is useful to write it in the
matrix form: 
\begin{equation}
\frac{d \mathds{W}}{dr}=\left( \frac{1}{r}\mathds{R}+\Theta\right)\mathds{W}\,,
\end{equation}
where
\begin{equation}
\begin{aligned}\mathds{W} & =\left[\begin{array}{c}
\Psi_{\mathsf{p}} \\
\Psi_{Q} 
\end{array}\right]\,, &  &  & \mathds{R} & =\left[\begin{array}{lr}
0 & 0\\
0 & -2
\end{array}\right]\,,\end{aligned}
\end{equation}
and

\begin{equation}\Theta=-\frac{2}{r\phi_{0}}\left[\begin{array}{lr}
\frac{\mu_{0}+p_{0}}{\phi_{0}}+2\mathcal{A}_{0}+\frac{\mathcal{A}_{0}}{f'\left(\mu_{0}\right)} & \qquad-\frac{\mu_{0}+p_{0}}{\phi_{0}}\left(\frac{1}{2}\phi_{0}+2\mathcal{A}_{0}\right)-\frac{\mathcal{A}_{0}^{2}}{f'\left(\mu_{0}\right)}-\frac{r\mathcal{A}_{0}\phi_{0}}{2\left(\mu_{0}+p_{0}\right)}\frac{d\mu_{0}}{dr}-\upsilon^{2}\\
\frac{1}{f'\left(\mu_{0}\right)} & 2\mathcal{A}_{0}-\frac{\mathcal{A}_{0}}{f'\left(\mu_{0}\right)}-\frac{r\phi_{0}}{2\left(\mu_{0}+p_{0}\right)}\frac{d\mu_{0}}{dr}-\frac{\mu_{0}+p_{0}}{\phi_{0}}
\end{array}\right]\,.
\end{equation}

The regularity conditions that we have imposed on the thermodynamic
variables of the equilibrium configuration, $\mu_{0}$ and $p_{0}$,
and on the equation of state of the perturbed fluid in the comoving
frame, $f$, guarantee that the matrix $\Theta$ is analytic at $r=0$
and $r\phi_{0}$ does not vanish in the interior or boundary of the
star. These conditions imply that $r=0$ is a regular singular point
of the system.

To solve the system
of ordinary differential equations, we will follow
the formalism in Ref.~\citep{Coddington_Levinson_Book}. Since the $\Theta$ 
is assumed to be a real analytic matrix at the center of the star, 
it can be expanded in a convergent
power series of the form 
\begin{equation}
\Theta\left(r\right)=\sum_{n=0}^{+\infty}\Theta_{n}r^{n}\,.
\label{eq:Theta_matrix_power_series}
\end{equation}
Then, the solution matrix $\mathds{W}$ can be written in a power series guaranteed
to converge to the solution in a neighborhood of $r=0$. The radius
of convergence of the power series solution is equal, except for possibly
at $r=0$, to the radius of convergence of the power series of $\Theta$,
which, of course, depends on the equilibrium background spacetime considered.

Before proceeding, we remark that the general
form of the solutions of the boundary value problem~(\ref{eq:Static_Radial_Adiabatic_pdot_hat})--(\ref{eq:Static_Radial_Adiabatic_boundary_condition_coefficients})
for a general static, spherically symmetric spacetime can be rather
complicated. However, imposing the previous regularity conditions on
$\mu_{0}$ and $p_{0}$, and assuming $f'\left(\mu_{0}\right)$ is
not zero in a neighborhood of the center of the star, it can be shown
that various entries of the $\Theta_{0}$ and $\Theta_{1}$ coefficient-matrices of the series~\eqref{eq:Theta_matrix_power_series} are
zero, which greatly simplifies the general family of solutions.

Taking into consideration the regularity of the background yields the power series solution
\begin{equation}
\left[\begin{array}{c}
\Psi_{\mathsf{p}} \\
\Psi_{Q} 
\end{array}\right]=\left[\begin{array}{cc}
-\frac{1}{r}\left(\Theta_{0}\right)_{12} & \quad1\\
\frac{1}{r^{2}} & 0
\end{array}\right]\mathds{P}_{\mathds{W}}\left[\begin{array}{c}
c_{1}\\
c_{2}
\end{array}\right]\,,\label{eq:Radial_Adiabatic_static_matrix_system_solutions}
\end{equation}
where $c_{1}$ and $c_{2}$ are integration constants, which might
depend on the eigenfrequencies $\upsilon$, or equivalently on $\lambda$,
defined in Eq.~(\ref{eq:Radial_Adiabatic_eigenfrequencies_v-sigma_relation}).
The notation $\left(\Theta_{n}\right)_{ij}$ is to be interpreted
as the $ij$-entry of the $n$th order coefficient of the power
series of $\Theta$, and $\mathds{P}_{\mathds{W}}$ is a real analytic matrix,
such that
\begin{equation}
\begin{aligned}
\mathds{P}_{\mathds{W}}\left(r\right)&=\sum_{n=0}^{+\infty}\mathds{P}_{n}r^{n}\\
\mathds{P}_{0} & =\mathds{I}_{2}\\
\mathds{P}_{k} & =\frac{1}{k}\sum_{j=0}^{k-1}\mathds{A}_{k-1-j}\mathds{P}_{j}\,,\quad\text{for }k\geq1
\end{aligned}
\label{eq:Radial_Adiabatic_static_matrix_system_recurrence_relation}
\end{equation}
where $\mathds{I}_{2}$ represents the $2\times2$ identity matrix, the matrix
$\mathds{A}$ is given by
\begin{equation}
\mathds{A}=\left[\begin{array}{lr}
\Theta_{22}-r\left(\Theta_{0}\right)_{12}\Theta_{21} & r^{2}\Theta_{21}\\
\frac{\Theta_{12}-\left(\Theta_{0}\right)_{12}}{r^{2}}+\frac{\left(\Theta_{0}\right)_{12}\left(\Theta_{22}-\Theta_{11}\right)}{r}-\left(\Theta_{0}\right)_{12}^{2}\Theta_{21} & \quad\Theta_{11}+r\left(\Theta_{0}\right)_{12}\Theta_{21}
\end{array}\right]\,,
\end{equation}
and $\mathds{A}_{n}$ represents the $n$th order coefficient of its power
series, that is, $\mathds{A}\left(r\right)=\sum_{n=0}^{+\infty}\mathds{A}_{n}r^{n}$.

The general family of solutions of the system, for $0<r<a$, where
$a\in\mathbb{R}_{>0}$, is given by Eqs.~(\ref{eq:Radial_Adiabatic_static_matrix_system_solutions})
and (\ref{eq:Radial_Adiabatic_static_matrix_system_recurrence_relation}).
To select the physically acceptable solutions and extend the domain to $r=0$, we will impose the
boundary conditions~\ref{enu:general_boundary_condition_1} and \ref{enu:general_boundary_condition_2} of Sec.~\ref{Com_Obs}.
Considering $\mathds{P}_{0}=\mathds{I}_{2}$, we can directly compute the lower order
coefficients of the power series expansion of $\mathds{W}$. Imposing
the boundary condition at the center sets the coefficient $c_{1}$
to be zero so that the perturbations do not diverge at $r=0$ at all times. Then, we find
\begin{equation}
\left[\begin{aligned}\Psi_{\mathsf{p}} \\
\Psi_{Q} 
\end{aligned}
\right]=\left[\begin{aligned}c_{2}+\mathcal{O}\left(r^{2}\right)\\
\mathcal{O}\left(r\right)
\end{aligned}
\right]\,.
\end{equation}

Provided the background spacetime, the equation of state of the perturbed fluid, and the values of the eigenfrequencies, these results allow us to find analytic solutions for
the perturbations that verify the boundary conditions. 
The eigenfrequencies themselves cannot be restricted directly
from these results. Nonetheless, in the following subsection, we will show that 
general useful properties for the eigenfrequencies can be derived by relating $\upsilon$
with the eigenvalues of a Sturm-Liouville problem.

\subsection{Sturm-Liouville eigenvalue problem}

We do not need to break covariance for the following result. Therefore, to keep the discussion independent of a local coordinate system, we
will consider the hat derivatives without associating them with
derivatives in a specific coordinate system. Let $\ell$ be an affine parameter of the
$e$ congruence, and without loss of generality, we set that $\ell=0$ 
represents the center of the star and $\ell=\ell_{\mathfrak B}$ the boundary
of the equilibrium star.

Taking the hat derivative of Eq.~(\ref{eq:Static_Radial_Adiabatic_Q_hat})
we find~
\begin{equation}
D_{e}\left[\exp{\left(\int_{ \ell_{0}}^{\ell}F\left(x\right)dx\right)}\widehat{\Psi}_{Q} \right]+\exp{\left(\int_{\ell_{0}}^{\ell}F\left(x\right)dx\right)}G\left(\ell\right)\Psi_{Q} =-\frac{1}{f'\left(\mu_{0}\right)}\exp{\left(\int_{\ell_{0}}^{\ell}F\left(x\right)dx\right)}\lambda^{2}\Psi_{Q} \,,\label{eq:Static_Radial_Adiabatic_Sturm-Liouville_form}
\end{equation}
where, for notational convenience, we have indicated the hat derivatives as $D_{e}\equiv e^{\alpha}D_{\alpha}$, $\ell_{0}\in\mathbb{R}$,
$\lambda$ is an integration constant following from Eq.~(\ref{eq:upsilon_hat_A_relation})
and
\begin{equation}
\begin{aligned}F\left(\ell\right) & =\phi_{0}+4\mathcal{A}_{0}-\frac{\widehat{\mu}_{0}}{\mu_{0}+p_{0}}+\frac{f''\left(\mu_{0}\right)\widehat{\mu}_{0}}{f'\left(\mu_{0}\right)}\,,\\
G\left(\ell\right) & =\frac{f''\left(\mu_{0}\right)\widehat{\mu}_{0}}{f'\left(\mu_{0}\right)}\left(\phi_{0}+2\mathcal{A}_{0}-\frac{\widehat{\mu}_{0}}{\mu_{0}+p_{0}}-\frac{\mu_{0}+p_{0}}{\phi_{0}}\right)+\frac{1}{f'\left(\mu_{0}\right)}\left(2\mathcal{A}_{0}\phi_{0}-p_{0}+\mathcal{A}_{0}^{2}\right)\\
 & +\mu_{0}+3p_{0}+\mathcal{A}_{0}\phi_{0}+2\mathcal{A}_{0}^{2}-\frac{1}{2}\phi_{0}^{2}-\frac{2\mathcal{A}_{0}\widehat{\mu}_{0}}{\mu_{0}+p_{0}}-\frac{\mu_{0}+p_{0}}{\phi_{0}}\left(\frac{\widehat{\mu}_{0}}{\mu_{0}+p_{0}}+\frac{\mu_{0}+p_{0}}{\phi_{0}}\right)\\
 & -\left[D_{e}\left(\frac{\widehat{\mu}_{0}}{\mu_{0}+p_{0}}\right)+D_{e}\left(\frac{\mu_{0}+p_{0}}{\phi_{0}}\right)\right]\,.
\end{aligned}
\end{equation}
Moreover, conditions~\ref{enu:general_boundary_condition_1} and
\ref{enu:general_boundary_condition_2} of Sec.~\ref{Com_Obs} imply the following set of
separated self-adjoint boundary conditions:
\begin{equation}
\left\{ \begin{aligned}\left.\Psi_{Q} \right|_{\mathfrak C} & =0\\
\left.\widehat{\Psi}_{Q} +\left(\phi_{0}+2\mathcal{A}_{0}-\frac{\widehat{\mu}_{0}}{\mu_{0}+p_{0}}-\frac{\mu_{0}+p_{0}}{\phi_{0}}\right)\Psi_{Q} \right|_{\mathfrak B} & =0
\end{aligned}
\right.
\label{eq:Static_Radial_Adiabatic_boundary_condition_Sturm-Liouville-1}
\end{equation}
Therefore, provided the background spacetime and $f'$ are sufficiently
regular, $\Psi_{Q} $ verifies a formal
Sturm-Liouville eigenvalue problem with weight function
\begin{equation}
w\left(\ell\right)=\frac{1}{f'\left(\mu_{0}\right)}\exp{\left(\int_{\ell_{0}}^{\ell}F\left(x\right)dx\right)}\,,
\end{equation}
and eigenvalues $\lambda^{2}$. This last assumption, however, is not
trivial, and we have to specify the conditions under which the associated
Sturm-Liouville operator is self-adjoint.

Considering the discussion in the previous subsection, in particular
the regularity conditions in subsec~\ref{enu:regularity_conditions},
the function $\phi_{0}$ has a simple pole at $\mathfrak{C}$.
This endpoint, at $\ell=0$, is limit-point since we have found solutions
that are not in $L^{2}\left(\left]0,\delta\right[,w\right)$, for
any $\delta\in\mathbb{R}_{>0}$. However, in the conditions considered
in the previous subsection, we have explicitly shown that all solutions
for $\Psi_{Q} $ are real analytic in $\left]0,\varepsilon\right[$,
for some $\varepsilon\in\mathbb{R}_{>0}$, hence there are no non-trivial
solutions that have an infinite number of zeros in a right-neighborhood
of $\ell=0$. We can then conclude that the endpoint $\ell=0$ is
LPNO. Moreover, the regularity conditions guarantee that the coefficients
of the differential equation verify
\begin{equation}
w(\ell)>0, \quad  \exp{\left(\int_{\ell_{0}}^{\ell}F\left(x\right)dx\right)} >0\,,
\end{equation} 
almost everywhere and 
\begin{equation}
\exp{\left(-\int_{\ell_{0}}^{\ell}F\left(x\right)dx\right)}, \; \exp{\left(\int_{\ell_{0}}^{\ell}F\left(x\right)dx\right)}G\left(\ell\right),\; w\left(\ell\right) \; \in L_{\text{loc}}\left(\left]0,\ell_{\mathfrak B}\right[,\mathbb{R}\right)
\end{equation}
where $ L_{\text{loc}}\left(\left]0,\ell_{\mathfrak B}\right[,\mathbb{R}\right)$ is the set of locally Lebesgue integrable real functions on $\left]0,\ell_{\mathfrak B}\right[$.  Given the previous considerations, we can apply the results in Refs.~\citep{Zettl_book_2005, Zhang_Sun_Zettl_2014}
to Eq.~(\ref{eq:Static_Radial_Adiabatic_Sturm-Liouville_form}) together
with the boundary conditions~(\ref{eq:Static_Radial_Adiabatic_boundary_condition_Sturm-Liouville-1}),
concluding that the eigenvalues verify several very useful properties.
In particular, the eigenvalues $\lambda^{2}$ are (i) real, (ii) simple, (iii) countable, and (iv) have a minimum and are unbounded from above, that is $\lambda^{2}=\left\{ \lambda_{0},\lambda_{1},\lambda_{2},...\right\} $, with $\lambda_{0}<\lambda_{1}<\lambda_{2}<...\to+\infty$, and $\lambda_{0}\in\mathbb{R}$. Indeed, in the following subsection, we show explicitly that, in the considered
setup, the eigenfrequencies are such that $\upsilon^{2}$ are bounded from below and derive necessary constraints for the minimum value.

\subsection{Lower bound for the square of the eigenfrequencies}

The result that $\Psi_{Q} $ verifies a
Sturm-Liouville eigenvalue problem is of great importance for the
description of the behavior of the perturbed star. The fact that the
eigenvalues $\lambda^{2}$, hence $\upsilon^{2}$, are bounded from
below sets that either in the oscillating case or in the continuous
collapse or expanding scenarios, the eigenfrequencies of the excitable
eigenmodes are not arbitrarily small. In addition to those results,
in the considered setup, we will prove the following:

\begin{prop}
\label{Proposition:minimum_bound_eigenvalues_1}
Let the boundary conditions~\ref{enu:general_boundary_condition_1} and \ref{enu:general_boundary_condition_2} of Sec.~\ref{Com_Obs} and the regularity conditions in Sec.~\ref{enu:regularity_conditions} hold. If, in addition, $\widehat{\mathcal{A}}_0 \left( \ell_\mathfrak B \right)\geq0$, then non-trivial $\mathcal{C}^{1}$ solutions of the system (\ref{eq:Static_Radial_Adiabatic_pdot_hat}) and (\ref{eq:Static_Radial_Adiabatic_Q_hat}) exist only if
\begin{equation}
\upsilon^{2}\geq\left.-\mathcal{A}_{0}\phi_{0}-\frac{1}{2}\mu_{0}\right|_{\ell_{\mathfrak B}}\,.
\label{eq:minimum_bound_eigenvalues_1}
\end{equation}
\end{prop}

The premises of Proposition~\ref{Proposition:minimum_bound_eigenvalues_1} can be significantly relaxed, leading to the following result:

\begin{prop}
\label{Proposition:minimum_bound_eigenvalues_2}
Let the boundary conditions~\ref{enu:general_boundary_condition_1} and \ref{enu:general_boundary_condition_2} of Sec.~\ref{Com_Obs} hold and consider the following regularity conditions
\begin{itemize} 
\item the equilibrium fluid verifies the weak energy condition;
\item the background spacetime is a solution of the Tolman-Oppenheimer-Volkoff equation characterized by non-trivial, $\mathcal{C}^{1}$ functions $\mathcal{A}_{0}$, $\mu_{0}$ and $p_{0}$, and $\phi_{0}$ has a simple pole at the center, but is otherwise of class $\mathcal{C}^{1}$ within the domain of the solution; 
\item $f'\left(\mu_{0}\right)$ is positive in the interior of the perturbed star. 
\end{itemize}
Then, non-trivial $\mathcal{C}^{1}$ solutions of the system \eqref{eq:Static_Radial_Adiabatic_pdot_hat} and (\ref{eq:Static_Radial_Adiabatic_Q_hat}) exist only if
\begin{equation} 
\max_{\ell\in\left]0,\ell_{\mathfrak B}\right[}\upsilon^{2}>-\max_{\ell\in\left]0,\ell_{\mathfrak B}\right[}\left[\frac{\mu_{0}+p_{0}}{\phi_{0}}\left(\frac{1}{2}\phi_{0}+2\mathcal{A}_{0}\right)+\frac{\mathcal{A}_{0}^{2}}{f'\left(\mu_{0}\right)}+\frac{\mathcal{A}_{0}\widehat{\mu}_{0}}{\mu_{0}+p_{0}}\right]\,.
\label{eq:Sturm_Liouville_bound_freq} 
\end{equation}
\end{prop}

\begin{lem}
\label{Lemma:root_Psi_pDot}If the premises of Proposition~\ref{Proposition:minimum_bound_eigenvalues_2}
hold, a non-trivial $\mathcal{C}^{1}$ solution $\Psi_{\mathsf{p}} $
has at least one root.
\end{lem}

Notice that \eqref{eq:minimum_bound_eigenvalues_1} is much simpler than
\eqref{eq:Sturm_Liouville_bound_freq}. However, even under the conditions where both 
inequalities can be used, it should be stressed that without specifying the 
background spacetime and the equation of state of the perturbed fluid, there is no 
way to infer which provides a tighter bound. Indeed, the determination of the tighter bound depends critically of the background spacetime.
Nonetheless, finding a constraint on the minimum allowed values of $\upsilon^{2}$
is of great importance when searching for solutions to the system.
Since the results in Ref.~\citep{Zhang_Sun_Zettl_2014} also associate
the number of nodes of the eigenfunctions with the ordinal number
of the eigenvalues in the sequence $\left(\lambda_{n}\right)_{n\in\mathbb{N}}$,
these constraints provide a baseline to search for
the eigenfrequencies algorithmically. Lemma~\ref{Lemma:root_Psi_pDot} provides a further criterion useful to find the fundamental eigenfrequency.

As in the previous subsection, in the following proofs, we will consider $\ell$ to be an affine
parameter of the congruence associated with the integral curves of the $e$ vector field, and without loss of generality, we set that $\ell=0$ represents the center of the star and $\ell=\ell_{\mathfrak B}$
the boundary of the equilibrium star. 

\begin{proof}
[Proof for Proposition~\ref{Proposition:minimum_bound_eigenvalues_1}]

The function $\phi_{0}$ has a simple pole at $\ell=0$. Therefore, imposing regularity of $\Psi_{Q}$ and $\Psi_{\mathsf{p}}$ and their derivatives in the interior and boundary of the star, Eq.~(\ref{eq:Static_Radial_Adiabatic_Q_hat})
implies
\begin{equation}
\Psi_{Q} \left(0\right)=0\,.
\end{equation}
In that case, we can disregard the solution with $\Psi_{\mathsf{p}} \left(0\right)=0$ since this corresponds to the trivial solution. Then, depending on the initial conditions, either $\Psi_{\mathsf{p}}\left(0\right)>0$ or $\Psi_{\mathsf{p}}\left(0\right)<0$. As expected, the reasoning for one case applies similarly to the other. Therefore, we will explicitly treat only the case $\Psi_{\mathsf{p}}\left(0\right)>0$. Then, assume $\Psi_{\mathsf{p}}\left(0\right)>0$.
Imposing $f'\left(\mu_{0}\right)>0$ and given that $\phi_{0}$ is positive, Eq.~\eqref{eq:Static_Radial_Adiabatic_Q_hat} further implies
\begin{equation}
\widehat{\Psi}_{Q} \left(0\right)\Psi_{\mathsf{p}} \left(0\right)<0\,.
\end{equation}
Therefore, for all $\ell\in\left]0,\epsilon\right[$, for some
$0<\epsilon<\ell_{\mathfrak B}$, we have 
$\Psi_{\mathsf{p}} \left(\ell\right)\Psi_{Q} \left(\ell\right)<0$, that is, for our choice of initial data
\begin{equation}
\Psi_{Q} \left(\ell\right)<0\;,\text{for all }\ell\in\left]0,\epsilon\right[\,.
\end{equation}

Now, the values of the eigenfrequencies are those that verify the boundary conditions~\ref{enu:general_boundary_condition_1} and \ref{enu:general_boundary_condition_2}, in particular Eq.~(\ref{eq:Static_Radial_Adiabatic_boundary_condition_coefficients}).
Imposing that the regularity conditions in Sec.~\ref{enu:regularity_conditions}
hold, then $\mathcal{A}_{0}\ge0$ within the equilibrium star,
being zero only at the center: $\ell=0$.
Therefore, if $\mathcal{C}^1$ solutions exist, for the fundamental mode, that is, for the lowest value of $\upsilon^{2}$, choosing $\Psi_{\mathsf{p}}\left(0\right)>0$, we must have 
\begin{equation}
\Psi_{\mathsf{p}}-\mathcal{A}_{0}\Psi_{Q}>0\;, \text{for }\ell\in\left[0,\ell_{\mathfrak B}\right[\,,
\end{equation}
and
\begin{equation}
\Psi_{\mathsf{p}}-\mathcal{A}_{0}\Psi_{Q}=0\;, \text{at }\ell=\ell_{\mathfrak B}\,.
\end{equation}
These results then imply that at the boundary
\begin{equation}
\left. D_e \left( \Psi_{\mathsf{p}}-\mathcal{A}_0\Psi_Q \right) \right|_{\ell_\mathfrak B}=
\left. \widehat{\Psi}_\mathsf{p}-\mathcal{A}_0 \widehat{\Psi}_Q -\widehat{\mathcal{A}}_0\Psi_Q \right|_{\ell_{\mathfrak B}}\leq0\,,
\end{equation}
where, for notational convenience, we have indicated the hat derivative as $D_{e}\equiv e^{\alpha}D_{\alpha}$.
Since $\Psi_{Q}$ verifies a Sturm-Liouville eigenvalue problem, for the fundamental eigenmode $\Psi_{Q}$ has no zeros. Therefore,  $\Psi_{Q}\left(\ell\right)<0$, for all $\ell\in\left]0,\ell_{\mathfrak B}\right]$. Then, further imposing that $\widehat{\mathcal{A}}_0 \left( \ell_\mathfrak B \right)\geq0$, the previous inequality is verified if
\begin{equation}
\left.\widehat{\Psi}_{\mathsf{p}}-\mathcal{A}_{0}\widehat{\Psi}_{Q} \right|_{\ell_{\mathfrak B}}\leq0\,.
\label{eq:minimum1_p_Q_derivative_values_lB}
\end{equation}
At the boundary, $\ell=\ell_{\mathfrak B}$, Eqs.~\eqref{eq:Static_Radial_Adiabatic_pdot_hat} and \eqref{eq:Static_Radial_Adiabatic_Q_hat} reduce to
\begin{equation}
\begin{aligned}
\widehat{\Psi}_{\mathsf{p}} & =\left[\frac{\mu_{0}}{\phi_{0}}\left(\frac{1}{2}\phi_{0}+\mathcal{A}_{0}\right)+\frac{\mathcal{A}_{0}\widehat{\mu}_{0}}{\mu_{0}}-2\mathcal{A}_{0}^{2}+\upsilon^{2}\right]\Psi_{Q}\,,\\
\widehat{\Psi}_{Q} & =-\left(\phi_{0}+2\mathcal{A}_{0}-\frac{\widehat{\mu}_{0}}{\mu_{0}}-\frac{\mu_{0}}{\phi_{0}}\right)\Psi_{Q}\,.
\end{aligned}
\label{eq:minimum1_general_eqs}
\end{equation}
Substituting Eq.~\eqref{eq:minimum1_general_eqs} in the inequality~\eqref{eq:minimum1_p_Q_derivative_values_lB} we find
\begin{equation}
\left.\left(\frac{1}{2}\mu_{0}+\phi_{0}\mathcal{A}_{0}+\upsilon^{2}\right)\Psi_{Q}\right|_{\ell_{\mathfrak B}}\leq0\,.
\label{eq:minimum1_inter}
\end{equation}
	Remarking once again that for the fundamental eigenmode $\Psi_{Q}\left(\ell\right)<0$, for all $\ell\in\left]0,\ell_{\mathfrak B}\right]$, inequality~\eqref{eq:minimum1_inter} implies inequality~\eqref{eq:minimum_bound_eigenvalues_1}.

\end{proof}

\begin{proof}[Proof for Proposition~\ref{Proposition:minimum_bound_eigenvalues_2} and Lemma~\ref{Lemma:root_Psi_pDot}]

For clarity, we will repeat some of the intermediate results found in the proof of Proposition~\ref{Proposition:minimum_bound_eigenvalues_1}.

Consider the premises of Proposition~\ref{Proposition:minimum_bound_eigenvalues_2}. In those conditions, $\mathcal{A}_{0}\ge0$ within the equilibrium star, being zero only at the center: $\ell=0$. Now, if $\mathcal{C}^{1}$ solutions of the boundary value problem~\eqref{eq:Static_Radial_Adiabatic_pdot_hat}, \eqref{eq:Static_Radial_Adiabatic_Q_hat} and \eqref{eq:Static_Radial_Adiabatic_boundary_condition_coefficients} exist for all $\ell\in\left[0,\ell_{\mathfrak B}\right]$, then
\begin{equation}
\Psi_{\mathsf{p}} \left(\ell_{\mathfrak B}\right)\Psi_{Q} \left(\ell_{\mathfrak B}\right)\geq0\,.\label{eq:Sturm_Liouville_condition_stars_surf}
\end{equation}
On the other hand, since $\phi_{0}$ has a simple pole at $\ell=0$, from Eq.~(\ref{eq:Static_Radial_Adiabatic_Q_hat}),
regularity of $\widehat{\Psi}_{Q} \left(0\right)$
implies 
\begin{equation}
\Psi_{Q} \left(0\right)=0\,.
\end{equation}
In particular, this allows us to disregard the solution with $\Psi_{\mathsf{p}} \left(0\right)=0$, since this corresponds to the trivial solution. 
Moreover, imposing $f'\left(\mu_{0}\right)>0$,
given that $\phi_{0}$ is positive, Eq.~\eqref{eq:Static_Radial_Adiabatic_Q_hat} leads to the conclusion that
\begin{equation}
\widehat{\Psi}_{Q} \left(0\right)\Psi_{\mathsf{p}} \left(0\right)<0\,.
\end{equation}
Therefore, for all $\ell\in\left]0,\epsilon\right[$, for some
$0<\epsilon<\ell_{\mathfrak B}$, we have 
\begin{equation}
\Psi_{\mathsf{p}} \left(\ell\right)\Psi_{Q} \left(\ell\right)<0\,.
\label{eq:minimum2_behavior_center_general}
\end{equation}
Considering Bolzano's theorem, from Eqs.~\eqref{eq:Sturm_Liouville_condition_stars_surf} and \eqref{eq:minimum2_behavior_center_general} we conclude that there is a point $0<a\leq\ell_{\mathfrak B}$ where either $\Psi_{\mathsf{p}}\left( \ell \right) $
or $\Psi_{Q}\left( \ell \right) $ vanishes.
Now, either $\Psi_{\mathsf{p}} \left(0\right)>0$
or $\Psi_{\mathsf{p}} \left(0\right)<0$.
As expected, the reasoning for one case applies similarly
to the other. Hence, we will explicitly treat only the case
$\Psi_{\mathsf{p}} \left(0\right)>0$. 
In that case, Eq.~\eqref{eq:minimum2_behavior_center_general} implies that for all $\ell\in\left]0,\epsilon\right[$, 
$\Psi_{Q} \left(\ell\right)<0$.
Continuing, assume $a\in\left]0,\ell_{\mathfrak B}\right[$ is such that
\begin{equation}
\begin{aligned}
\Psi_{Q} \left(a\right)&=0\,,\\
\widehat{\Psi}_{Q} \left(a\right)&\geq0\,,\\
\Psi_{\mathsf{p}} \left(\ell\leq a\right)&>0\,.
\end{aligned}
\label{eq:minimum2_Q_root_only}
\end{equation}
If several such values exist, we take $a$ to be the smallest one that verifies Eq.~\eqref{eq:minimum2_Q_root_only}. Notice that if $\Psi_{\mathsf{p}} \left(a\right)=0$,
we would recover the trivial solution, hence in light of Eq.~\eqref{eq:Static_Radial_Adiabatic_boundary_condition_coefficients}, we can exclude the case $a=\ell_\mathfrak B$. At $\ell=a$, Eq.~\eqref{eq:Static_Radial_Adiabatic_Q_hat}
reads
\begin{equation}
\widehat{\Psi}_{Q} =-\frac{1}{f'\left(\mu_{0}\right)}\Psi_{\mathsf{p}} \,,
\end{equation}
however, further assuming $f'\left(\mu_{0}\right)>0$, the above equation
contradicts the hypothesis. This result, in particular, proves Lemma~\ref{Lemma:root_Psi_pDot}, that is, either $\Psi_{Q} $ or $\Psi_{\mathsf{p}} $
must have a root in the interior of the star, but independently of $\upsilon^{2}$, we have shown
that $\Psi_{Q} $ can only have a root
at $\ell=a$ if $\Psi_{\mathsf{p}} $ has
a root at some $\ell<a$. 

From the previous result, we conclude that for
a valid solution of the boundary value problem to exist, there must
be either a point $\ell=b<a$ or, if there is no value $a$ that verifies Eq.~\eqref{eq:minimum2_Q_root_only}, a value $b\in\text{\ensuremath{\left]0,\ell_{\mathfrak B}\right[}}$, such that 
\begin{equation}
\begin{aligned}
\Psi_{\mathsf{p}} \left(b\right)&=0\,,\\
\widehat{\Psi}_{\mathsf{p}} \left(b\right)&\leq0\,,\\
\Psi_{Q} \left(\ell\leq b\right)&<0\,.
\end{aligned}
\end{equation}
Once again, notice that if $\Psi_{Q} \left(b\right)=0$,
we would recover the trivial solution, hence in light of Eq.~\eqref{eq:Static_Radial_Adiabatic_boundary_condition_coefficients}, we can exclude the case $b=\ell_\mathfrak B$.
At $\ell=b$, Eq.~(\ref{eq:Static_Radial_Adiabatic_pdot_hat}) then
reads
\begin{equation}
\widehat{\Psi}_{\mathsf{p}} =\left[\frac{\mu_{0}+p_{0}}{\phi_{0}}\left(\frac{1}{2}\phi_{0}+2\mathcal{A}_{0}\right)+\frac{\mathcal{A}_{0}^{2}}{f'\left(\mu_{0}\right)}+\frac{\mathcal{A}_{0}\widehat{\mu}_{0}}{\mu_{0}+p_{0}}+\upsilon^{2}\right]\Psi_{Q} \,.
\end{equation}
However, for $\upsilon^{2}$ negative and sufficiently small we would have 
\begin{equation}
\left.\frac{\mu_{0}+p_{0}}{\phi_{0}}\left(\frac{1}{2}\phi_{0}+2\mathcal{A}_{0}\right)+\frac{\mathcal{A}_{0}^{2}}{f'\left(\mu_{0}\right)}+\frac{\mathcal{A}_{0}\widehat{\mu}_{0}}{\mu_{0}+p_{0}}+\upsilon^{2}\right|_{ \ell=b}<0\,,
\end{equation}
hence $\widehat{\Psi}_{\mathsf{p}} \left(b\right)\geq0$, which would contradict the hypothesis.

The case where $\Psi_{\mathsf{p}} \left(0\right)<0$
follows exactly the same reasoning, confirming that in the considered
setup, regular solutions of the system with the considered boundary
conditions exist only if the value of $\upsilon^{2}$ is bounded from
below.

Based on these results, we can find a lower bound for the value of $\upsilon^{2}$.
Since the quantity 
\begin{equation}
\frac{\mu_{0}+p_{0}}{\phi_{0}}\left(\frac{1}{2}\phi_{0}+2\mathcal{A}_{0}\right)+\frac{\mathcal{A}_{0}^{2}}{f'\left(\mu_{0}\right)}+\frac{\mathcal{A}_{0}\widehat{\mu}_{0}}{\mu_{0}+p_{0}}+\upsilon^{2}\,,
\label{eq:minimum2_quantity}
\end{equation}
cannot be negative at all interior points of the perturbed star, non-trivial
$\mathcal{C}^{1}$ solutions of the boundary value problem necessarily verify
inequality~\eqref{eq:Sturm_Liouville_bound_freq}. We remark, however, that inequality~\eqref{eq:Sturm_Liouville_bound_freq} is rather conservative, so much so that even if $\upsilon^2$ verifies the inequality, the quantity in \eqref{eq:minimum2_quantity} might still be negative at all points within the star.
\end{proof}

\section{\label{Conclusions}Conclusions}

We developed a framework to describe general first-order perturbations
of static, locally rotationally symmetric of class II spacetimes within
the theory of general relativity. The new system of equations is
completely general and applicable to any equilibrium configuration. Moreover,
by construction, the framework is covariant and identification gauge
invariant. This last point is of pivotal importance when compared
with perturbation theory available in the literature. The classical metric-based approaches
are intrinsically dependent on a coordinate system and the gauge,
which makes it challenging to understand the dynamics of the perturbed spacetime unambiguously.
Indeed, the adopted covariant spacetime decomposition formalism
provides a manifest advantage over the standard theory: all quantities
are geometrically and physically motivated, which clearly shows the dependencies and
the source of the various physical properties of the perturbations.
In addition, the framework leads to a natural separation between even
and odd parity components of the perturbations, such that the systems
of differential equations for each parity are completely
decoupled at the linear level.

As a first application, we considered the study of linear, isotropic,
and adiabatic perturbations. To do so, we have explicitly shown the importance
of choosing frames to describe the unperturbed and the perturbed
spacetimes. To our knowledge, this cannot
be done in general in the metric-based linearized theory available in the literature and makes the covariant gauge-invariant approach all the
more powerful. Given the symmetries of the problem, we considered
two meaningful frames: a frame associated with observers locally comoving with the elements of volume of the fluid and a frame associated
with static observers with respect to an observer at spatial infinity.
These frames represent the classical Lagrangian
and Eulerian pictures, respectively.
The adopted formalism makes it
quite simple to change between frames and evaluate the advantages
of one frame over the other. In particular, we were able to derive
the relation between the equations of state of the comoving and static
frames in the perturbed spacetime, showing that, even at linear level,
we have to account for phase changes between the equilibrium
and the perturbed fluids.
Moreover, we have shown that, since a perfect fluid in the comoving frame will not be perceived
as perfect in the static frame, and the net momentum flow
must be included in the equation of state
to account for the correction to the rate of change of the energy
density and the pressure when compared to those measured in the comoving
frame.

Focusing on analyzing the general properties of linear, isotropic,
and adiabatic perturbations, we have rigorously shown that the problem
can be cast in the form of a singular Sturm-Liouville eigenvalue problem, inheriting
the standard properties for the eigenfunctions and eigenvalues. Moreover,
under rather general regularity conditions, we have found
lower bounds for the values of the eigenfrequencies. The application of the methods developed in this article, the determination of the analytical solutions
to the system of equations, and the discussion regarding the stability
of selected solutions of the theory will be done elsewhere.

\begin{acknowledgments}
PL acknowledges financial support provided under the European Union\textquoteright s
H2020 ERC Advanced Grant \textquotedblleft Black holes: gravitational
engines of discovery\textquotedblright{} grant agreement no. Gravitas--101052587.
Views and opinions expressed are, however, those of the author only
and do not necessarily reflect those of the European Union or the
European Research Council. Neither the European Union nor the granting
authority can be held responsible for them. PL acknowledges support
from the Villum Investigator program supported by the VILLUM Foundation
(grant no. VIL37766) and the DNRF Chair program (grant no. DNRF162)
by the Danish National Research Foundation.

This project has received funding from the European Union's Horizon
2020 research and innovation programme under the Marie Sklodowska-Curie
grant agreement No 101007855 and No 101131233.

The work of SC has been carried out in the framework of activities
of the INFN Research Project QGSKY.
\end{acknowledgments}

\appendix

\section{Field equations in the 1+1+2 formalism\label{Appendix:General_1p1p2_eqs}}

Here, we present the Einstein field equations for the theory of General
Relativity, written in the 1+1+2 covariant spacetime decomposition
formalism introduced in Section~\ref{sec:1p1p2_decomposition}. Although
the complete set of equations has already been presented in literature, these expressions are known to
contain several typographical errors. To our knowledge,
these errors do not affect published works. However, the corrected equations are essential in studying general linear perturbations of static LRS II
spacetimes, hence the importance of presenting them here.

\subsection{Equations for the kinematical quantities associated with the timelike
congruence}

We find the following evolution and propagation equations for the
kinematical quantities associated with the $u$ congruence.

Scalar equations:
\begin{equation}
\begin{aligned}\widehat{\mathcal{A}}-\dot{\theta} & =\frac{1}{2}\left(\mu+3p\right)-\Lambda+\frac{1}{3}\theta^{2}+\frac{3}{2}\Sigma^{2}-2\Omega^{2}+2\Sigma_{\mu}\Sigma^{\mu}-\mathcal{A}\left(\mathcal{A}+\phi\right)\\
 & -2\Omega_{\mu}\Omega^{\mu}+\mathcal{A}_{\mu}\left(a^{\mu}-\mathcal{A}^{\mu}\right)+\Sigma_{\mu\nu}\Sigma^{\mu\nu}-\delta_{\mu}\mathcal{A}^{\mu}\,,
\end{aligned}
\label{1p1p2_eqs:Raychaudhuri}
\end{equation}

\begin{equation}
\begin{aligned}\dot{\Sigma}-\frac{2}{3}\widehat{\mathcal{A}} & =\frac{1}{2}\Pi-\mathcal{E}+\frac{1}{3}\left(2\mathcal{A}-\phi\right)\mathcal{A}-\left(\frac{2}{3}\theta+\frac{1}{2}\Sigma\right)\Sigma-\frac{2}{3}\Omega^{2}+\Sigma_{\mu}\left(2\alpha^{\mu}-\frac{1}{3}\Sigma^{\mu}\right)\\
 & -\frac{1}{3}\mathcal{A}_{\mu}\left(\mathcal{A}^{\mu}+2a^{\mu}\right)+\frac{1}{3}\Omega_{\mu}\Omega^{\mu}+\frac{1}{3}\Sigma_{\mu\nu}\Sigma^{\mu\nu}-\frac{1}{3}\delta_{\mu}\mathcal{A}^{\mu}\,,
\end{aligned}
\label{1p1p2_eqs:SigmaS_dot}
\end{equation}

\begin{equation}
\dot{\Omega}=\frac{1}{2}\varepsilon^{\mu\nu}\delta_{\mu}\mathcal{A}_{\nu}+\mathcal{A}\xi+\left(\Sigma-\frac{2}{3}\theta\right)\Omega+\Omega_{\mu}\left(\Sigma^{\mu}+\alpha^{\mu}\right)\,,\label{1p1p2_eqs:OmegaS_dot}
\end{equation}

\begin{equation}
\widehat{\Omega}=\Omega\left(\mathcal{A}-\phi\right)+\Omega_{\mu}\left(\mathcal{A}^{\mu}+a^{\mu}\right)-\delta_{\mu}\Omega^{\mu}\,,\label{1p1p2_eqs:OmegaS_hat}
\end{equation}
\begin{equation}
\frac{2}{3}\widehat{\theta}-\widehat{\Sigma}=Q+\frac{3}{2}\Sigma\phi+2\Omega\xi+\delta_{\mu}\Sigma^{\mu}+\varepsilon_{\mu\nu}\delta^{\mu}\Omega^{\nu}-2\Sigma_{\mu}a^{\mu}+2\varepsilon_{\mu\nu}\mathcal{A}^{\mu}\Omega^{\nu}-\zeta_{\mu\nu}\Sigma^{\mu\nu}\,;\label{1p1p2_eqs:SigmaS_hat_theta_hat}
\end{equation}

Vector equations:
\begin{equation}
\begin{aligned}N_{\alpha}{}^{\mu}\dot{\Omega}_{\mu}+\frac{1}{2}\varepsilon_{\alpha}{}^{\mu}\widehat{\mathcal{A}}_{\mu} & =\frac{1}{2}\xi\mathcal{A}_{\alpha}-\left(\frac{2}{3}\theta+\frac{1}{2}\Sigma\right)\Omega_{\alpha}+\Omega\left(\Sigma_{\alpha}-\alpha_{\alpha}\right)\\
 & -\frac{1}{2}\varepsilon_{\alpha}{}^{\mu}\left(\frac{1}{2}\phi\mathcal{A}_{\mu}+\mathcal{A}a_{\mu}-\delta_{\mu}\mathcal{A}\right)-\frac{1}{2}\varepsilon_{\alpha}{}^{\mu}\zeta_{\mu\nu}\mathcal{A}^{\nu}+\Sigma_{\alpha\mu}\Omega^{\mu}\,,
\end{aligned}
\label{1p1p2_eqs:OmegaV_dot_curlyAV_hat}
\end{equation}
\begin{equation}
\begin{aligned}N_{\alpha}{}^{\mu}\dot{\Sigma}_{\mu}-\frac{1}{2}N_{\alpha}{}^{\mu}\widehat{\mathcal{A}}_{\mu} & =\frac{1}{2}\delta_{\alpha}\mathcal{A}+\left(\mathcal{A}-\frac{1}{4}\phi\right)\mathcal{A}_{\alpha}-\left(\frac{2}{3}\theta+\frac{1}{2}\Sigma\right)\Sigma_{\alpha}+\frac{1}{2}\mathcal{A}a_{\alpha}-\frac{3}{2}\Sigma\alpha_{\alpha}\\
 & -\Omega\Omega_{\alpha}-\frac{1}{2}\left(\zeta_{\alpha\mu}+\varepsilon_{\alpha\mu}\xi\right)\mathcal{A}^{\mu}+\Sigma_{\alpha\mu}\left(\alpha^{\mu}-\Sigma^{\mu}\right)-\mathcal{E}_{\alpha}+\frac{1}{2}\Pi_{\alpha}\,,
\end{aligned}
\label{1p1p2_eqs:SigmaV_dot_AV_hat}
\end{equation}

\begin{equation}
\begin{aligned}N_{\alpha}{}^{\mu}\widehat{\Sigma}_{\mu}-\varepsilon_{\alpha}{}^{\mu}\widehat{\Omega}_{\mu} & =\frac{1}{2}\delta_{\alpha}\Sigma+\frac{2}{3}\delta_{\alpha}\theta-\varepsilon_{\alpha}{}^{\mu}\delta_{\mu}\Omega-\frac{3}{2}\phi\Sigma_{\alpha}+\varepsilon_{\alpha}{}^{\mu}\Sigma_{\mu}\xi-\Omega_{\alpha}\xi+\left(\frac{1}{2}\phi+2\mathcal{A}\right)\varepsilon_{\alpha}{}^{\mu}\Omega_{\mu}\\
 & -\frac{3}{2}\Sigma a_{\alpha}+\varepsilon_{\alpha}{}^{\mu}\Omega\left(a_{\mu}-2\mathcal{A}_{\mu}\right)-\delta_{\mu}\Sigma_{\alpha}{}^{\mu}-\zeta_{\alpha}{}^{\mu}\Sigma_{\mu}+\Sigma_{\alpha}{}^{\mu}a_{\mu}+\varepsilon_{\alpha\mu}\zeta^{\mu\nu}\Omega_{\nu}-Q_{\alpha}\,;
\end{aligned}
\label{1p1p2_eqs:SigmaV_hat_OmegaV_hat}
\end{equation}

Tensor equations:
\begin{equation}
\begin{aligned}N^{\mu}{}_{\{\alpha}N_{\beta\}}{}^{\nu}\dot{\Sigma}_{\mu\nu} & =\delta_{\{\alpha}\mathcal{A}_{\beta\}}+\mathcal{A}_{\{\alpha}\mathcal{A}_{\beta\}}-\Sigma_{\{\alpha}\Sigma_{\beta\}}-2\Sigma_{\{\alpha}\alpha_{\beta\}}-\Omega_{\{\alpha}\Omega_{\beta\}}+\mathcal{A}\zeta_{\alpha\beta}\\
 & -\left(\frac{2}{3}\theta-\Sigma\right)\Sigma_{\alpha\beta}-\Sigma_{\{\alpha|}{}^{\mu}\Sigma_{\mu|\beta\}}-\mathcal{E}_{\alpha\beta}+\frac{1}{2}\Pi_{\alpha\beta}\,,
\end{aligned}
\label{1p1p2_eqs:SigmaT_dot}
\end{equation}
\begin{equation}
\begin{aligned}N^{\mu}{}_{\{\alpha}N_{\beta\}}{}^{\nu}\widehat{\Sigma}_{\mu\nu} & =\delta_{\{\alpha}\Sigma_{\beta\}}+\frac{1}{2}\varepsilon_{\{\alpha|}{}^{\mu}\delta_{\mu}\Omega_{|\beta\}}+\frac{1}{2}\varepsilon_{\{\alpha|}{}^{\mu}\delta_{|\beta\}}\Omega_{\mu}-\frac{1}{2}\Sigma_{\alpha\beta}\phi+\varepsilon^{\mu}{}_{\{\alpha}\Sigma_{\beta\}\mu}\xi+\frac{3}{2}\Sigma\zeta_{\alpha\beta}\\
 & -\varepsilon_{\mu\{\alpha}\zeta_{\beta\}}{}^{\mu}\Omega-2a_{\{\alpha}\Sigma_{\beta\}}+\varepsilon_{\{\alpha|}{}^{\mu}\Omega_{\mu}\mathcal{A}_{|\beta\}}+\varepsilon_{\{\alpha|}{}^{\mu}\Omega_{|\beta\}}\mathcal{A}_{\mu}-\Sigma_{\mu\{\alpha}\zeta_{\beta\}}{}^{\mu}-\varepsilon_{\mu\{\alpha}\mathcal{H}_{\beta\}}{}^{\mu}\,;
\end{aligned}
\label{1p1p2_eqs:SigmaT_hat}
\end{equation}

Constraint equations:
\begin{equation}
\delta_{\mu}\Omega^{\mu}+\varepsilon^{\mu\nu}\delta_{\mu}\Sigma_{\nu}=\left(2\mathcal{A}-\phi\right)\Omega-3\Sigma\xi+\mathcal{H}+\varepsilon^{\mu\nu}\zeta_{\mu}{}^{\gamma}\Sigma_{\gamma\nu}\,,\label{1p1p2_eqs:deltaOmegaV_deltaSigmaV}
\end{equation}
\begin{equation}
\begin{aligned}\delta_{\alpha}\Sigma-\frac{2}{3}\delta_{\alpha}\theta+2\varepsilon_{\alpha}{}^{\mu}\delta_{\mu}\Omega+2\delta_{\mu}\Sigma_{\alpha}{}^{\mu} & =\phi\left(\varepsilon_{\alpha}{}^{\mu}\Omega_{\mu}-\Sigma_{\alpha}\right)-2\xi\left(\Omega_{\alpha}-3\varepsilon_{\alpha}{}^{\mu}\Sigma_{\mu}\right)-4\varepsilon_{\alpha}{}^{\mu}\Omega\mathcal{A}_{\mu}\\
 & +2\zeta_{\alpha\mu}\Sigma^{\mu}+2\varepsilon_{\alpha}{}^{\mu}\Omega^{\nu}\zeta_{\mu\nu}-2\varepsilon_{\alpha}{}^{\mu}\mathcal{H}_{\mu}-Q_{\alpha}\,.
\end{aligned}
\label{1p1p2_eqs:deltaSigmaS_deltaTheta_deltaOmegaS_deltaSigmaT}
\end{equation}

\subsection{Equations for the kinematical quantities associated with the spacelike
congruence}

We also find the following evolution and propagation equations for
the kinematical quantities associated with the $e$ congruence.

Scalar equations:
\begin{equation}
\begin{aligned}\dot{\phi} & =Q+\left(\frac{1}{3}\theta-\frac{1}{2}\Sigma\right)\left(2\mathcal{A}-\phi\right)+2\Omega\xi+\delta_{\mu}\alpha^{\mu}-\alpha_{\mu}\left(a^{\mu}-\mathcal{A}^{\mu}\right)\\
 & -\left(\mathcal{A}^{\nu}+a^{\nu}\right)\left(\Sigma_{\nu}+\varepsilon_{\mu\nu}\Omega^{\mu}\right)-\Sigma^{\mu\nu}\zeta_{\mu\nu}\,,
\end{aligned}
\label{1p1p2_eqs:phi_dot}
\end{equation}

\begin{equation}
\begin{aligned}\widehat{\phi} & =\left(\frac{1}{3}\theta+\Sigma\right)\left(\frac{2}{3}\theta-\Sigma\right)-\frac{1}{2}\phi^{2}+2\xi^{2}-\frac{2}{3}\left(\mu+\Lambda\right)-\frac{1}{2}\Pi-\mathcal{E}\\
 & +\delta_{\mu}a^{\mu}-a_{\mu}a^{\mu}-\Sigma_{\mu}\Sigma^{\mu}+\Omega_{\mu}\Omega^{\mu}+2\varepsilon_{\mu\nu}\alpha^{\mu}\Omega^{\nu}-\zeta_{\mu\nu}\zeta^{\mu\nu}\,,
\end{aligned}
\label{1p1p2_eqs:phi_hat}
\end{equation}
\begin{equation}
\begin{aligned}\dot{\xi} & =\left(\frac{1}{2}\Sigma-\frac{1}{3}\theta\right)\xi+\frac{1}{2}\mathcal{H}+\Omega\left(\mathcal{A}-\frac{1}{2}\phi\right)+\frac{1}{2}\varepsilon^{\mu\nu}\delta_{\mu}\alpha_{\nu}\\
 & +\frac{1}{2}\left(\mathcal{A}_{\mu}+a_{\mu}\right)\left[\Omega^{\mu}+\varepsilon^{\mu\nu}\left(\Sigma_{\nu}+\alpha_{\nu}\right)\right]-\frac{1}{2}\varepsilon^{\mu\nu}\Sigma_{\mu\gamma}\zeta^{\gamma}{}_{\nu}\,,
\end{aligned}
\label{1p1p2_eqs:xi_dot}
\end{equation}

\begin{equation}
\widehat{\xi}=\left(\frac{1}{3}\theta+\Sigma\right)\Omega-\phi\xi+\frac{1}{2}\varepsilon^{\mu\nu}\delta_{\mu}a_{\nu}+\Omega^{\mu}\left(\alpha_{\mu}+\Sigma_{\mu}\right)-\frac{1}{2}\varepsilon^{\mu\nu}\zeta_{\mu\gamma}\zeta^{\gamma}{}_{\nu}\,;\label{1p1p2_eqs:xi_hat}
\end{equation}

Vector equations:
\begin{equation}
\begin{aligned}N^{\mu}{}_{\alpha}\widehat{\alpha}_{\mu}-N^{\mu}{}_{\alpha}\dot{a}_{\mu} & =\varepsilon_{\mu\alpha}\mathcal{H}^{\mu}-\left(\mathcal{A}+\frac{1}{2}\phi\right)\alpha_{\alpha}+\varepsilon_{\alpha\mu}\alpha^{\mu}\xi+\left(\frac{1}{3}\theta+\Sigma\right)\left(\mathcal{A}_{\alpha}+a_{\alpha}\right)\\
 & +\left(\frac{1}{2}\phi-\mathcal{A}\right)\left(\Sigma_{\alpha}+\varepsilon_{\alpha\mu}\Omega^{\mu}\right)+\left(\Omega_{\alpha}-\varepsilon_{\alpha\mu}\Sigma^{\mu}\right)\xi\\
 & +\left(\Sigma^{\mu}+\varepsilon^{\mu\nu}\Omega_{\nu}-\alpha^{\mu}\right)\zeta_{\mu\alpha}+\frac{1}{2}Q_{\alpha}\,;
\end{aligned}
\label{1p1p2_eqs:alpha_hat_a_dot}
\end{equation}

Tensor equations:
\begin{equation}
\begin{aligned}N^{\mu}{}_{\{\alpha}N_{\beta\}}{}^{\nu}\dot{\zeta}_{\mu\nu} & =\left(\frac{1}{2}\Sigma-\frac{1}{3}\theta\right)\zeta_{\alpha\beta}+\varepsilon_{\mu\{\alpha|}\zeta^{\mu}{}_{|\beta\}}\Omega+\left(\mathcal{A}-\frac{1}{2}\phi\right)\Sigma_{\alpha\beta}-\Sigma_{\{\alpha|\mu}\left(\zeta^{\mu}{}_{|\beta\}}+\varepsilon^{\mu}{}_{|\beta\}}\xi\right)\\
 & +\delta_{\{\alpha}\alpha_{\beta\}}+\left(\mathcal{A}_{\{\alpha}-a_{\{\alpha}\right)\alpha_{\beta\}}-\left(\Sigma_{\{\alpha}-\varepsilon_{\{\alpha|\mu}\Omega^{\mu}\right)\left(\mathcal{A}_{\beta\}}+a_{\beta\}}\right)-\varepsilon^{\mu}{}_{\{\alpha}\mathcal{H}_{\beta\}\mu}\,,
\end{aligned}
\label{1p1p2_eqs:zeta_dot}
\end{equation}
\begin{equation}
\begin{aligned}N^{\mu}{}_{\{\alpha}N_{\beta\}}{}^{\nu}\widehat{\zeta}_{\mu\nu} & =\delta_{\{\alpha}a_{\beta\}}-\phi\zeta_{\alpha\beta}-\zeta_{\{\alpha|\mu}\zeta^{\mu}{}_{|\beta\}}-a_{\{\alpha}a_{\beta\}}-\Sigma_{\{\alpha}\Sigma_{\beta\}}-\Omega_{\{\alpha}\Omega_{\beta\}}\\
 & +2\alpha_{\{\alpha}\varepsilon_{\beta\}\mu}\Omega^{\mu}+\left(\frac{1}{3}\theta+\Sigma\right)\Sigma_{\alpha\beta}-\mathcal{E}_{\alpha\beta}-\frac{1}{2}\Pi_{\alpha\beta}\,;
\end{aligned}
\label{1p1p2_eqs:zeta_hat}
\end{equation}

Constraint equation:
\begin{equation}
\begin{aligned}\frac{1}{2}\delta_{\alpha}\phi-\delta_{\mu}\zeta_{\alpha}{}^{\mu}-\varepsilon_{\alpha}{}^{\mu}\delta_{\mu}\xi= & -2\varepsilon_{\alpha\mu}a^{\mu}\xi-\Omega\left(\Omega_{\alpha}+\varepsilon_{\alpha\mu}\Sigma^{\mu}-2\varepsilon_{\alpha\mu}\alpha^{\mu}\right)+\left(\frac{1}{3}\theta-\frac{1}{2}\Sigma\right)\left(\Sigma_{\alpha}-\varepsilon_{\alpha\mu}\Omega^{\mu}\right)\\
 & -\mathcal{E}_{\alpha}-\frac{1}{2}\Pi_{\alpha}-\Sigma_{\alpha}{}^{\mu}\left(\Sigma_{\mu}-\varepsilon_{\mu\nu}\Omega^{\nu}\right)\,.
\end{aligned}
\label{1p1p2_eqs:deltaPhi_deltaZeta_deltaXi}
\end{equation}

\subsection{Electric and magnetic parts of the Weyl tensor}

Moreover, we have the following evolution and propagation equations
the electric and magnetic part of the Weyl tensor.

Scalar equations:
\begin{equation}
\begin{aligned}\dot{\mathcal{E}}+\frac{1}{2}\dot{\Pi}+\frac{1}{3}\widehat{Q} & =3\mathcal{H}\xi+\mathcal{E}\left(\frac{3}{2}\Sigma-\theta\right)-\frac{1}{2}\Pi\left(\frac{1}{3}\theta+\frac{1}{2}\Sigma\right)+\frac{1}{3}Q\left(\frac{1}{2}\phi-2\mathcal{A}\right)-\frac{1}{2}\left(\mu+p\right)\Sigma\\
 & +\frac{1}{6}\delta_{\mu}Q^{\mu}+\varepsilon^{\mu\nu}\delta_{\mu}\mathcal{H}_{\nu}+\frac{1}{3}Q_{\mu}\left(a^{\mu}+\mathcal{A}^{\mu}\right)+2\varepsilon_{\mu\nu}\mathcal{A}^{\mu}\mathcal{H}^{\nu}\\
 & +\mathcal{E}{}^{\mu}\left(\Sigma_{\mu}+2\alpha_{\mu}-\varepsilon_{\mu\nu}\Omega^{\nu}\right)-\frac{1}{2}\left(\frac{1}{3}\Sigma_{\mu}-2\alpha_{\mu}+\varepsilon_{\mu\nu}\Omega^{\nu}\right)\Pi^{\mu}\\
 & -\left(\mathcal{E}_{\mu\nu}-\frac{1}{6}\Pi_{\mu\nu}\right)\Sigma^{\mu\nu}+\varepsilon_{\mu\nu}\mathcal{H}_{\alpha}{}^{\mu}\zeta^{\nu\alpha}\,,
\end{aligned}
\label{1p1p2_eqs:ES_dot_PiS_dot_Q_hat}
\end{equation}
\begin{equation}
\begin{aligned}\widehat{\mathcal{E}}-\frac{1}{3}\widehat{\mu}+\frac{1}{2}\widehat{\Pi}= & -\delta_{\mu}\mathcal{E}^{\mu}-\frac{1}{2}\delta_{\mu}\Pi^{\mu}-\frac{3}{2}\phi\left(\mathcal{E}+\frac{1}{2}\Pi\right)+\left(\frac{1}{2}\Sigma-\frac{1}{3}\theta\right)Q+3\mathcal{H}\Omega\\
 & +2\left(\mathcal{E}_{\mu}+\frac{1}{2}\Pi_{\mu}\right)a^{\mu}+\frac{1}{2}\Sigma_{\mu}Q^{\mu}+3\mathcal{H}_{\mu}\Omega^{\mu}+\frac{3}{2}\varepsilon_{\mu\nu}Q^{\mu}\Omega^{\nu}\\
 & +\varepsilon^{\mu\nu}\Sigma_{\mu}\mathcal{H}_{\nu}+\varepsilon^{\mu\nu}\Sigma_{\mu\gamma}\mathcal{H}_{\nu}{}^{\gamma}+\left(\mathcal{E}^{\mu\nu}+\frac{1}{2}\Pi^{\mu\nu}\right)\zeta_{\mu\nu}\,,
\end{aligned}
\label{1p1p2_eqs:ES_hat_mu_hat_PiS_hat}
\end{equation}

\begin{equation}
\begin{aligned}\dot{\mathcal{H}} & =\frac{1}{2}\varepsilon^{\mu\nu}\delta_{\mu}\Pi_{\nu}-\varepsilon^{\mu\nu}\delta_{\mu}\mathcal{E}_{\nu}+\Omega Q-3\left(\mathcal{E}-\frac{1}{2}\Pi\right)\xi-\left(\theta-\frac{3}{2}\Sigma\right)\mathcal{H}\\
 & -2\varepsilon^{\mu\nu}\mathcal{A}_{\mu}\mathcal{E}_{\nu}+\mathcal{H}^{\mu}\left(2\alpha_{\mu}+\Sigma_{\mu}-\varepsilon_{\mu\nu}\Omega^{\nu}\right)-\mathcal{H}^{\mu\nu}\Sigma_{\mu\nu}\\
 & -\frac{1}{2}\left(\Omega_{\mu}+\varepsilon_{\mu\nu}\Sigma^{\nu}\right)Q^{\mu}+\varepsilon^{\mu\nu}\mathcal{E}_{\nu\gamma}\zeta_{\mu}{}^{\gamma}+\frac{1}{2}\varepsilon_{\mu}{}^{\nu}\Pi^{\gamma\mu}\zeta_{\nu\gamma}\,,
\end{aligned}
\label{1p1p2_eqs:HS_dot}
\end{equation}
\begin{equation}
\begin{aligned}\widehat{\mathcal{H}}= & -\delta_{\mu}\mathcal{H}^{\mu}-\frac{1}{2}\varepsilon_{\mu\nu}\delta^{\mu}Q^{\nu}-\frac{3}{2}\mathcal{H}\phi-\left(3\mathcal{E}+\mu+p-\frac{1}{2}\Pi\right)\Omega-Q\xi\\
 & +2\mathcal{H}^{\mu}a_{\mu}-\left(3\mathcal{E}^{\mu}-\frac{1}{2}\Pi^{\mu}\right)\Omega_{\mu}+\varepsilon_{\mu\nu}\Sigma^{\nu}\left(\mathcal{E}^{\mu}+\frac{1}{2}\Pi^{\mu}\right)\\
 & -\varepsilon_{\nu\gamma}\Sigma^{\nu\mu}\left(\mathcal{E}_{\mu}{}^{\gamma}+\frac{1}{2}\Pi_{\mu}{}^{\gamma}\right)+\mathcal{H}^{\mu\nu}\zeta_{\mu\nu}\,;
\end{aligned}
\label{1p1p2_eqs:HS_hat}
\end{equation}

Vector equations:
\begin{equation}
\begin{aligned}N_{\alpha}{}^{\mu}\dot{\mathcal{E}}_{\mu}+\frac{1}{2}N_{\alpha}{}^{\mu}\dot{\Pi}_{\mu} & =\left(\frac{1}{2}\phi-\mathcal{A}\right)\left(\frac{1}{2}Q_{\alpha}+\varepsilon_{\alpha\mu}\mathcal{H}^{\mu}\right)-\frac{1}{2}Q\mathcal{A}_{\alpha}+\frac{3}{2}\mathcal{E}_{\alpha}\Sigma+3\mathcal{H}_{\alpha}\xi-\frac{3}{2}\left(\mathcal{E}+\frac{1}{2}\Pi\right)\alpha_{\alpha}\\
 & +\varepsilon_{\alpha\mu}\Omega\left(\mathcal{E}^{\mu}-\frac{1}{2}\Pi^{\mu}\right)-\left(\mathcal{E}_{\alpha}+\frac{1}{6}\Pi_{\alpha}\right)\theta+\frac{3}{2}\varepsilon_{\alpha\mu}\mathcal{A}^{\mu}\mathcal{H}+\frac{1}{4}\Sigma\Pi_{\alpha}\\
 & -\frac{1}{2}\left(\mu+p+\Pi\right)\left(\Sigma_{\alpha}-\varepsilon_{\alpha\mu}\Omega^{\mu}\right)+\frac{1}{2}\left(\zeta_{\alpha\mu}+\varepsilon_{\alpha\mu}\xi\right)Q^{\mu}\\
 & +\mathcal{E}_{\alpha\mu}\left(2\Sigma^{\mu}+\alpha^{\mu}-2\varepsilon^{\mu\nu}\Omega_{\nu}\right)+\Sigma_{\alpha\mu}\left(\mathcal{E}^{\mu}-\frac{1}{2}\Pi^{\mu}\right)\\
 & -\varepsilon^{\mu\nu}\mathcal{H}_{\alpha\mu}\mathcal{A}_{\nu}-\varepsilon^{\mu\nu}\mathcal{H}_{\mu}\zeta_{\nu\alpha}+\frac{1}{2}\Pi_{\alpha\mu}\alpha^{\mu}\\
 & -\frac{1}{2}\delta_{\alpha}Q+\frac{1}{2}\varepsilon_{\alpha}{}^{\mu}\delta_{\mu}\mathcal{H}+\varepsilon^{\mu\nu}\delta_{\mu}\mathcal{H}_{\nu\alpha}\,,
\end{aligned}
\label{1p1p2_eqs:EV_dot_PiV_dot}
\end{equation}

\begin{equation}
\begin{aligned}N^{\mu}{}_{\alpha}\widehat{\mathcal{E}}_{\mu}+\frac{1}{2}N^{\mu}{}_{\alpha}\widehat{\Pi}_{\mu} & =\frac{1}{2}Q\Sigma_{\alpha}+3\mathcal{H}_{\alpha}\Omega-\frac{3}{2}\mathcal{H}\Omega_{\alpha}-\left(\frac{1}{3}\theta+\frac{1}{4}\Sigma\right)Q_{\alpha}-\frac{3}{2}\left(\mathcal{E}_{\alpha}+\frac{1}{2}\Pi_{\alpha}\right)\phi\\
 & -\frac{3}{2}\left(\mathcal{E}+\frac{1}{2}\Pi\right)a_{\alpha}+\frac{3}{2}\varepsilon_{\alpha\mu}\left(\Sigma^{\mu}\mathcal{H}-\Sigma\mathcal{H}^{\mu}\right)+\varepsilon_{\alpha\mu}\left(\mathcal{E}^{\mu}+\frac{1}{2}\Pi^{\mu}\right)\xi\\
 & +\frac{3}{2}\varepsilon_{\alpha\mu}\left(Q^{\mu}\Omega-Q\Omega^{\mu}\right)+\frac{1}{2}\Sigma_{\alpha\mu}Q^{\mu}+\varepsilon_{\alpha\nu}\left(\mathcal{H}_{\mu}\Sigma^{\mu\nu}-\Sigma_{\mu}\mathcal{H}^{\mu\nu}\right)\\
 & +\left(\mathcal{E}_{\alpha\mu}+\frac{1}{2}\Pi_{\alpha\mu}\right)a^{\mu}-\left(\mathcal{E}^{\mu}+\frac{1}{2}\Pi^{\mu}\right)\zeta_{\mu\alpha}+3\mathcal{H}_{\alpha\mu}\Omega^{\mu}\\
 & +\frac{1}{2}\delta_{\alpha}\mathcal{E}+\frac{1}{3}\delta_{\alpha}\mu+\frac{1}{4}\delta_{\alpha}\Pi-\delta_{\mu}\mathcal{E}_{\alpha}{}^{\mu}-\frac{1}{2}\delta_{\mu}\Pi_{\alpha}{}^{\mu}\,,
\end{aligned}
\label{1p1p2_eqs:EV_hat_PiV_hat}
\end{equation}

\begin{equation}
\begin{aligned}\frac{1}{2}\varepsilon_{\alpha}{}^{\mu}\widehat{\mathcal{E}}_{\mu}-N_{\alpha}{}^{\mu}\dot{\mathcal{H}}_{\mu}-\frac{1}{4}\varepsilon_{\alpha}{}^{\mu}\widehat{\Pi}_{\mu} & =\frac{3}{4}\varepsilon_{\alpha}{}^{\mu}\delta_{\mu}\mathcal{E}-\frac{3}{8}\varepsilon_{\alpha}{}^{\mu}\delta_{\mu}\Pi+\frac{1}{2}\varepsilon^{\mu\nu}\delta_{\mu}\mathcal{E}_{\nu\alpha}-\frac{1}{4}\varepsilon^{\mu\nu}\delta_{\mu}\Pi_{\nu\alpha}-\frac{3}{4}\mathcal{H}\left(\Sigma_{\alpha}-2\alpha_{\alpha}\right)\\
 & +\frac{3}{2}\varepsilon_{\alpha\mu}\mathcal{A}^{\mu}\mathcal{E}+\frac{3}{8}\varepsilon_{\alpha\mu}Q^{\mu}\Sigma-\frac{1}{4}\varepsilon_{\alpha\mu}\Sigma^{\mu}Q-\frac{3}{4}\varepsilon_{\alpha\mu}\Omega^{\mu}\mathcal{H}+\frac{5}{2}\xi\left(\mathcal{E}_{\alpha}-\frac{1}{2}\Pi_{\alpha}\right)\\
 & -\frac{3}{4}\left(Q_{\alpha}\Omega+Q\Omega_{\alpha}\right)-\frac{3}{4}\varepsilon_{\alpha\mu}a^{\mu}\left(\mathcal{E}-\frac{1}{2}\Pi\right)-\varepsilon_{\alpha\mu}\mathcal{E}^{\mu}\mathcal{A}+\mathcal{H}_{\alpha}\left(\theta-\frac{3}{4}\Sigma\right)\\
 & -\frac{1}{4}\varepsilon_{\alpha\mu}\left(\mathcal{E}^{\mu}-\frac{1}{2}\Pi^{\mu}\right)\phi+\frac{1}{2}\varepsilon_{\alpha\mu}\mathcal{H}^{\mu}\Omega+\frac{1}{2}\varepsilon_{\alpha\mu}a_{\nu}\left(\mathcal{E}^{\mu\nu}-\frac{1}{2}\Pi^{\mu\nu}\right)\\
 & -\frac{1}{2}\varepsilon^{\mu\nu}\zeta_{\nu\alpha}\left(\mathcal{E}_{\mu}-\frac{1}{2}\Pi_{\mu}\right)-\varepsilon_{\alpha\mu}\zeta^{\mu\nu}\left(\mathcal{E}_{\nu}-\frac{1}{2}\Pi_{\nu}\right)+\varepsilon^{\mu\nu}\mathcal{A}_{\mu}\mathcal{E}_{\nu\alpha}\\
 & -\frac{1}{2}\Sigma_{\alpha\mu}\left(3\mathcal{H}^{\mu}+\frac{1}{2}\varepsilon^{\mu\nu}Q_{\nu}\right)-\mathcal{H}_{\alpha\mu}\left(\frac{3}{2}\Sigma^{\mu}+\alpha^{\mu}-\frac{1}{2}\varepsilon^{\mu\nu}\Omega_{\nu}\right)\,,
\end{aligned}
\label{1p1p2_eqs:EV_hat_HV_dot_PiV_hat}
\end{equation}

\begin{equation}
\begin{aligned}N^{\mu}{}_{\alpha}\widehat{\mathcal{H}}_{\mu}-\frac{1}{2}\varepsilon_{\alpha}{}^{\mu}\widehat{Q}_{\mu} & =\frac{1}{2}\delta_{\alpha}\mathcal{H}-\delta_{\mu}\mathcal{H}_{\alpha}{}^{\mu}-\frac{1}{2}\varepsilon_{\alpha}{}^{\mu}\delta_{\mu}Q-\frac{3}{2}\varepsilon_{\alpha\mu}\Sigma^{\mu}\left(\mathcal{E}+\frac{1}{2}\Pi\right)-\frac{3}{2}\mathcal{H}a_{\alpha}\\
 & +\frac{1}{2}\varepsilon_{\alpha\mu}Qa^{\mu}-\Omega\left(3\mathcal{E}_{\alpha}-\frac{1}{2}\Pi_{\alpha}\right)+\frac{3}{2}\varepsilon_{\alpha\mu}\Sigma\left(\mathcal{E}^{\mu}+\frac{1}{2}\Pi^{\mu}\right)\\
 & +\left(\varepsilon_{\alpha\mu}\mathcal{H}^{\mu}-\frac{1}{2}Q_{\alpha}\right)\xi-\Omega_{\alpha}\left(\mu+p+\frac{1}{4}\Pi-\frac{3}{2}\mathcal{E}\right)\\
 & -\frac{3}{2}\left(\mathcal{H}_{\alpha}-\frac{1}{6}\varepsilon_{\alpha\mu}Q^{\mu}\right)\phi-\Omega^{\mu}\left(3\mathcal{E}_{\alpha\mu}-\frac{1}{2}\Pi_{\alpha\mu}\right)\\
 & +\mathcal{H}_{\alpha\mu}a^{\mu}-\mathcal{H}^{\mu}\zeta_{\alpha\mu}+\varepsilon_{\alpha\nu}\Sigma_{\mu}\left(\mathcal{E}^{\mu\nu}+\frac{1}{2}\Pi^{\mu\nu}\right)\\
 & -\varepsilon_{\alpha\nu}\Sigma_{\mu}{}^{\nu}\left(\mathcal{E}^{\mu}+\frac{1}{2}\Pi^{\mu}\right)+\frac{1}{2}\varepsilon_{\alpha\mu}Q_{\nu}\zeta^{\mu\nu}\,;
\end{aligned}
\label{1p1p2_eqs:HV_hat_QV_hat}
\end{equation}

Tensor equations:
\begin{equation}
\begin{aligned}N^{\mu}{}_{\{\alpha}N_{\beta\}}{}^{\nu}\dot{\mathcal{E}}_{\mu\nu}+\frac{1}{2}N^{\mu}{}_{\{\alpha}N_{\beta\}}{}^{\nu}\dot{\Pi}_{\mu\nu}+N^{\mu}{}_{\{\alpha}\varepsilon_{\beta\}}{}^{\nu}\widehat{\mathcal{H}}_{\mu\nu} & =\varepsilon_{\{\alpha|}{}^{\nu}\delta_{\nu}\mathcal{H}_{|\beta\}}-\frac{1}{2}\delta_{\{\alpha}Q_{\beta\}}-\frac{1}{2}\left(\mu+p+3\mathcal{E}-\frac{1}{2}\Pi\right)\Sigma_{\alpha\beta}\\
 & -\frac{1}{2}Q\zeta_{\alpha\beta}-\frac{3}{2}\mathcal{H}\varepsilon^{\nu}{}_{\{\alpha}\zeta_{\beta\}\nu}-\left(\theta+\frac{3}{2}\Sigma\right)\mathcal{E}_{\alpha\beta}+\mathcal{H}_{\alpha\beta}\xi\\
 & +\varepsilon_{\mu\{\alpha}\mathcal{E}_{\beta\}}{}^{\mu}\Omega-\left(\frac{1}{6}\theta-\frac{1}{4}\Sigma\right)\Pi_{\alpha\beta}+\frac{1}{2}\varepsilon^{\mu}{}_{\{\alpha}\Pi_{\beta\}\mu}\Omega\\
 & +\left(\frac{1}{2}\phi+2\mathcal{A}\right)\varepsilon_{\mu\{\alpha}\mathcal{H}_{\beta\}}{}^{\mu}+\left(3\mathcal{E}_{\{\alpha}-\frac{1}{2}\Pi_{\{\alpha}\right)\Sigma_{\beta\}}\\
 & -\mathcal{A}_{\{\alpha}Q_{\beta\}}-\left(2\mathcal{E}_{\{\alpha}+\Pi_{\{\alpha}\right)\left(\alpha_{\beta\}}-\frac{1}{2}\varepsilon_{\beta\}}{}^{\mu}\Omega_{\mu}\right)\\
 & +\mathcal{H}_{\{\alpha}\varepsilon_{\beta\}}{}^{\mu}\left(2\mathcal{A}_{\mu}-a_{\mu}\right)+\mathcal{H}_{\mu}\varepsilon^{\mu}{}_{\{\alpha}a_{\beta\}}\\
 & -\mathcal{H}^{\mu}{}_{\{\alpha}\varepsilon_{\beta\}}{}^{\nu}\zeta_{\mu\nu}\,,
\end{aligned}
\label{1p1p2_eqs:ET_dot_PiT_dot_HT_hat}
\end{equation}

\begin{equation}
\begin{aligned}N^{\mu}{}_{\{\alpha}\varepsilon_{\beta\}}{}^{\nu}\widehat{\mathcal{E}}_{\mu\nu}-\frac{1}{2}N^{\mu}{}_{\{\alpha}\varepsilon_{\beta\}}{}^{\nu}\widehat{\Pi}_{\mu\nu}-\dot{\mathcal{H}}_{\left\{ \alpha\beta\right\} } & =\varepsilon_{\{\alpha|}{}^{\mu}\delta_{\mu}\mathcal{E}_{|\beta\}}-\frac{1}{2}\varepsilon_{\{\alpha|}{}^{\mu}\delta_{\mu}\Pi_{|\beta\}}+\xi\left(\mathcal{E}_{\alpha\beta}-\frac{1}{2}\Pi_{\alpha\beta}\right)\\
 & +\frac{3}{2}\mathcal{H}\Sigma_{\alpha\beta}+\frac{1}{2}Q\varepsilon_{\mu\{\alpha}\Sigma_{\beta\}}{}^{\mu}+\left(\theta+\frac{3}{2}\Sigma\right)\mathcal{H}_{\alpha\beta}\\
 & -\varepsilon_{\mu\{\alpha}\mathcal{H}_{\beta\}}{}^{\mu}\Omega-\frac{3}{2}Q_{\{\alpha}\Omega_{\beta\}}-\frac{1}{4}\phi\varepsilon_{\mu\{\alpha}\Pi_{\beta\}}{}^{\mu}\\
 & -\frac{1}{2}a^{\mu}\varepsilon_{\mu\{\alpha}\Pi_{\beta\}}-\left(3\mathcal{H}_{\{\alpha}+\frac{1}{2}\varepsilon_{\mu\{\alpha}Q^{\mu}\right)\Sigma_{\beta\}}\\
 & +\Omega^{\mu}\varepsilon_{\mu\{\alpha}\mathcal{H}_{\beta\}}-\frac{3}{2}\left(\mathcal{E}-\frac{1}{2}\Pi\right)\varepsilon_{\mu\{\alpha}\zeta_{\beta\}}{}^{\mu}\\
 & +\varepsilon_{\mu\{\alpha}a_{\beta\}}\left(\mathcal{E}^{\mu}-\frac{1}{2}\Pi^{\mu}\right)-3\mathcal{H}_{\mu\{\alpha}\Sigma_{\beta\}}{}^{\mu}\\
 & +2\mathcal{H}_{\{\alpha}\alpha_{\beta\}}+\left(\frac{1}{2}\phi+2\mathcal{A}\right)\varepsilon_{\mu\{\alpha}\mathcal{E}_{\beta\}}{}^{\mu}\\
 & +\zeta_{\mu\nu}\varepsilon^{\mu}{}_{\{\alpha}\left(\mathcal{E}^{\nu}{}_{\beta\}}-\frac{1}{2}\Pi^{\nu}{}_{\beta\}}\right)\\
 & -\varepsilon_{\mu\{\alpha}\mathcal{E}_{\beta\}}\left(2\mathcal{A}^{\mu}-a^{\mu}\right)\,,
\end{aligned}
\label{1p1p2_eqs:ET_hat_PiT_hat_HT_dot}
\end{equation}
where we have used the fact that for any two symmetric and traceless
2-tensors on the sheet, $A$ and $B$, that is $A_{\mu\nu}=N_{\{\mu}{}^{\alpha}N_{\nu\}}{}^{\beta}A_{\alpha\beta}$
and $B_{\mu\nu}=N_{\{\mu}{}^{\alpha}N_{\nu\}}{}^{\beta}B_{\alpha\beta}$,
the contraction $A^{\delta}{}_{\{a}B_{b\}\delta}=0$.

\subsection{Matter conservation laws}

Lastly, the energy and momentum conservation equations are given by

\begin{equation}
\dot{\mu}+\widehat{Q}=Q^{\mu}\left(a_{\mu}-2\mathcal{A}_{\mu}\right)-2\Pi^{\mu}\Sigma_{\mu}-\left(\phi+2\mathcal{A}\right)Q-\frac{3}{2}\Pi\Sigma-\theta\left(\mu+p\right)-\delta_{\mu}Q^{\mu}-\Pi^{\mu\nu}\Sigma_{\mu\nu}\,,\label{1p1p2_eqs:Conservation_energy}
\end{equation}
\begin{equation}
\begin{aligned}\dot{Q}+\widehat{p}+\hat{\Pi}= & -\delta_{\mu}\Pi^{\mu}-\left(\frac{3}{2}\phi+\mathcal{A}\right)\Pi-\left(\frac{4}{3}\theta+\Sigma\right)Q-\left(\mu+p\right)\mathcal{A}\\
 & +\left(\alpha_{\mu}-\Sigma_{\mu}+\varepsilon_{\mu\nu}\Omega^{\nu}\right)Q^{\mu}+\Pi^{\mu}\left(2a_{\mu}-\mathcal{A}_{\mu}\right)+\Pi_{\mu\nu}\zeta^{\mu\nu}\,,
\end{aligned}
\label{1p1p2_eqs:Momentum_conservation_scalar}
\end{equation}
\begin{equation}
\begin{aligned}N_{\alpha}{}^{\mu}\dot{Q}_{\mu}+N_{\alpha}{}^{\mu}\widehat{\Pi}_{\mu}= & -\delta_{\alpha}p+\frac{1}{2}\delta_{\alpha}\Pi-\delta_{\mu}\Pi_{\alpha}{}^{\mu}-Q\left(\alpha_{\alpha}+\Sigma_{\alpha}+\varepsilon_{\alpha\beta}\Omega^{\beta}\right)+\varepsilon_{\alpha\mu}Q^{\mu}\Omega\\
 & -\frac{3}{2}\Pi a_{\alpha}-\left(\frac{4}{3}\theta-\frac{1}{2}\Sigma\right)Q_{\alpha}-\left(\frac{3}{2}\phi+\mathcal{A}\right)\Pi_{\alpha}-\Pi^{\mu}\left(\zeta_{\alpha\mu}-\varepsilon_{\alpha\mu}\xi\right)\\
 & -\left(\mu+p-\frac{1}{2}\Pi\right)\mathcal{A}_{\alpha}-Q^{\mu}\Sigma_{\alpha\mu}+\Pi_{\alpha\mu}\left(a^{\mu}-\mathcal{A}^{\mu}\right)\,.
\end{aligned}
\label{1p1p2_eqs:Momentum_conservation_vector}
\end{equation}

\section{Linearized 1+1+2 equations for the angular gradients variables\label{Appendix:Linear_eqs_angular_variables}}

In this appendix, we present a set of covariant, gauge invariant
equations for the linear order perturbations of a background spacetime
assumed to be static LRS II and permeated by a general matter
fluid, such that the equilibrium configuration is characterized by
the scalars $\left\{ \phi_{0},\mathcal{A}_{0},\mathcal{E}_{0},\mu_{0},p_{0},\Pi_{0},\Lambda\right\} $.
Considering the variables in Eq.~(\ref{eq:GI_angular_gradients_definition})
and using the commutation relations (\ref{1p1p2_eqs:Commutation_relations_scalars}),
we find the following set of linearized equations for the various
quantities that characterize the perturbed spacetime.

\subsection{Linearized equations for the kinematical quantities associated with
the timelike congruence}
\begin{itemize}

\item Evolution and propagation equations for scalars and gradients of scalar
quantities:

\begin{equation}
\begin{aligned}\left(\delta_{\alpha}\theta\right)^{\cdot}-\widehat{\mathbb{A}}_{\alpha} & =\left[\frac{1}{2}\left(\mu_{0}+3p_{0}\right)-\Lambda-\mathcal{A}_{0}\left(\mathcal{A}_{0}+\phi_{0}\right)\right]a_{\alpha}-\frac{1}{2}\left(\mathfrak{m}_{\alpha}+3\mathfrak{p}_{\alpha}\right)\\
 & +\mathcal{A}_{0}\mathbb{F}_{\alpha}+\left(\frac{3}{2}\phi_{0}+2\mathcal{A}_{0}\right)\mathbb{A}_{\alpha}+\delta_{\alpha}\left(\delta_{\mu}\mathcal{A}^{\mu}\right)\,,
\end{aligned}
\end{equation}
\begin{equation}
\begin{aligned}\frac{2}{3}\left(\delta_{\alpha}\theta\right)^{\cdot}-\left(\delta_{\alpha}\Sigma\right)^{\cdot} & =\phi_{0}\mathbb{A}_{\alpha}+\mathcal{A}_{0}\left(\mathbb{F}_{\alpha}+\phi_{0}\mathcal{A}_{\alpha}\right)-\left(\frac{1}{3}\mu_{0}+p_{0}-\frac{2}{3}\Lambda+\frac{1}{2}\Pi_{0}-\mathcal{E}_{0}\right)\mathcal{A}_{\alpha}\\
 & -\frac{1}{3}\mathfrak{m}_{\alpha}-\mathfrak{p}_{\alpha}-\frac{1}{2}\mathbb{P}_{\alpha}+\mathbb{E}_{\alpha}+\delta_{\alpha}\left(\delta_{\mu}\mathcal{A}^{\mu}\right)\,,
\end{aligned}
\end{equation}
\begin{equation}
\begin{aligned}\frac{3}{2}\widehat{\delta_{\alpha}\Sigma}-\widehat{\delta_{\alpha}\theta} & =\left(\mu_{0}+3p_{0}-2\Lambda+\frac{3}{2}\Pi_{0}-3\mathcal{E}_{0}-3\mathcal{A}_{0}\phi_{0}\right)\varepsilon_{\alpha\mu}\Omega^{\mu}\\
 & +\frac{1}{2}\phi_{0}\left(\delta_{\alpha}\theta-6\delta_{\alpha}\Sigma\right)-\frac{3}{2}\delta_{\alpha}Q-\frac{3}{2}\delta_{\alpha}\left(\delta_{\mu}\Sigma^{\mu}\right)-\frac{3}{2}\varepsilon_{\mu\nu}\delta_{\alpha}\left(\delta^{\mu}\Omega^{\nu}\right)\,,
\end{aligned}
\end{equation}

\begin{equation}
\dot{\Omega}=\frac{1}{2}\varepsilon^{\mu\nu}\delta_{\mu}\mathcal{A}_{\nu}+\mathcal{A}_{0}\xi\,,
\end{equation}

\begin{equation}
\hat{\Omega}=\left(\mathcal{A}_{0}-\phi_{0}\right)\Omega-\delta_{\mu}\Omega^{\mu}\,;
\end{equation}

\item Evolution and propagation equations for vector quantities:
\begin{equation}
\dot{\Omega}_{\alpha}+\frac{1}{2}\varepsilon_{\alpha}{}^{\mu}\widehat{\mathcal{A}}_{\mu}=-\frac{1}{2}\varepsilon_{\alpha}{}^{\mu}\left(\frac{1}{2}\phi_{0}\mathcal{A}_{\mu}+\mathcal{A}_{0}a_{\mu}-\mathbb{A}_{\mu}\right)\,,
\end{equation}
\begin{equation}
\begin{aligned}\dot{\Sigma}_{\alpha}-\frac{1}{2}\widehat{\mathcal{A}}_{\alpha}= & \left(\mathcal{A}_{0}-\frac{1}{4}\phi_{0}\right)\mathcal{A}_{\alpha}+\frac{1}{2}\mathcal{A}_{0}a_{\alpha}+\frac{1}{2}\mathbb{A}_{\alpha}-\mathcal{E}_{\alpha}+\frac{1}{2}\Pi_{\alpha}\,,\end{aligned}
\end{equation}

\begin{equation}
\begin{aligned}\widehat{\Sigma}_{\alpha}-\varepsilon_{\alpha}{}^{\mu}\widehat{\Omega}_{\mu} & =\frac{1}{2}\delta_{\alpha}\Sigma+\frac{2}{3}\delta_{\alpha}\theta-\frac{3}{2}\phi_{0}\Sigma_{\alpha}+\left(\frac{1}{2}\phi_{0}+2\mathcal{A}_{0}\right)\varepsilon_{\alpha}{}^{\mu}\Omega_{\mu}\\
 & -\varepsilon_{\alpha}{}^{\mu}\delta_{\mu}\Omega-\delta_{\mu}\Sigma_{\alpha}{}^{\mu}-Q_{\alpha}\,;
\end{aligned}
\end{equation}

\item Evolution and propagation equations for tensor quantities:
\begin{equation}
\dot{\Sigma}_{\alpha\beta}=\delta_{\{\alpha}\mathcal{A}_{\beta\}}+\mathcal{A}_{0}\zeta_{\alpha\beta}-\mathcal{E}_{\alpha\beta}+\frac{1}{2}\Pi_{\alpha\beta}\,,
\end{equation}
\begin{equation}
\widehat{\Sigma}_{\alpha\beta}=\delta_{\{\alpha}\Sigma_{\beta\}}+\frac{1}{2}\varepsilon_{\{\alpha|}{}^{\mu}\delta_{\mu}\Omega_{|\beta\}}+\frac{1}{2}\varepsilon_{\{\alpha|}{}^{\mu}\delta_{|\beta\}}\Omega_{\mu}-\frac{1}{2}\phi_{0}\Sigma_{\alpha\beta}-\varepsilon_{\mu\{\alpha}\mathcal{H}_{\beta\}}{}^{\mu}\,;
\end{equation}

\item Constraint equations:

\begin{equation}
\varepsilon^{\mu\nu}\delta_{\mu}\mathbb{A}_{\nu}=-2\widehat{\mathcal{A}}_{0}\xi\,,
\end{equation}
\begin{equation}
\delta_{\mu}\Omega^{\mu}+\varepsilon^{\mu\nu}\delta_{\mu}\Sigma_{\nu}=\left(2\mathcal{A}_{0}-\phi_{0}\right)\Omega+\mathcal{H}\,,
\end{equation}
\begin{equation}
\varepsilon_{\alpha}{}^{\mu}\delta_{\mu}\Omega+\delta_{\mu}\Sigma_{\alpha}{}^{\mu}=\frac{1}{2}\phi_{0}\left(\varepsilon_{\alpha}{}^{\mu}\Omega_{\mu}-\Sigma_{\alpha}\right)-\varepsilon_{\alpha}{}^{\mu}\mathcal{H}_{\mu}+\frac{1}{3}\delta_{\alpha}\theta-\frac{1}{2}\delta_{\alpha}\Sigma-\frac{1}{2}Q_{\alpha}\,.
\end{equation}

\end{itemize}\hfill\hfill

\subsection{Linearized equations for the kinematical quantities associated with
the spacelike congruence}
\begin{itemize}

\item Evolution and propagation equations for scalars and gradients of scalar
quantities:

\begin{equation}
\begin{aligned}\dot{\mathbb{F}}_{\alpha} & =\left[\frac{1}{2}\phi_{0}^{2}+\frac{2}{3}\left(\mu_{0}+\Lambda\right)+\frac{1}{2}\Pi_{0}+\mathcal{E}_{0}\right]\left(\Sigma_{\alpha}+\alpha_{\alpha}-\varepsilon_{\alpha\mu}\Omega^{\mu}\right)\\
 & +\delta_{\alpha}Q+\left(\mathcal{A}_{0}-\frac{1}{2}\phi_{0}\right)\left(\frac{2}{3}\delta_{\alpha}\theta-\delta_{\alpha}\Sigma\right)+\delta_{\alpha}\left(\delta_{\mu}\alpha^{\mu}\right)\,,
\end{aligned}
\end{equation}
\begin{equation}
\begin{aligned}\widehat{\mathbb{F}}_{\alpha} & =\left[\frac{1}{2}\phi_{0}^{2}+\frac{2}{3}\left(\mu_{0}+\Lambda\right)+\frac{1}{2}\Pi_{0}+\mathcal{E}_{0}\right]a_{\alpha}+\delta_{\alpha}\left(\delta_{\mu}a^{\mu}\right)\\
 & -\frac{2}{3}\mathfrak{m}_{\alpha}-\frac{1}{2}\mathbb{P}_{\alpha}-\mathbb{E}_{\alpha}-\frac{3}{2}\phi_{0}\mathbb{F}_{\alpha}\,,
\end{aligned}
\end{equation}
\begin{equation}
\dot{\xi}=\frac{1}{2}\mathcal{H}+\left(\mathcal{A}_{0}-\frac{1}{2}\phi_{0}\right)\Omega+\frac{1}{2}\varepsilon^{\mu\nu}\delta_{\mu}\alpha_{\nu}\,,
\end{equation}

\begin{equation}
\widehat{\xi}=-\phi_{0}\xi+\frac{1}{2}\varepsilon^{\mu\nu}\delta_{\mu}a_{\nu}\,;
\end{equation}

\item Evolution and propagation equations for vector quantities:

\begin{equation}
\widehat{\alpha}_{\alpha}-\dot{a}_{\alpha}=
\varepsilon_{\mu\alpha}\mathcal{H}^{\mu}-\left(\mathcal{A}_{0}+
\frac{1}{2}\phi_{0}\right)\alpha_{\alpha}+
\frac{1}{2}Q_{\alpha}+
\left(\frac{1}{2}\phi_{0}-
\mathcal{A}_{0}\right)\left(\Sigma_{\alpha}+
\varepsilon_{\alpha\mu}\Omega^{\mu}\right)\,;
\end{equation}

\item Evolution and propagation equations for tensor quantities:
\begin{equation}
\dot{\zeta}_{\alpha\beta}=\left(\mathcal{A}_{0}-\frac{1}{2}\phi_{0}\right)\Sigma_{\alpha\beta}+\delta_{\{\alpha}\alpha_{\beta\}}-\varepsilon^{\mu}{}_{\{\alpha}\mathcal{H}_{\beta\}\mu}\,,
\end{equation}
\begin{equation}
\widehat{\zeta}_{\alpha\beta}=\delta_{\{\alpha}a_{\beta\}}-\phi_{0}\zeta_{\alpha\beta}-\mathcal{E}_{\alpha\beta}-\frac{1}{2}\Pi_{\alpha\beta}\,;
\end{equation}

\item Constraint equations:

\begin{equation}
\varepsilon^{\mu\nu}\delta_{\mu}\mathbb{F}_{\nu}=\left[\phi_{0}^{2}+\frac{4}{3}\left(\mu_{0}+\Lambda\right)+\Pi_{0}+2\mathcal{E}_{0}\right]\xi\,,
\end{equation}
\begin{equation}\label{delta_zeta_pert_eq}
\delta_{\mu}\zeta_{\alpha}{}^{\mu}+\varepsilon_{\alpha}{}^{\mu}\delta_{\mu}\xi=\frac{1}{2}\mathbb{F}_{\alpha}+\mathcal{E}_{\alpha}+\frac{1}{2}\Pi_{\alpha}\,.
\end{equation}

\end{itemize}\hfill\hfill

\subsection{Linearized equations for the Weyl tensor components and the matter
variables}
\begin{itemize}

\item Evolution and propagation equations for scalars and gradients of scalar
quantities:

\begin{equation}
\begin{aligned}\widehat{\delta_{\alpha}Q}+\dot{\mathfrak{m}}_{\alpha} & =\widehat{\mu}_{0}\left(\varepsilon_{\alpha\mu}\Omega^{\mu}-\Sigma_{\alpha}-\alpha_{\alpha}\right)-\frac{3}{2}\Pi_{0}\delta_{\alpha}\Sigma-\delta_{\alpha}\left(\delta_{\mu}Q^{\mu}\right)\\
 & -\left(\mu_{0}+p_{0}\right)\delta_{\alpha}\theta-\left(\frac{3}{2}\phi_{0}+2\mathcal{A}_{0}\right)\delta_{\alpha}Q\,,
\end{aligned}
\end{equation}

\begin{equation}
\begin{aligned}\frac{1}{3}\dot{\mathfrak{m}}_{\alpha}-\frac{1}{2}\dot{\mathbb{P}}_{\alpha}-\dot{\mathbb{E}}_{\alpha} & =\frac{3}{2}\phi_{0}\left(\mathcal{E}_{0}+\frac{1}{2}\Pi_{0}\right)\left(\varepsilon_{\alpha\mu}\Omega^{\mu}-\Sigma_{\alpha}-\alpha_{\alpha}\right)-\frac{1}{2}\phi_{0}\delta_{\alpha}Q\\
 & +\left(\mu_{0}+p_{0}-\frac{1}{2}\Pi_{0}-3\mathcal{E}_{0}\right)\left(\frac{1}{2}\delta_{\alpha}\Sigma-\frac{1}{3}\delta_{\alpha}\theta\right)\\
 & -\frac{1}{2}\delta_{\alpha}\left(\delta_{\mu}Q^{\mu}\right)-\varepsilon^{\mu\nu}\delta_{\alpha}\left(\delta_{\mu}\mathcal{H}_{\nu}\right)\,,
\end{aligned}
\end{equation}

\begin{equation}
\begin{aligned}\left(\delta_{\alpha}Q\right)^{\cdot}+\widehat{\mathbb{P}}_{\alpha}+\widehat{\mathfrak{p}}_{\alpha} & =\left(\mu_{0}+p_{0}\right)\left(\mathcal{A}_{0}a_{\alpha}-\mathbb{A}_{\alpha}\right)-\mathcal{A}_{0}\left(\mathfrak{m}_{\alpha}+\mathfrak{p}_{\alpha}+\mathbb{P}_{\alpha}\right)\\
 & -\frac{1}{2}\phi_{0}\left(\mathfrak{p}_{\alpha}+4\mathbb{P}_{\alpha}\right)+\left(\frac{3}{2}\phi_{0}+\mathcal{A}_{0}\right)\Pi_{0}a_{\alpha}\\
 & -\Pi_{0}\left(\frac{3}{2}\mathbb{F}_{\alpha}+\mathbb{A}_{\alpha}\right)-\delta_{\alpha}\left(\delta_{\mu}\Pi^{\mu}\right)\,,
\end{aligned}
\end{equation}

\begin{equation}
\begin{aligned}\frac{3}{2}\widehat{\mathbb{P}}_{\alpha}+3\widehat{\mathbb{E}}_{\alpha}-\widehat{\mathfrak{m}}_{\alpha}= & -6\phi_{0}\left(\mathbb{E}_{\alpha}+\frac{1}{2}\mathbb{P}_{\alpha}-\frac{1}{12}\mathfrak{m}_{\alpha}\right)-\frac{9}{2}\left(\mathcal{E}_{0}+\frac{1}{2}\Pi_{0}\right)\left(\mathbb{F}_{\alpha}-\phi_{0}a_{\alpha}\right)\\
 & -3\delta_{\alpha}\left(\delta_{\mu}\mathcal{E}^{\mu}\right)-\frac{3}{2}\delta_{\alpha}\left(\delta_{\mu}\Pi^{\mu}\right)\,,
\end{aligned}
\end{equation}

\begin{equation}
\dot{\mathcal{H}}=\frac{1}{2}\varepsilon^{\mu\nu}\delta_{\mu}\Pi_{\nu}-\varepsilon^{\mu\nu}\delta_{\mu}\mathcal{E}_{\nu}-3\left(\mathcal{E}_{0}-\frac{1}{2}\Pi_{0}\right)\xi\,,
\end{equation}
\begin{equation}
\widehat{\mathcal{H}}=-\delta_{\mu}\mathcal{H}^{\mu}-\frac{1}{2}\varepsilon_{\mu\nu}\delta^{\mu}Q^{\nu}-\frac{3}{2}\phi_{0}\mathcal{H}-\left(3\mathcal{E}_{0}+\mu_{0}+p_{0}-\frac{1}{2}\Pi_{0}\right)\Omega\,;
\end{equation}

\item Evolution and propagation equations for vector quantities:
\begin{equation}
\begin{aligned}\dot{\mathcal{E}}_{\alpha}+\frac{1}{2}\dot{\Pi}_{\alpha} & =\left(\frac{1}{2}\phi_{0}-\mathcal{A}_{0}\right)\left(\frac{1}{2}Q_{\alpha}+\varepsilon_{\alpha\mu}\mathcal{H}^{\mu}\right)-\frac{3}{2}\left(\mathcal{E}_{0}+\frac{1}{2}\Pi_{0}\right)\alpha_{\alpha}\\
 & -\frac{1}{2}\left(\mu_{0}+p_{0}+\Pi_{0}\right)\left(\Sigma_{\alpha}-\varepsilon_{\alpha\mu}\Omega^{\mu}\right)-\frac{1}{2}\delta_{\alpha}Q\\
 & +\frac{1}{2}\varepsilon_{\alpha}{}^{\mu}\delta_{\mu}\mathcal{H}+\varepsilon^{\mu\nu}\delta_{\mu}\mathcal{H}_{\nu\alpha}\,,
\end{aligned}
\end{equation}

\begin{equation}
\begin{aligned}\widehat{\mathcal{E}}_{\alpha}+\frac{1}{2}\widehat{\Pi}_{\alpha} & =-\frac{3}{2}\phi_{0}\left(\mathcal{E}_{\alpha}+\frac{1}{2}\Pi_{\alpha}\right)-\frac{3}{2}\left(\mathcal{E}_{0}+\frac{1}{2}\Pi_{0}\right)a_{\alpha}-\frac{1}{2}\delta_{\mu}\Pi_{\alpha}{}^{\mu}\\
 & +\frac{1}{2}\mathbb{E}_{\alpha}+\frac{1}{3}\mathfrak{m}_{\alpha}+\frac{1}{4}\mathbb{P}_{\alpha}-\delta_{\mu}\mathcal{E}_{\alpha}{}^{\mu}\,,
\end{aligned}
\end{equation}

\begin{equation}
\begin{aligned}\frac{1}{2}\varepsilon_{\alpha}{}^{\mu}\widehat{\mathcal{E}}_{\mu}-\dot{\mathcal{H}}_{\alpha}-\frac{1}{4}\varepsilon_{\alpha}{}^{\mu}\widehat{\Pi}_{\mu} & =\frac{3}{4}\varepsilon_{\alpha}{}^{\mu}\mathbb{E}_{\mu}-\frac{3}{8}\varepsilon_{\alpha}{}^{\mu}\mathbb{P}_{\mu}+\frac{1}{2}\varepsilon^{\mu\nu}\delta_{\mu}\mathcal{E}_{\nu\alpha}-\frac{1}{4}\varepsilon^{\mu\nu}\delta_{\mu}\Pi_{\nu\alpha}\\
 & +\frac{3}{2}\mathcal{E}_{0}\varepsilon_{\alpha\mu}\mathcal{A}^{\mu}-\frac{3}{4}\left(\mathcal{E}_{0}-\frac{1}{2}\Pi_{0}\right)\varepsilon_{\alpha\mu}a^{\mu}-\mathcal{A}_{0}\varepsilon_{\alpha\mu}\mathcal{E}^{\mu}\\
 & -\frac{1}{4}\phi_{0}\varepsilon_{\alpha\mu}\left(\mathcal{E}^{\mu}-\frac{1}{2}\Pi^{\mu}\right)\,,
\end{aligned}
\end{equation}

\begin{equation}
\begin{aligned}\widehat{\mathcal{H}}_{\alpha}-\frac{1}{2}\varepsilon_{\alpha}{}^{\mu}\widehat{Q}_{\mu} & =\frac{1}{2}\delta_{\alpha}\mathcal{H}-\delta_{\mu}\mathcal{H}_{\alpha}{}^{\mu}-\frac{1}{2}\varepsilon_{\alpha}{}^{\mu}\delta_{\mu}Q-\frac{3}{2}\left(\mathcal{E}_{0}+\frac{1}{2}\Pi_{0}\right)\varepsilon_{\alpha\mu}\Sigma^{\mu}\\
 & -\left(\mu_{0}+p_{0}+\frac{1}{4}\Pi_{0}-\frac{3}{2}\mathcal{E}_{0}\right)\Omega_{\alpha}-\frac{3}{2}\phi_{0}\left(\mathcal{H}_{\alpha}-\frac{1}{6}\varepsilon_{\alpha\mu}Q^{\mu}\right)\,,
\end{aligned}
\end{equation}
\begin{equation}
\begin{aligned}\dot{Q}_{\alpha}+\widehat{\Pi}_{\alpha} & =\frac{1}{2}\mathbb{P}_{\alpha}-\mathfrak{p}_{\alpha}-\frac{3}{2}\Pi_{0}a_{\alpha}-\left(\frac{3}{2}\phi_{0}+\mathcal{A}_{0}\right)\Pi_{\alpha}\\
 & -\delta_{\mu}\Pi_{\alpha}{}^{\mu}-\left(\mu_{0}+p_{0}-\frac{1}{2}\Pi_{0}\right)\mathcal{A}_{\alpha}\,;
\end{aligned}
\end{equation}

\item Evolution and propagation equations for tensor quantities:
\begin{equation}
\begin{aligned}\dot{\mathcal{E}}_{\alpha\beta}+\frac{1}{2}\dot{\Pi}_{\alpha\beta}+N^{\mu}{}_{\{\alpha}\varepsilon_{\beta\}}{}^{\nu}\widehat{\mathcal{H}}_{\mu\nu} & =\varepsilon_{\{\alpha|}{}^{\nu}\delta_{\nu}\mathcal{H}_{|\beta\}}-\frac{1}{2}\delta_{\{\alpha}Q_{\beta\}}+\left(\frac{1}{2}\phi_{0}+2\mathcal{A}_{0}\right)\varepsilon_{\mu\{\alpha}\mathcal{H}_{\beta\}}{}^{\mu}\\
 & -\frac{1}{2}\left(\mu_{0}+p_{0}+3\mathcal{E}_{0}-\frac{1}{2}\Pi_{0}\right)\Sigma_{\alpha\beta}\,,
\end{aligned}
\end{equation}

\begin{equation}
\begin{aligned}N^{\mu}{}_{\{\alpha}\varepsilon_{\beta\}}{}^{\nu}\widehat{\mathcal{E}}_{\mu\nu}-\frac{1}{2}N^{\mu}{}_{\{\alpha}\varepsilon_{\beta\}}{}^{\nu}\widehat{\Pi}_{\mu\nu}-\dot{\mathcal{H}}_{\alpha\beta} & =\varepsilon_{\{\alpha|}{}^{\mu}\delta_{\mu}\mathcal{E}_{|\beta\}}-\frac{1}{2}\varepsilon_{\{\alpha|}{}^{\mu}\delta_{\mu}\Pi_{|\beta\}}-\frac{3}{2}\left(\mathcal{E}_{0}-\frac{1}{2}\Pi_{0}\right)\varepsilon_{\mu\{\alpha}\zeta_{\beta\}}{}^{\mu}\\
 & +\left(\frac{1}{2}\phi_{0}+2\mathcal{A}_{0}\right)\varepsilon_{\mu\{\alpha}\mathcal{E}_{\beta\}}{}^{\mu}-\frac{1}{4}\phi_{0}\varepsilon_{\mu\{\alpha}\Pi_{\beta\}}{}^{\mu}\,;
\end{aligned}
\end{equation}

\item Constraint equations:
\begin{equation}
\frac{1}{3}\varepsilon^{\mu\nu}\delta_{\mu}\mathfrak{m}_{\nu}-\varepsilon^{\mu\nu}\delta_{\mu}\mathbb{E}_{\nu}-\frac{1}{2}\varepsilon^{\mu\nu}\delta_{\mu}\mathbb{P}_{\nu}=-3\phi_{0}\left(\mathcal{E}_{0}+\frac{1}{2}\Pi_{0}\right)\xi\,,
\end{equation}

\begin{equation}
\varepsilon^{\mu\nu}\delta_{\mu}\mathfrak{p}_{\nu}=-2\widehat{p}_{0}\xi\,,
\end{equation}

\end{itemize}

\section{Linearized 1+1+2 equations for the dot-derivatives variables \label{Appendix:Linear_eqs_dot-derivatives_variables}}

In this appendix, we present a set of covariant, gauge invariant
equations for the linear order perturbations of a background spacetime
assumed to be static LRS II and permeated by a general matter
fluid, such that the equilibrium configuration is characterized by
the scalars $\left\{ \phi_{0},\mathcal{A}_{0},\mathcal{E}_{0},\mu_{0},p_{0},\Pi_{0},\Lambda\right\} $.
Considering the variables in (\ref{eq:GI_dot_derivatives_definition}) and using the commutation
relations (\ref{1p1p2_eqs:Commutation_relations_scalars}), we find
the following set of linearized equations for the various quantities
that characterize the perturbed spacetime. Here, we also show that the variables in Eq.~(\ref{eq:GI_angular_gradients_definition})
and (\ref{eq:GI_dot_derivatives_definition}) are not fully independent.

\subsection{Linearized equations for the kinematical quantities associated with
the timelike congruence}
\begin{itemize}

\item Evolution and propagation equations for scalars and dot-derivatives
of scalar quantities:
\begin{equation}
\widehat{\mathsf{A}}-\ddot{\theta}=
\frac{1}{2}\left(\mathsf{m}+3\mathsf{p}\right)+
\widehat{\mathcal{A}}_{0}\left(\frac{1}{3}\theta+
\Sigma\right)-\left(3\mathcal{A}_{0}+\phi_{0}\right)\mathsf{A} 
-\mathcal{A}_{0}\mathsf{F}-
\left(\delta_{\mu}\mathcal{A}^{\mu}\right)^{\cdot}\,,
\end{equation}

\begin{equation}
\ddot{\Sigma}-\frac{2}{3}\ddot{\theta}=\frac{1}{3}\left(\mathsf{m}+3\mathsf{p}\right)+\frac{1}{2}\mathcal{P}-\mathsf{E}-\mathcal{A}_{0}\mathsf{F}-\phi_{0}\mathsf{A}-\left(\delta_{\mu}\mathcal{A}^{\mu}\right)^{.}\,,
\end{equation}

\begin{equation}
\dot{\Omega}=\frac{1}{2}\varepsilon^{\mu\nu}\delta_{\mu}\mathcal{A}_{\nu}+\mathcal{A}_{0}\xi\,,
\end{equation}

\begin{equation}
\widehat{\Omega}=\left(\mathcal{A}_{0}-\phi_{0}\right)\Omega-\delta_{\mu}\Omega^{\mu}\,,
\end{equation}
\begin{equation}
\frac{2}{3}\widehat{\theta}-\widehat{\Sigma}=Q+\frac{3}{2}\phi_{0}\Sigma+\delta_{\mu}\Sigma^{\mu}+\varepsilon_{\mu\nu}\delta^{\mu}\Omega^{\nu}\,;
\end{equation}

\item Evolution and propagation equations for vector quantities:
\begin{equation}
\dot{\Omega}_{\alpha}+\frac{1}{2}\varepsilon_{\alpha}{}^{\mu}\widehat{\mathcal{A}}_{\mu}=-\frac{1}{2}\varepsilon_{\alpha}{}^{\mu}\left(\frac{1}{2}\phi_{0}\mathcal{A}_{\mu}+\mathcal{A}_{0}a_{\mu}-\mathbb{A}_{\mu}\right)\,,
\end{equation}
\begin{equation}
\dot{\Sigma}_{\alpha}-\frac{1}{2}N_{\alpha}{}^{\mu}\widehat{\mathcal{A}}_{\alpha} =
\frac{1}{2}\mathbb{A}_{\alpha}+\left(\mathcal{A}_{0}-
\frac{1}{4}\phi_{0}\right)\mathcal{A}_{\alpha}+\frac{1}{2}\mathcal{A}_{0}a_{\alpha}
-\mathcal{E}_{\alpha}+\frac{1}{2}\Pi_{\alpha}\,,
\end{equation}

\begin{equation}
\begin{aligned}\hat{\Sigma}_{\alpha}-\varepsilon_{\alpha}{}^{\mu}\widehat{\Omega}_{\mu} & =\frac{1}{2}\delta_{\alpha}\Sigma+\frac{2}{3}\delta_{\alpha}\theta-\varepsilon_{\alpha}{}^{\mu}\delta_{\mu}\Omega+\left(\frac{1}{2}\phi_{0}+2\mathcal{A}_{0}\right)\varepsilon_{\alpha}{}^{\mu}\Omega_{\mu}\\
 & -\frac{3}{2}\phi_{0}\Sigma_{\alpha}-\delta_{\mu}\Sigma_{\alpha}{}^{\mu}-Q_{\alpha}\,;
\end{aligned}
\end{equation}

\item Evolution and propagation equations for tensor quantities:
\begin{equation}
\dot{\Sigma}_{\alpha\beta}=\delta_{\{\alpha}\mathcal{A}_{\beta\}}+\mathcal{A}_{0}\zeta_{\alpha\beta}-\mathcal{E}_{\alpha\beta}+\frac{1}{2}\Pi_{\alpha\beta}\,,
\end{equation}
\begin{equation}
\widehat{\Sigma}_{\alpha\beta} =
\delta_{\{\alpha}\Sigma_{\beta\}}+
\frac{1}{2}\varepsilon_{\{\alpha|}{}^{\mu}\delta_{\mu}\Omega_{|\beta\}}+
\frac{1}{2}\varepsilon_{\{\alpha|}{}^{\mu}\delta_{|\beta\}}\Omega_{\mu}-
\frac{1}{2}\phi_{0}\Sigma_{\alpha\beta}
-\varepsilon_{\mu\{\alpha}\mathcal{H}_{\beta\}}{}^{\mu}\,;
\end{equation}

\item Constraint equations:
\begin{equation}
\delta_{\mu}\Omega^{\mu}+\varepsilon^{\mu\nu}\delta_{\mu}\Sigma_{\nu}=\left(2\mathcal{A}_{0}-\phi_{0}\right)\Omega+\mathcal{H}\,,
\end{equation}
\begin{equation}
\delta_{\alpha}\Sigma-\frac{2}{3}\delta_{\alpha}\theta+2\varepsilon_{\alpha}{}^{\mu}\delta_{\mu}\Omega+2\delta_{\mu}\Sigma_{\alpha}{}^{\mu}=\phi_{0}\left(\varepsilon_{\alpha}{}^{\mu}\Omega_{\mu}-\Sigma_{\alpha}\right)-2\varepsilon_{\alpha}{}^{\mu}\mathcal{H}_{\mu}-Q_{\alpha}\,.
\end{equation}

\end{itemize}\hfill\hfill

\subsection{Linearized equations for the kinematical quantities associated with
the spacelike congruence}
\begin{itemize}

\item Evolution and propagation equations for scalars and dot-derivatives
of scalar quantities:
\begin{equation}
\mathsf{F}=Q+\left(2\mathcal{A}_{0}-\phi_{0}\right)\left(\frac{1}{3}\theta-\frac{1}{2}\Sigma\right)+\delta_{\mu}\alpha^{\mu}\,,
\end{equation}

\begin{equation}
\widehat{\mathsf{F}}=\hat{\phi}_{0}\left(\frac{1}{3}\theta+\Sigma\right)-\left(\mathcal{A}_{0}+\phi_{0}\right)\mathsf{F}-\frac{2}{3}\mathsf{m}-\frac{1}{2}\mathcal{P}-\mathsf{E}+\left(\delta_{\mu}a^{\mu}\right)^{.}\,,
\end{equation}
\begin{equation}
\dot{\xi}=\frac{1}{2}\mathcal{H}+\left(\mathcal{A}_{0}-\frac{1}{2}\phi_{0}\right)\Omega+\frac{1}{2}\varepsilon^{\mu\nu}\delta_{\mu}\alpha_{\nu}\,,
\end{equation}

\begin{equation}
\widehat{\xi}=-\phi_{0}\xi+\frac{1}{2}\varepsilon^{\mu\nu}\delta_{\mu}a_{\nu}\,;
\end{equation}

\item Evolution and propagation equations for vector quantities:
\begin{equation}
\widehat{\alpha}_{\alpha}-\dot{a}_{\alpha} =
\varepsilon_{\mu\alpha}\mathcal{H}^{\mu}-
\left(\mathcal{A}_{0}+\frac{1}{2}\phi_{0}\right)\alpha_{\alpha}+\frac{1}{2}Q_{\alpha}
+\left(\frac{1}{2}\phi_{0}-\mathcal{A}_{0}\right)\left(\Sigma_{\alpha}+\varepsilon_{\alpha\mu}\Omega^{\mu}\right)\,;
\end{equation}

\item Evolution and propagation equations for tensor quantities:
\begin{equation}
\dot{\zeta}_{\alpha\beta}=\left(\mathcal{A}_{0}-\frac{1}{2}\phi_{0}\right)\Sigma_{\alpha\beta}+\delta_{\{\alpha}\alpha_{\beta\}}-\varepsilon^{\mu}{}_{\{\alpha}\mathcal{H}_{\beta\}\mu}\,,
\end{equation}
\begin{equation}
\widehat{\zeta}_{\alpha\beta}=\delta_{\{\alpha}a_{\beta\}}-\phi_{0}\zeta_{\alpha\beta}-\mathcal{E}_{\alpha\beta}-\frac{1}{2}\Pi_{\alpha\beta}\,;
\end{equation}

\item Constraint equation:
\begin{equation}
\delta_{\mu}\zeta_{\alpha}{}^{\mu}+\varepsilon_{\alpha}{}^{\mu}\delta_{\mu}\xi-\frac{1}{2}\mathbb{F}_{\alpha}=\mathcal{E}_{\alpha}+\frac{1}{2}\Pi_{\alpha}\,.
\end{equation}

\end{itemize}\hfill\hfill

\subsection{Linearized equations for the Weyl tensor components and the matter
variables}
\begin{itemize}

\item Evolution and propagation equations for scalars and dot-derivatives
of scalar quantities:
\begin{equation}
\begin{aligned}\mathsf{E}+\frac{1}{2}\mathcal{P}+\frac{1}{3}\widehat{Q} & =\mathcal{E}_{0}\left(\frac{3}{2}\Sigma-\theta\right)-\frac{1}{2}\Pi_{0}\left(\frac{1}{3}\theta+\frac{1}{2}\Sigma\right)+\frac{1}{3}\left(\frac{1}{2}\phi_{0}-2\mathcal{A}_{0}\right)Q\\
 & -\frac{1}{2}\left(\mu_{0}+p_{0}\right)\Sigma+\frac{1}{6}\delta_{\mu}Q^{\mu}+\varepsilon^{\mu\nu}\delta_{\mu}\mathcal{H}_{\nu}\,,
\end{aligned}
\end{equation}
\begin{equation}
\begin{aligned}\widehat{\mathsf{E}}-\frac{1}{3}\widehat{\mathsf{m}}+\frac{1}{2}\widehat{\mathcal{P}} & =\left(\widehat{\mathcal{E}}_{0}-\frac{1}{3}\hat{\mu}_{0}+\frac{1}{2}\widehat{\Pi}_{0}\right)\left(\frac{1}{3}\theta+\Sigma\right)-\left(\delta_{\mu}\mathcal{E}^{\mu}\right)^{.}-\frac{1}{2}\left(\delta_{\mu}\Pi^{\mu}\right)^{.}\\
 & +\frac{1}{3}\mathcal{A}_{0}\mathsf{m}-\frac{3}{2}\left(\mathcal{E}_{0}+\frac{1}{2}\Pi_{0}\right)\mathsf{F}-\left(\mathcal{A}_{0}+\frac{3}{2}\phi_{0}\right)\left(\mathsf{E}+\frac{1}{2}\mathcal{P}\right)\,,
\end{aligned}
\end{equation}

\begin{equation}
\dot{\mathcal{H}}=\frac{1}{2}\varepsilon^{\mu\nu}\delta_{\mu}\Pi_{\nu}-\varepsilon^{\mu\nu}\delta_{\mu}\mathcal{E}_{\nu}-3\left(\mathcal{E}_{0}-\frac{1}{2}\Pi_{0}\right)\xi\,,
\end{equation}
\begin{equation}
\widehat{\mathcal{H}}=-\delta_{\mu}\mathcal{H}^{\mu}-\frac{1}{2}\varepsilon_{\mu\nu}\delta^{\mu}Q^{\nu}-\frac{3}{2}\phi_{0}\mathcal{H}-\left(3\mathcal{E}_{0}+\mu_{0}+p_{0}-\frac{1}{2}\Pi_{0}\right)\Omega\,,
\end{equation}

\begin{equation}
\mathsf{m}+\widehat{Q}=-\left(\phi_{0}+2\mathcal{A}_{0}\right)Q-\frac{3}{2}\Pi_{0}\Sigma-\left(\mu_{0}+p_{0}\right)\theta-\delta_{\mu}Q^{\mu}\,,
\end{equation}
\begin{equation}
\begin{aligned}\ddot{Q}+\widehat{\mathsf{p}}+\widehat{\mathcal{P}} & =\left(\widehat{p}_{0}+\widehat{\Pi}_{0}\right)\left(\frac{1}{3}\theta+\Sigma\right)-\Pi_{0}\left(\frac{3}{2}\mathsf{F}+\mathsf{A}\right)-\left(\frac{3}{2}\phi_{0}+2\mathcal{A}_{0}\right)\mathcal{P}\\
 & -\mathcal{A}_{0}\left(\mathsf{m}+2\mathsf{p}\right)-\left(\mu_{0}+p_{0}\right)\mathsf{A}-\left(\delta_{\mu}\Pi^{\mu}\right)^{.}\,;
\end{aligned}
\end{equation}

\item Evolution and propagation equations for vector quantities:
\begin{equation}
\begin{aligned}\dot{\mathcal{E}}_{\alpha}+\frac{1}{2}\dot{\Pi}_{\alpha} & =\left(\frac{1}{2}\phi_{0}-\mathcal{A}_{0}\right)\left(\frac{1}{2}Q_{\alpha}+\varepsilon_{\alpha\mu}\mathcal{H}^{\mu}\right)-\frac{3}{2}\left(\mathcal{E}_{0}+\frac{1}{2}\Pi_{0}\right)\alpha_{\alpha}\\
 & -\frac{1}{2}\left(\mu_{0}+p_{0}+\Pi_{0}\right)\left(\Sigma_{\alpha}-\varepsilon_{\alpha\mu}\Omega^{\mu}\right)\\
 & -\frac{1}{2}\delta_{\alpha}Q+\frac{1}{2}\varepsilon_{\alpha}{}^{\mu}\delta_{\mu}\mathcal{H}+\varepsilon^{\mu\nu}\delta_{\mu}\mathcal{H}_{\nu\alpha}\,,
\end{aligned}
\end{equation}

\begin{equation}
\begin{aligned}\widehat{\mathcal{E}}_{\alpha}+\frac{1}{2}\widehat{\Pi}_{\alpha} & =-\frac{3}{2}\phi_{0}\left(\mathcal{E}_{\alpha}+\frac{1}{2}\Pi_{\alpha}\right)-\frac{3}{2}\left(\mathcal{E}_{0}+\frac{1}{2}\Pi_{0}\right)a_{\alpha}\\
 & +\frac{1}{2}\mathbb{E}_{\alpha}+\frac{1}{3}\mathfrak{m}_{\alpha}+\frac{1}{4}\mathbb{P}_{\alpha}-\delta_{\mu}\mathcal{E}_{\alpha}{}^{\mu}-\frac{1}{2}\delta_{\mu}\Pi_{\alpha}{}^{\mu}\,,
\end{aligned}
\end{equation}

\begin{equation}
\begin{aligned}\frac{1}{2}\varepsilon_{\alpha}{}^{\mu}\widehat{\mathcal{E}}_{\mu}-\dot{\mathcal{H}}_{\alpha}-\frac{1}{4}\varepsilon_{\alpha}{}^{\mu}\widehat{\Pi}_{\mu} & =\frac{3}{4}\varepsilon_{\alpha}{}^{\mu}\mathbb{E}_{\mu}-\frac{3}{8}\varepsilon_{\alpha}{}^{\mu}\mathbb{P}_{\mu}+\frac{1}{2}\varepsilon^{\mu\nu}\delta_{\mu}\mathcal{E}_{\nu\alpha}-\frac{1}{4}\varepsilon^{\mu\nu}\delta_{\mu}\Pi_{\nu\alpha}\\
 & +\frac{3}{2}\varepsilon_{\alpha\mu}\mathcal{E}_{0}\mathcal{A}^{\mu}-\frac{3}{4}\varepsilon_{\alpha\mu}\left(\mathcal{E}_{0}-\frac{1}{2}\Pi_{0}\right)a^{\mu}-\varepsilon_{\alpha\mu}\mathcal{E}^{\mu}\mathcal{A}\\
 & -\frac{1}{4}\varepsilon_{\alpha\mu}\phi_{0}\left(\mathcal{E}^{\mu}-\frac{1}{2}\Pi^{\mu}\right)\,,
\end{aligned}
\end{equation}

\begin{equation}
\begin{aligned}\widehat{\mathcal{H}}_{\alpha}-\frac{1}{2}\varepsilon_{\alpha}{}^{\mu}\widehat{Q}_{\mu} & =\frac{1}{2}\delta_{\alpha}\mathcal{H}-\delta_{\mu}\mathcal{H}_{\alpha}{}^{\mu}-\frac{1}{2}\varepsilon_{\alpha}{}^{\mu}\delta_{\mu}Q-\frac{3}{2}\varepsilon_{\alpha\mu}\left(\mathcal{E}_{0}+\frac{1}{2}\Pi_{0}\right)\Sigma^{\mu}\\
 & -\left(\mu_{0}+p_{0}+\frac{1}{4}\Pi_{0}-\frac{3}{2}\mathcal{E}_{0}\right)\Omega_{\alpha}-\frac{3}{2}\phi_{0}\left(\mathcal{H}_{\alpha}-\frac{1}{6}\varepsilon_{\alpha\mu}Q^{\mu}\right)\,,
\end{aligned}
\end{equation}

\begin{equation}
\dot{Q}_{\alpha}+\widehat{\Pi}_{\alpha} =
-\mathfrak{p}_{\alpha}+\frac{1}{2}\mathbb{P}_{\alpha}-
\delta_{\mu}\Pi_{\alpha}{}^{\mu}-\frac{3}{2}\Pi_{0}a_{\alpha}
-\left(\frac{3}{2}\phi_{0}+\mathcal{A}_{0}\right)\Pi_{\alpha}-
\left(\mu_{0}+p_{0}-\frac{1}{2}\Pi_{0}\right)\mathcal{A}_{\alpha}\,;
\end{equation}

\item Evolution and propagation equations for tensor quantities:
\begin{equation}
\begin{aligned}\dot{\mathcal{E}}_{\alpha\beta}+\frac{1}{2}\dot{\Pi}_{\alpha\beta}+\varepsilon_{\{\alpha|}{}^{\nu}\widehat{\mathcal{H}}_{|\beta\}\nu} & =\varepsilon_{\{\alpha|}{}^{\nu}\delta_{\nu}\mathcal{H}_{|\beta\}}-\frac{1}{2}\left(\mu_{0}+p_{0}+3\mathcal{E}_{0}-\frac{1}{2}\Pi_{0}\right)\Sigma_{\alpha\beta}\\
 & +\left(\frac{1}{2}\phi_{0}+2\mathcal{A}_{0}\right)\varepsilon_{\mu\{\alpha}\mathcal{H}_{\beta\}}{}^{\mu}-\frac{1}{2}\delta_{\{\alpha}Q_{\beta\}}\,,
\end{aligned}
\end{equation}

\begin{equation}
\begin{aligned}\varepsilon_{\{\alpha|}{}^{\nu}\widehat{\mathcal{E}}_{|\beta\}\nu}-\frac{1}{2}\varepsilon_{\{\alpha|}{}^{\nu}\widehat{\Pi}_{|\beta\}\nu}-\dot{\mathcal{H}}_{\left\{ \alpha\beta\right\} } & =\varepsilon_{\{\alpha|}{}^{\mu}\delta_{\mu}\mathcal{E}_{|\beta\}}-\frac{1}{2}\varepsilon_{\{\alpha|}{}^{\mu}\delta_{\mu}\Pi_{|\beta\}}-\frac{1}{4}\phi_{0}\varepsilon_{\mu\{\alpha}\Pi_{\beta\}}{}^{\mu}\\
 & -\frac{3}{2}\left(\mathcal{E}_{0}-\frac{1}{2}\Pi_{0}\right)\varepsilon_{\mu\{\alpha}\zeta_{\beta\}}{}^{\mu}+\left(\frac{1}{2}\phi_{0}+2\mathcal{A}_{0}\right)\varepsilon_{\mu\{\alpha}\mathcal{E}_{\beta\}}{}^{\mu}\,.
\end{aligned}
\end{equation}

\end{itemize}\hfill\hfill

\subsection{Extra equations}

The variables $\left\{ \mathfrak{m},\mathfrak{p},\mathbb{P},\mathbb{F},\mathbb{A},\mathbb{E}\right\} $
and $\left\{ \mathsf{m},\mathsf{p},\mathcal{P},\mathsf{F},\mathsf{A},\mathsf{E}\right\} $
are not fully independent. Using the commutation relations~(\ref{1p1p2_eqs:Commutation_relations_scalars})
we find
\begin{equation}\label{Var_Sets_Conn}
\begin{aligned}\delta_{\alpha}\mathsf{m}-\dot{\mathfrak{m}}_{\alpha} & =\widehat{\mu}_{0}\left(\Sigma_{\alpha}-\varepsilon_{\alpha\mu}\Omega^{\mu}+\alpha_{\alpha}\right)\,,\\
\delta_{\alpha}\mathsf{p}-\dot{\mathfrak{p}}_{\alpha} & =\widehat{p}_{0}\left(\Sigma_{\alpha}-\varepsilon_{\alpha\mu}\Omega^{\mu}+\alpha_{\alpha}\right)\,,\\
\delta_{\alpha}\mathcal{P}-\dot{\mathbb{P}}_{\alpha} & =\widehat{\Pi}_{0}\left(\Sigma_{\alpha}-\varepsilon_{\alpha\mu}\Omega^{\mu}+\alpha_{\alpha}\right)\,,\\
\delta_{\alpha}\mathsf{F}-\dot{\mathbb{F}}_{\alpha} & =\widehat{\phi}_{0}\left(\Sigma_{\alpha}-\varepsilon_{\alpha\mu}\Omega^{\mu}+\alpha_{\alpha}\right)\,,\\
\delta_{\alpha}\mathsf{A}-\dot{\mathbb{A}}_{\alpha} & =\widehat{\mathcal{A}}_{0}\left(\Sigma_{\alpha}-\varepsilon_{\alpha\mu}\Omega^{\mu}+\alpha_{\alpha}\right)\,,\\
\delta_{\alpha}\mathsf{E}-\dot{\mathbb{E}}_{\alpha} & =\widehat{\mathcal{E}}_{0}\left(\Sigma_{\alpha}-\varepsilon_{\alpha\mu}\Omega^{\mu}+\alpha_{\alpha}\right)\,.
\end{aligned}
\end{equation}
Moreover, we also have
\begin{equation}
\begin{aligned}\varepsilon^{\alpha\beta}\delta_{\alpha}\mathfrak{m}_{\beta} & =-2\widehat{\mu}_{0}\xi\,,\\
\varepsilon^{\alpha\beta}\delta_{\alpha}\mathfrak{p}_{\beta} & =-2\widehat{p}_{0}\xi\,,\\
\varepsilon^{\alpha\beta}\delta_{\alpha}\mathbb{P}_{\beta} & =-2\widehat{\Pi}_{0}\xi\,,\\
\varepsilon^{\alpha\beta}\delta_{\alpha}\mathbb{F}_{\beta} & =-2\widehat{\phi}_{0}\xi\,,\\
\varepsilon^{\alpha\beta}\delta_{\alpha}\mathbb{A}_{\beta} & =-2\widehat{\mathcal{A}}_{0}\xi\,,\\
\varepsilon^{\alpha\beta}\delta_{\alpha}\mathbb{E}_{\beta} & =-2\widehat{\mathcal{E}}_{0}\xi\,.
\end{aligned}
\end{equation}

\section{\label{Appendix:Harmonics_properties}Scalar, vector and tensor harmonics}

We list some of the properties of the scalar, vector, and tensor eigenfunctions
of the covariantly defined Laplace-Beltrami operator on 2-hypersurfaces.
For concreteness, we consider a locally rotationally symmetric of
class II spacetime, such that it admits the existence of spatial sections
that can be described by 2-hypersurfaces.

\subsection{Scalar harmonics}

Let $\mathcal{Q}^{\left(k\right)}$ represent the scalar eigenfunctions
of the covariantly defined Laplace-Beltrami operator $\delta^{2}\equiv\delta^{\mu}\delta_{\mu}$,
where the $\delta$ operator was defined in Eq.~(\ref{Def_eq:delta_operator_definition}),
such that
\begin{equation}
\begin{aligned}\delta^{2}\mathcal{Q}^{\left(k\right)} & =-\frac{k^{2}}{r^{2}}\mathcal{Q}^{\left(k\right)}\,,\\
\widehat{\mathcal{Q}}^{\left(k\right)}=\dot{\mathcal{Q}}^{\left(k\right)} & =0\,,
\end{aligned}
\end{equation}
where we have assumed that the harmonic index verifies $k^{2}\geq0$
and the function $r$ is covariantly defined as
\begin{equation}
\begin{aligned}\frac{\hat{r}}{r} & =\frac{1}{2}\phi\,,\\
\frac{\dot{r}}{r} & =\frac{1}{3}\theta-\frac{1}{2}\Sigma\,,\\
\delta_{\alpha}r & =0\,,
\end{aligned}
\label{Harmonics_eq:r_definitions}
\end{equation}
or, using the 1+1+2 equations in Appendix~\ref{Appendix:General_1p1p2_eqs}
in the particular case of an LRS II spacetime, we have:
\begin{equation}
\frac{1}{r^{2}}=\frac{\phi^{2}}{4}-\mathcal{E}+\frac{1}{3}\left(\mu+\Lambda\right)-\frac{\Pi}{2}-\left(\frac{1}{3}\theta-\frac{1}{2}\Sigma\right)^{2}\,.\label{Harmonics_eq:r_definitions_algebraic}
\end{equation}

\subsection{Vector harmonics}

From the scalar harmonics $\mathcal{Q}^{\left(k\right)}$, we may
define a complete set of 1-tensors that can be used as a basis for
sufficiently smooth 1-tensor fields defined on the 2-hypersurfaces. The vector
harmonics are defined as
\begin{equation}
\begin{aligned}\mathcal{Q}_{\alpha}^{\left(k\right)} & =r\delta_{\alpha}\mathcal{Q}^{\left(k\right)}\,,\\
\bar{\mathcal{Q}}_{\alpha}^{\left(k\right)} & =r\varepsilon_{\alpha\mu}\delta^{\mu}\mathcal{Q}^{\left(k\right)}\,,
\end{aligned}
\end{equation}
where $\mathcal{Q}_{\alpha}^{\left(k\right)}$ are referred as the
``even'' vector harmonics and $\bar{\mathcal{Q}}_{\alpha}^{\left(k\right)}$
as the ``odd'' vector harmonics. These have the following
properties
\begin{equation}
\begin{aligned}
u^{\mu}\nabla_{\mu}{\mathcal{Q}}_{\alpha}^{\left(k\right)} & =0\,,\\
e^{\mu}D_{\mu}{\mathcal{Q}}_{\alpha}^{\left(k\right)} & =0\,,\\
\delta^{\mu}\mathcal{Q}_{\mu}^{\left(k\right)} & =-\frac{k^{2}}{r}\mathcal{Q}^{\left(k\right)}\,,\\
\varepsilon^{\mu\nu}\delta_{\mu}\mathcal{Q}_{\nu}^{\left(k\right)} & =0\,,\\
\delta^{2}\mathcal{Q}_{\alpha}^{\left(k\right)} & =\frac{1-k^{2}}{r^{2}}\mathcal{Q}_{\alpha}^{\left(k\right)}\,;
\end{aligned}
\end{equation}
and 
\begin{equation}
\begin{aligned}
u^{\mu}\nabla_{\mu}\bar{\mathcal{Q}}_{\alpha}^{\left(k\right)} & =0\,,\\
e^{\mu}D_{\mu}\bar{\mathcal{Q}}_{\alpha}^{\left(k\right)} & =0\,,\\
\delta^{\mu}\bar{\mathcal{Q}}_{\mu}^{\left(k\right)} & =0\,,\\
\varepsilon^{\mu\nu}\delta_{\mu}\bar{\mathcal{Q}}_{\nu}^{\left(k\right)} & =\frac{k^{2}}{r}\mathcal{Q}^{\left(k\right)}\,,\\
\delta^{2}\bar{\mathcal{Q}}_{\alpha}^{\left(k\right)} & =\frac{1-k^{2}}{r^{2}}\bar{\mathcal{Q}}_{\alpha}^{\left(k\right)}\,.
\end{aligned}
\end{equation}
Moreover, we have
\begin{equation}
\begin{aligned}\bar{Q}_{\alpha}^{\left(k\right)} & =\varepsilon_{\alpha}{}^{\mu}Q_{\mu}^{\left(k\right)}\,,\\
Q_{\alpha}^{\left(k\right)} & =-\varepsilon_{\alpha}{}^{\mu}\bar{Q}_{\mu}^{\left(k\right)}\,,
\end{aligned}
\end{equation}
which can be used to readily verify $N^{\mu\nu}\mathcal{Q}_{\mu}^{\left(k\right)}\bar{\mathcal{Q}}_{\nu}^{\left(k\right)}=0$,
valid for each value of $k$.

\subsection{Tensor harmonics}

The scalar and vector harmonics introduced in the previous subsection
can themselves be used to define a complete set of symmetric, traceless
2-tensors that form a basis for sufficiently smooth, symmetric, traceless
2-tensor fields defined on the 2-hypersurfaces. The tensor harmonics are defined
as
\begin{equation}
\begin{aligned}\mathcal{Q}_{\alpha\beta}^{\left(k\right)} & =r^{2}\delta_{\{\alpha}\delta_{\beta\}}\mathcal{Q}^{\left(k\right)}\,,\\
\bar{\mathcal{Q}}_{\alpha\beta}^{\left(k\right)} & =
r^{2}\varepsilon_{\mu\{\alpha|}\delta^{\mu}\delta_{|\beta\}}\mathcal{Q}^{\left(k\right)}\,,
\end{aligned}
\end{equation}
where, similarly to the vector harmonics above, $\mathcal{Q}_{\alpha\beta}^{\left(k\right)}$
are referred as the ``even'' tensor harmonics and $\bar{\mathcal{Q}}_{\alpha\beta}^{\left(k\right)}$
are referred to as the ``odd'' tensor harmonics. These have the following
properties
\begin{equation}
\begin{aligned}
u^{\mu}\nabla_{\mu}{\mathcal{Q}}_{\alpha\beta}^{\left(k\right)} & =0\,,\\
e^{\mu}D_{\mu}{\mathcal{Q}}_{\alpha\beta}^{\left(k\right)} & =0\,,\\
\delta^{\beta}\mathcal{Q}_{\alpha\beta}^{\left(k\right)} & =\frac{2-k^{2}}{2r}\mathcal{Q}_{\alpha}^{\left(k\right)}\,,\\
\varepsilon^{\mu\nu}\delta_{\mu}\mathcal{Q}_{\nu\alpha}^{\left(k\right)} & =\frac{2-k^{2}}{2r}\bar{\mathcal{Q}}_{\alpha}^{\left(k\right)}\,;
\end{aligned}
\end{equation}
and 
\begin{equation}
\begin{aligned}
u^{\mu}\nabla_{\mu}\mathcal{\bar{Q}}_{\alpha\beta}^{\left(k\right)} & =0\,,\\
e^{\mu}D_{\mu}\mathcal{\bar{Q}}_{\alpha\beta}^{\left(k\right)} & =0\,,\\
\delta^{\mu}\mathcal{\bar{Q}}_{\alpha\mu}^{\left(k\right)} & =-\frac{2-k^{2}}{2r}\bar{\mathcal{Q}}_{\alpha}^{\left(k\right)}\,,\\
\varepsilon^{\mu\nu}\delta_{\mu}\mathcal{\bar{Q}}_{\nu\alpha}^{\left(k\right)} & =\frac{2-k^{2}}{2r}\mathcal{Q}_{\alpha}^{\left(k\right)}\,.
\end{aligned}
\end{equation}
Moreover, we have
\begin{equation}
\begin{aligned}\mathcal{Q}_{\alpha\beta}^{\left(k\right)} & =-\varepsilon^{\mu}{}_{\{\alpha}\bar{\mathcal{Q}}_{\beta\}\mu}^{\left(k\right)}\,,\\
\bar{\mathcal{Q}}_{\alpha\beta}^{\left(k\right)} & =\varepsilon^{\mu}{}_{\{\alpha}\mathcal{Q}_{\beta\}\mu}^{\left(k\right)}\,,
\end{aligned}
\end{equation}
which can be used to readily prove that $N^{\alpha\mu}N^{\beta\nu}\mathcal{Q}_{\alpha\beta}^{\left(k\right)}\bar{\mathcal{Q}}_{\mu\nu}^{\left(k\right)}=0$,
valid for each value of $k$.

\section{Change in the equation of state under isotropic frame transformations\label{Appendix:isotropic frame transformations}}
In this Appendix, we make use of generalized Lorentz boosts in the 1+1+2 formalism to derive the transformation law for the equation of state when changing between the comoving frame and the static frame.

\subsection{Isotropic frame transformations and projectors}

Let the dyad $\left(u,e\right)$ be formed, respectively, by a timelike
and a spacelike vector field in a spacetime, such that $u^{\alpha}u_{\alpha}=-1$
and $e^{\alpha}e_{\alpha}=1$, and let
\begin{equation}
\begin{aligned}\bar{u}^{\alpha} & =\left(1+a\right)u^{\alpha}+be^{\alpha}+m^{\alpha}\,,\\
\bar{e}^{\alpha} & =cu^{\alpha}+\left(1+d\right)e^{\alpha}+k^{\alpha}\,,
\end{aligned}
\end{equation}
represent the components in some local coordinate system of a dyad
$\left(\bar{u},\bar{e}\right)$, with $\left\{ a,b,c,d\right\} $
being scalar coefficients and
\begin{equation}
\begin{aligned}u^{\alpha}m_{\alpha} & =0\,, & u^{\alpha}k_{\alpha} & =0\,,\\
e^{\alpha}m_{\alpha} & =0\,, & e^{\alpha}k_{\alpha} & =0\,.
\end{aligned}
\end{equation}
In the expressions above, we can assume, without loss of generality, that the vector fields $m$ and $k$
are associated with rotational displacements. Then, for our purposes,
it suffices to consider that the frame transformation is isotropic,
such that $m^{\alpha}=0$ and $k^{\alpha}=0$.

Imposing that the new dyad verifies
\begin{equation}
\begin{aligned}\bar{u}^{\alpha}\bar{u}_{\alpha} & =-1\,,\\
\bar{e}^{\alpha}\bar{e}_{\alpha} & =+1\,,\\
\bar{u}^{\alpha}\bar{e}_{\alpha} & =0\,,
\end{aligned}
\end{equation}
we find that the transformation coefficients verify
\begin{equation}
\begin{aligned}\left(1+d\right)^{2} & =\left(1+a\right)^{2}\\
\left(1+a\right)^{2} & =1+c^{2}\\
b^{2} & =c^{2}\,.
\end{aligned}
\label{Isotropic_frame_transformations_eq:frame_coefficients_relation}
\end{equation}
Assuming, $b,c>0$, hence $a,d>0$, that is, the transformation does
not swap the direction between the corresponding vector fields, i.e., imposing that
$u^{\alpha}\bar{u}_{\alpha}\leq0$ and $e^{\alpha}\bar{e}_{\alpha}\geq0$,
we see that the relation between $a$ and $c$ can be made obvious
by introducing a single hyperbolic angle $\beta$ such that
\begin{equation}
\begin{aligned}1+a & =\cosh\beta\,,\\
c & =\sinh\beta\,.
\end{aligned}
\end{equation}
Then, we have the following relation between the two dyads 
\begin{equation}
\begin{aligned}\bar{u}^{\alpha} & =u^{\alpha}\cosh\beta+e^{\alpha}\sinh\beta\,,\\
\bar{e}^{\alpha} & =u^{\alpha}\sinh\beta+e^{\alpha}\cosh\beta\,.
\end{aligned}
\label{Isotropic_frame_transformations_eq:dyads_relation_simp}
\end{equation}

Given relations~(\ref{Isotropic_frame_transformations_eq:dyads_relation_simp}),
the projectors $h$ and $N$ transform accordingly as~
\begin{align}
\bar{h}_{\alpha\beta} & =h_{\alpha\beta}+\left(u_{\alpha}u_{\beta}+e_{\alpha}e_{\beta}\right)\sinh^{2}\beta+\frac{1}{2}\left(u_{\alpha}e_{\beta}+e_{\alpha}u_{\beta}\right)\sinh\left(2\beta\right)\,,\\
\bar{N}_{\alpha\beta} & =N_{\alpha\beta}\,,
\end{align}
which verify $\bar{h}_{\alpha\beta}\bar{u}^{\beta}=0$, $\bar{N}_{\alpha\beta}\bar{u}^{\beta}=0$
and $\bar{N}_{\alpha\beta}\bar{e}^{\beta}=0$, as expected.

\subsection{Fluid variables transformations}

Consider now a general symmetric stress-energy tensor decomposed accordingly
with Eq.~(\ref{Def_eq:Stress-energy_tensor_decomposition}). Under
the transformation~(\ref{Isotropic_frame_transformations_eq:dyads_relation_simp})
we have~~~~
\begin{equation}
\begin{aligned}\bar{\mu} & =\mu-Q\sinh\left(2\beta\right)+\left(\mu+p+\Pi\right)\sinh^{2}\beta\,,\\
\bar{p} & =p-\frac{1}{3}Q\sinh\left(2\beta\right)+\frac{1}{3}\left(\mu+p+\Pi\right)\sinh^{2}\beta\,,\\
\bar{Q} & =Q\cosh\left(2\beta\right)-\frac{1}{2}\left(\mu+p+\Pi\right)\sinh\left(2\beta\right)\,,\\
\bar{\Pi} & =\Pi\left(1+\frac{2}{3}\sinh^{2}\beta\right)-\frac{2}{3}Q\sinh\left(2\beta\right)+\frac{2}{3}\left(\mu+p\right)\sinh^{2}\beta\,,
\end{aligned}
\label{Isotropic_frame_transformations_eq:mu_p_Q_frame_transformations_general}
\end{equation}
where we have used the intuitive notation of keeping an overline to
indicate variables measured in the frame associated with the dyad
$\left(\bar{u},\bar{e}\right)$. Moreover, we have the following relations
between the scalar kinematical quantities~~
\begin{equation}
\begin{aligned}\bar{\theta} & =\theta\cosh\beta+\left(\phi+\mathcal{A}\right)\sinh\beta+\nabla_{u}\cosh\beta+\nabla_{e}\sinh\beta\,,\\
\bar{\Sigma} & =\Sigma\cosh\beta-\frac{1}{3}\left(\phi-2\mathcal{A}\right)\sinh\beta+\frac{2}{3}\nabla_{u}\cosh\beta+\frac{2}{3}\nabla_{e}\sinh\beta\,,
\end{aligned}
\end{equation}
where $\nabla_{u}\equiv u^{\alpha}\nabla_{\alpha}$ and $\nabla_{e}\equiv e^{\alpha}\nabla_{\alpha}$.

\subsection{Relation between equations of state of the comoving and static observers}

The comoving and static observer frames considered in Sec.~\ref{sec:Adiabatic-isotropic-perturbations}
are related by an isotropic change of frame. Then, we can use the
results in the previous subsections of this appendix to find the equation
of state experienced by a static observer, knowing the equation of
state measured by the comoving observer.

Assuming the dyad $\left(u,e\right)$ to be associated to the comoving
frame and the dyad $\left(\bar{u},\bar{e}\right)$ to be associated
with the static frame, we can particularize Eq\@.~(\ref{Isotropic_frame_transformations_eq:mu_p_Q_frame_transformations_general})
to the case when $Q=0$ and $\Pi=0$, such that~

\begin{equation}
\begin{aligned}\bar{\mu} & =\mu+\left(\mu+p\right)\sinh^{2}\beta\,,\\
\bar{p} & =p+\frac{1}{3}\left(\mu+p\right)\sinh^{2}\beta\,,\\
\bar{Q} & =-\frac{1}{2}\left(\mu+p\right)\sinh\left(2\beta\right)\,,\\
\bar{\Pi} & =\frac{2}{3}\left(\mu+p\right)\sinh^{2}\beta\,,
\end{aligned}
\label{Isotropic_frame_transformations_eq:mu_p_Q_frame_transformations_particular}
\end{equation}
relating the nontrivial matter variables of the two frames. Moreover,
since in the static frame --- the one associated with the barred
variables --- we have $\frac{3}{2}\bar{\Sigma}=\bar{\theta}$, then~~
\begin{equation}
\tanh\beta=\frac{1}{\phi}\left(\Sigma-\frac{2}{3}\theta\right)\,,\label{Isotropic_frame_transformations_eq:c_Sigma_theta_relation_general}
\end{equation}
which upon substitution in Eq.~(\ref{Isotropic_frame_transformations_eq:mu_p_Q_frame_transformations_particular})
yields
\begin{equation}
\bar{Q}=-\frac{\mu+p}{\phi}\left(\Sigma-\frac{2}{3}\theta\right)\cosh^{2}\beta\,.
\end{equation}

Assuming the background spacetime to be static and LRS II, Eq.~(\ref{Isotropic_frame_transformations_eq:c_Sigma_theta_relation_general})
explicitly shows that $\beta$ is a first-order quantity with respect
to the background, i.e., $\beta$ vanishes in the background. This
is expected since comoving observers are also static observers for a static LRS II spacetime. Moreover, these results confirm the expectation
that the anisotropic pressure measured by the static observer, $\bar{\Pi}$,
is zero at linear level for adiabatic, radial perturbations.

We can now find the linear order correction with respect to the background
of the equation of state in the static frame. Taking the derivative
along $\bar{u}$ of $\bar{p}$ and $\bar{\mu}$, and using Eqs.~(\ref{Isotropic_frame_transformations_eq:dyads_relation_simp})
and (\ref{Isotropic_frame_transformations_eq:mu_p_Q_frame_transformations_particular})
we find, in general, ~
\begin{equation}
\begin{aligned}\nabla_{\bar{u}}\bar{\mu} & =\dot{\mu}\cosh\beta+\widehat{\mu}\sinh\beta+\left[\left(\nabla_{u}\sinh^{2}\beta\right)\left(\mu+p\right)+\sinh^{2}\beta\left(\dot{\mu}+\dot{p}\right)\right]\cosh\beta\\
 & +\left[\left(\nabla_{e}\sinh^{2}\beta\right)\left(\mu+p\right)+\sinh^{2}\beta\left(\widehat{\mu}+\widehat{p}\right)\right]\sinh\beta\,,\\
\nabla_{\bar{u}}\bar{p} & =\dot{p}\cosh\beta+\widehat{p}\sinh\beta+\frac{1}{3}\left[\left(\nabla_{u}\sinh^{2}\beta\right)\left(\mu+p\right)+\sinh^{2}\beta\left(\dot{\mu}+\dot{p}\right)\right]\cosh\beta\\
 & +\frac{1}{3}\left[\left(\nabla_{e}\sinh^{2}\beta\right)\left(\mu+p\right)+\sinh^{2}\beta\left(\widehat{\mu}+\widehat{p}\right)\right]\sinh\beta\,,
\end{aligned}
\label{Isotropic_frame_transformations_eq:relation_static_comoving_matter_variables_general}
\end{equation}
where $\dot{\mu}=u^{\alpha}\nabla_{\alpha}\mu$ and $\widehat{\mu}=e^{\alpha}\nabla_{\alpha}\mu$,
and similarly for $p$. Using the fact that $\beta$ is a first-order
quantity with respect to the background, in first-order perturbation
theory Eq.~(\ref{Isotropic_frame_transformations_eq:relation_static_comoving_matter_variables_general})
simplifies to

\begin{equation}
\begin{aligned}
	\nabla_{\bar{u}}\bar{\mu} & =\dot{\mu}+\widehat{\mu}_{0}\beta\,,\\
	\nabla_{\bar{u}}\bar{p} & =\dot{p}+\widehat{p}_{0}\beta\,.
\end{aligned}
\label{Isotropic_frame_transformations_eq:relation_static_comoving_matter_variables_final}
\end{equation}
Also, from Eq.~(\ref{Isotropic_frame_transformations_eq:mu_p_Q_frame_transformations_particular}),
at first order, we have 
\begin{equation}
\beta=-\frac{\bar{Q}}{\mu_{0}+p_{0}}\,.
\end{equation}

Gathering the intermediate results and using the equations that characterize
the background solution (cf. subsection~\ref{subsec:Background_spacetime_static_LRSII_perfect_fluid})
we find the transformed equation of state for the static frame
\begin{equation}
\nabla_{\bar{u}}\bar{\mu}=\frac{1}{f'\left(\mu_{0}\right)}\nabla_{\bar{u}}\bar{p}-
\frac{1}{\mu_{0}+p_{0}}\left(\widehat{\mu}_{0}-
\frac{\widehat{p}_{0}}{f'\left(\mu_{0}\right)}\right)Q\,,
\end{equation}
where $\dot{p}=f'\left(\mu_{0}\right)\dot{\mu}$, that is, $f'\left(\mu_{0}\right)$
represents the square of the adiabatic speed of sound measured in the comoving
frame.

\end{document}